\documentclass[10pt,letterpaper]{article}
\pdfoutput=1 

\usepackage[utf8]{inputenc} 
\usepackage{lipsum} 

\usepackage[top=2.5cm, bottom=3cm, left=3.5cm, right=3.5cm,
           heightrounded,marginparwidth=1.5cm, marginparsep=1cm]{geometry} 

\usepackage{changepage} 

\usepackage[shortlabels]{enumitem} 

\usepackage[square,numbers,merge,comma,sort&compress]{natbib} 
\makeatletter
\def\NAT@spacechar{\,}  
\makeatother

\usepackage{amsmath,amssymb,amsfonts,amsthm,amsbsy}
\usepackage{mathtools} 

\usepackage{fnpct}
\setfnpct{dont-mess-around} 
\usepackage{slashed,cancel}
\usepackage{old-arrows} 
\usepackage{comment}   
\usepackage{relsize}   
\usepackage{setspace}  
\usepackage{moresize}  
\usepackage{epsfig}
\usepackage{latexsym}
\usepackage{mathrsfs,calligra,aurical,bold-extra} 
\usepackage{calc}
\usepackage{float}
\usepackage{appendix}
\usepackage{bigints}
\usepackage{xargs}
\usepackage{extarrows}
\usepackage{empheq}
\usepackage{soul} 

\usepackage[svgnames]{xcolor}  
\definecolor{Green}{rgb}{0.05, 0.45, 0.25}
\xdefinecolor{dogwoodrose}{rgb}{0.8, 0.1, 0.55}
\xdefinecolor{RRed}{rgb}{0.7, 0.1, 0.525}

\usepackage{tikz}
\usetikzlibrary{scopes,decorations.pathmorphing,patterns,calc,arrows,
                shapes.geometric,shapes.arrows,decorations.markings,plotmarks}
\usepackage{pgfplots}

\usepackage[blocks]{authblk}  

\usepackage[fulladjust]{marginnote}%

\usepackage{graphicx} 
\graphicspath{{Figures/}} 

\usepackage[labelsep=colon]{caption}  
\usepackage[labelsep=colon,aboveskip=10pt,belowskip=10pt]{subcaption}
\DeclareCaptionSubType * [roman]{table}
\usepackage{array,multirow,makecell,booktabs}  
\newcolumntype{M}[2]{>{\centering\arraybackslash$}#1{#2\linewidth}<{$}}
\newcolumntype{T}[2]{>{\centering\arraybackslash}#1{#2\linewidth}<{}}
\newcolumntype{R}[1]{>{\raggedleft\arraybackslash}m{#1\linewidth}<{}}
\newcolumntype{L}[1]{>{\raggedright\arraybackslash}m{#1\linewidth}<{}}

\makeatletter
\renewcommand\mcell@classz{\@classx
   \@tempcnta \count@
   \prepnext@tok
   \@addtopreamble{
      \ifcase\@chnum
         \hfil
         \mcell@agape{\d@llarbegin\insert@column\d@llarend}\hfil \or
         \hskip1sp
         \mcell@agape{\d@llarbegin\insert@column\d@llarend}\hfil \or
         \hfil\hskip1sp
         \mcell@agape{\d@llarbegin \insert@column\d@llarend}\or
         \mcell@agape{$\vcenter
         \@startpbox{\@nextchar}\insert@column\@endpbox$}\or
         \mcell@agape{\vtop
         \@startpbox{\@nextchar}\insert@column\@endpbox}\or
         \mcell@agape{\vbox
         \@startpbox{\@nextchar}\insert@column\@endpbox}%
      \fi
      \global\let\mcell@left\relax\global\let\mcell@right\relax
    }\prepnext@tok}
\makeatother

\usepackage{titlesec}

\titleformat{\section}{\normalfont\fontsize{11}{11}\bfseries}{\thesection}{0.5em}{}
\titleformat{\subsection}{\normalfont\normalsize\bfseries}{\thesubsection}{0.5em}{}
\titleformat{\subsubsection}{\normalfont\normalsize\bfseries}{\thesubsubsection}{0.5em}{}

\titlespacing*{\section}{0pt}%
                {4ex plus 1ex minus .5ex}{1.75ex plus .25ex minus .25ex}
\titlespacing*{\subsection}{0pt}%
                {3.5ex plus 1ex minus .5ex}{1.25ex plus .2ex minus .2ex}
\titlespacing*{\subsubsection}{0pt}%
                {2.5ex plus 0.75ex minus .2ex}{0.75ex plus .15ex minus .15ex}
\titlespacing*{\paragraph}{0pt}%
                {1.85ex plus 0.5ex minus .15ex}{1em}

\usepackage{titletoc}

\addtocontents{toc}{\addvspace{-0.75em}}  

\titlecontents{section}
  [1.25em] {\addvspace{0.7em plus 0pt}\small}
  {\thecontentslabel\hspace{0.75em}}{}
  {\hspace{0.5em}\titlerule*[0.5em]{.}\contentspage}
  [\addvspace{0.0em plus 0pt}]

\titlecontents{subsection}
  [2.75em] {\addvspace{0.075em plus0pt}\fns}
  {\thecontentslabel\hspace{0.75em}}{\thecontentslabel\hspace{0.75em}}
  {\hspace{0.5em}\titlerule*[0.5em]{.}\small\contentspage}
  [\addvspace{0.075em plus 0pt}]

\usepackage{environ}
\makeatletter
\NewEnviron{subalign}[1]{
\begin{subequations}\label{#1}
\begin{align} \BODY \end{align}
\end{subequations}      }
\makeatother
\makeatletter
\newenvironment{subeqs}%
{\begingroup%
\setlength{\abovedisplayskip}{10pt plus 4pt minus 9pt}%
\setlength{\abovedisplayshortskip}{0pt plus 2pt minus 2pt}%
\setlength{\belowdisplayskip}{12pt plus 3pt minus 9pt}%
\setlength{\belowdisplayshortskip}{7pt plus 3pt minus 4pt}%
\begin{subequations}%
}%
{\end{subequations}\ignorespacesafterend%
\endgroup}%
\makeatother

\makeatletter
{\begingroup%
\setlength{\abovedisplayskip}{{#1}pt plus 3pt minus 3pt}%
\setlength{\abovedisplayshortskip}{{#2}pt plus 3pt}%
\setlength{\belowdisplayskip}{{#3}pt plus 3pt minus 3pt}%
\setlength{\belowdisplayshortskip}{{#4}pt plus 3pt minus 3pt}%
\begin{subequations}%
}%
{\end{subequations}\ignorespacesafterend%
\endgroup}%
\makeatother

\makeatletter
{\begingroup%
\setlength{\abovedisplayskip}{{#1}pt plus 3pt minus 3pt}%
\setlength{\abovedisplayshortskip}{{#2}pt plus 3pt}%
\setlength{\belowdisplayskip}{{#3}pt plus 3pt minus 3pt}%
\setlength{\belowdisplayshortskip}{{#4}pt plus 3pt minus 3pt}%
\begin{equation}%
}%
{\end{equation}\ignorespacesafterend%
\endgroup}%
\makeatother

\makeatletter
{%
\begin{equation}%
\begin{split}%
}%
{\end{split}%
\end{equation}\ignorespacesafterend%
}%
\makeatother

\usepackage{bm}  
\usepackage{dsfont}  

\DeclareMathAlphabet{\mathpzc}{OT1}{pzc}{m}{it}
\DeclareMathAlphabet{\mathcal}{OMS}{cmsy}{m}{n}
\DeclareSymbolFontAlphabet{\Scr}{rsfs}
\DeclareMathAlphabet{\mathbold}{U}{BOONDOX-ds}{m}{n}
\SetMathAlphabet{\mathbold}{bold}{U}{BOONDOX-ds}{b}{n}
\DeclareMathAlphabet{\mathcalboondox}{U}{BOONDOX-calo}{m}{n}
\SetMathAlphabet{\mathcalboondox}{bold}{U}{BOONDOX-calo}{b}{n}
\DeclareMathAlphabet{\mathbcalboondox}{U}{BOONDOX-calo}{b}{n}

\DeclareFontFamily{U}{matha}{\hyphenchar\font45}
\DeclareFontShape{U}{matha}{m}{n}{ <5> <6> <7> <8> <9> <10> gen * matha
                    <10.95> matha10 <12> <14.4> <17.28> <20.74> <24.88> matha12}{}
\DeclareSymbolFont{matha}{U}{matha}{m}{n}
\DeclareMathSymbol{\varleftarrow}{3}{matha}{"D0}
\DeclareMathSymbol{\varrightarrow}{3}{matha}{"D1}
\DeclareMathSymbol{\simeq}{3}{matha}{"14}
\DeclareMathSymbol{\sim}{3}{matha}{"12}
\DeclareMathSymbol{\ll}{3}{matha}{"21}
\DeclareMathSymbol{\gtrsim}{3}{matha}{"C1}
\DeclareMathSymbol{\lesssim}{3}{matha}{"C0}

\newcommand\linkcol{RRed}



\usepackage[breaklinks=true,backref=page]{hyperref}
\hypersetup{
    bookmarks=true,         
    bookmarksnumbered=true,
    pdftoolbar=true,        
    pdfmenubar=true,        
    pdffitwindow=false,     
    pdfauthor={},     
    pdfsubject={},   
    pdfcreator={},   
    pdfproducer={},  
    pdfpagemode={UseNone},
    pdfstartview={FitH},
    colorlinks=true,
    plainpages,
    linktoc=page,
    citecolor=blue,
    filecolor=black,
    linkcolor=\linkcol,
    urlcolor=Green,
}
\renewcommand*{\backref}[1]{}
\renewcommand*{\backrefalt}[4]{%
\ifcase #1 %
\relax
\or
~{\small [\textsc{p.~\fns{\!#2}}]}
\else
~{\small [\textsc{p.~\fns{\!#2}}]}%
\fi}

\usepackage{footnotebackref}
\usepackage{hypernat} 

\usepackage{cleveref} 

\makeatletter
\normalsize
\makeatother

\def\+{~+~}
\def\-{~-~}
\def\={\:=\:}
\newcommand\fns{\footnotesize}

\newcommand\qqquad{\quad\quad\quad}

\newcommand\qRq{\quad\Rightarrow\quad}

\newcommand\Real{\textrm{Re}}
\newcommand\Img{\textrm{Im}}

\newcommand\eps{\varepsilon}
\newcommand\epsz{\varepsilon_\ms{0}}
\newcommand\epsg{\varepsilon_\textrm{g}}
\newcommand\mug{\mu_\textrm{g}}
\newcommand\muz{\mu_\ms{0}}
\newcommand\w{\omega}

\newcommand\hgamma{\bar{\gamma}}
\newcommand\Tc{T_\textrm{c}}
\newcommand\ux{\vec{u}_x}
\newcommand\uy{\vec{u}_y}
\newcommand\uz{\vec{u}_z}
\newcommand\E{\mathbf{E}}
\newcommand\Ee{\mathbf{E}_\textrm{\text{e}}}
\newcommand\Eg{\mathbf{E}_\textrm{\text{g}}}
\newcommand\Ae{\mathbf{A}_\textrm{\text{e}}}
\newcommand\B{\mathbf{B}}
\newcommand\Be{\mathbf{B}_\textrm{\text{e}}}
\newcommand\Bg{\mathbf{B}_\textrm{\text{g}}}
\newcommand\A{\mathbf{A}}
\newcommand\Ag{\mathbf{A}_\textrm{\text{g}}}
\newcommand\Ve{V_{{}_{\!}\textrm{e}}}
\newcommand\Vg{V_{{}_{\!}\textrm{g}}}
\newcommand\g{\mathbf{g}}
\newcommand\Bc{B_\textsc{c}}
\newcommand\Bcone{B_{\textsc{c}{}_{1}}}
\newcommand\Bctwo{B_{\textsc{c}{}_{2}}}

\newcommand\jj{\mathbf{j}}
\newcommand\jg{\mathbf{j}_\textrm{g}}
\newcommand\js{\mathbf{j}_\text{s}}

\newcommand\jsk{\mathbf{j}_{\text{s}k}}

\newcommand\kk{\mathbf{k}}
\newcommand\x{\mathbf{x}}
\newcommand\ns{n_\textrm{s}}
\newcommand\gstar{g_\star}

\newcommand\lambdag{\lambda_\textrm{g}}
\newcommand\lambdae{\lambda_\textrm{e}}
\newcommand\rhog{\rho_\textrm{g}}

\newcommand\vvs{\mathbf{v}_\text{s}}
\newcommand\vs{v_\text{s}}
\newcommand\mt{\mathrm{m}}
\newcommand\cm{\mathrm{cm}}
\newcommand\nm{\mathrm{nm}}
\newcommand\s{\mathrm{s}}
\newcommand\Kelv{\mathrm{K}}
\newcommand\GN{\mathrm{G}}
\newcommand\kB{\mathrm{k}_\ms{\textsc{b}}}

\newcommand\Ang{\text{\AA}}
\newcommand\cbox{c_{{}_\Box}}

\providecommand{\abs}[1]{\left\lvert#1\right\rvert}

\newcommand{\ms}{\mathsmaller}

\newcommand{\dd}{\partial}
\newcommandx{\tcR}[1]{\textcolor{Crimson}{#1}}
\newcommandx{\tts}[1]{\text{\textsmaller{#1}}}
\newcommandx{\mlt}[1]{\mathlarger{\text{#1}}}
\newcommandx{\dt}[1][1=f,usedefault]{\frac{\partial{#1}}{\partial t}}
\newcommandx{\dtau}[1][1=f,usedefault]{\frac{\partial{#1}}{\partial\tau}}
\newcommandx{\dx}[1][1=f,usedefault]{\frac{\partial{#1}}{\partial x}}
\newcommandx{\ddx}[1][1=f,usedefault]{\frac{\partial^2{#1}}{{\partial x}^2}}
\newcommandx{\dm}[1][1=\mu,usedefault]{\partial_{#1}}
\newcommandx{\dmup}[1][1=\mu,usedefault]{\partial^{#1}}
\newcommandx{\subm}[2][1=p,2=A,usedefault]{{#1}_{\!\mathsmaller{#2}}}
\newcommandx{\subt}[2][1=p,2=A,usedefault]{{#1}_\text{\textsmaller{#2}}}
\newcommandx{\supm}[2][1=p,2=A,usedefault]{{#1}^{\!\mathsmaller{#2}}}
\newcommandx{\supt}[2][1=p,2=A,usedefault]{{#1}^\text{\textsmaller{#2}}}
\newcommandx{\subpt}[3][1=p,2=A,3=B,usedefault]{{#1}^\text{\textsmaller{#3}}_\text{\textsmaller{#2}}}
\newcommandx{\subpm}[3][1=p,2=A,3=B,usedefault]{{#1}^{\mathsmaller{#3}}_{\mathsmaller{#2}}}
\newcommandx{\sh}[1][1=\alpha,usedefault]{\sinh\left(#1\right)}
\newcommandx{\ch}[1][1=\alpha,usedefault]{\cosh\left(#1\right)}
\newcommandx{\sech}[1][1=\alpha,usedefault]{\mathrm{sech}\left(#1\right)}
\newcommandx{\cosech}[1][1=\alpha,usedefault]{\mathrm{cosech}\left(#1\right)} \newcommandx{\LCTd}[4][1=\mu,2=\nu,3=\rho,4=\sigma,usedefault]{\eps_{#1#2#3#4}}
\newcommandx{\LCTu}[4][1=\mu,2=\nu,3=\rho,4=\sigma,usedefault]{\eps^{#1#2#3#4}}

\newcommandx{\gmetr}[2][1=\mu,2=\nu,usedefault]{g_{{#1}{#2}}}
\newcommandx{\invgmetr}[2][1=\mu,2=\nu,usedefault]{g^{{#1}{#2}}}
\newcommandx{\spc}[3][1=\mu,2=a,3=b,usedefault]{{\w_{#1}}^{\!\!{#2}{#3}}}
\newcommandx{\Conn}[3][1=\mu,2=\nu,3=\lambda,usedefault]{{\Gamma_{{#1}{#2}}}^{\!\!#3}}
\newcommandx{\viel}[2][1=\mu,2=a,usedefault]{{e_{#1}}^{\!#2}}
\newcommandx{\inviel}[2][1=a,2=\mu,usedefault]{{e_{#1}}^{#2}}
\newcommandx{\vieluu}[2][1=\mu,2=a,usedefault]{e^{#1#2}}
\newcommandx{\Rdduu}[4][1=\mu,2=\nu,3=a,4=b,usedefault]{{R_{{#1}{#2}}}^{{#3}{#4}}}
\newcommandx{\hgamui}[1][1=0,usedefault]{\hgamma^{\mathsmaller{#1}}}
\newcommandx{\hgamdi}[1][1=0,usedefault]{\hgamma_{{}_{#1}}}
\newcommandx{\gamui}[1][1=0,usedefault]{\gamma^{\mathsmaller{#1}}}
\newcommandx{\gamdi}[1][1=0,usedefault]{\gamma_{{}_{#1}}}

\newcommandx{\emetr}[2][1=\mu,2=\nu,usedefault]{\eta_{{#1}{#2}}}
\newcommandx{\invemetr}[2][1=\mu,2=\nu,usedefault]{\eta^{{#1}{#2}}}
\newcommandx{\hmetr}[2][1=\mu,2=\nu,usedefault]{h_{{#1}{#2}}}
\newcommandx{\invhmetr}[2][1=\mu,2=\nu,usedefault]{h^{{#1}{#2}}}
\newcommandx{\bhmetr}[2][1=\mu,2=\nu,usedefault]{\bar{h}_{{#1}{#2}}}
\newcommandx{\binvhmetr}[2][1=\mu,2=\nu,usedefault]{\bar{h}^{{#1}{#2}}}
\newcommandx{\hud}[2][1=\mu,2=\nu,usedefault]{{h^{#1}}_{\!\!#2}}
\newcommandx{\Ruddd}[4][1=\sigma,2=\mu,3=\lambda,4=\nu,usedefault]{{R^{#1}}_{\!{#2}{#3}{#4}}}
\newcommandx{\Gam}[3][1=\lambda,2=\mu,3=\nu,usedefault]{{\Gamma^{#1}}_{\!{#2}{#3}}}
\newcommandx{\Gamd}[3][1=\mu,2=\nu,3=\lambda,usedefault]{\Gamma_{{#1}{#2}{#3}}}
\newcommandx{\Ricci}[2][1=\mu,2=\nu,usedefault]{R_{{#1}{#2}}}
\newcommandx{\GEinst}[2][1=\mu,2=\nu,usedefault]{G^{{}^\tts{(E)}}_{{#1}{#2}}}
\newcommandx{\Gscr}[3][1=\mu,2=\nu,3=\rho,usedefault]{\mathscr{G}_{{#1}{#2}{#3}}}

\DeclareFixedFont\trfont{OT1}{phv}{b}{sc}{11}

\title{%
       \vspace{-1.0cm}
       \centering\boldmath\LARGE\bfseries%
       Superconductors and gravity
       \bigskip
       }

\bigskip

\author{\textsc{Antonio Gallerati}
\vspace{0.1em}
    }
\affil{%
{Politecnico di Torino, Dipartimento di Scienza Applicata e Tecnologia, corso Duca degli Abruzzi 24, 10129 Torino, Italy}
    }
\affil{Istituto Nazionale di Fisica Nucleare, Sezione di Torino, via Pietro
       Giuria 1, 10125 Torino, Italy%
\vspace{-0.025em}
    }%
\affil{\href{mailto:antonio.gallerati@polito.it}{\texttt{antonio.gallerati@polito.it}}
    }

\author{\textsc{Giovanni Alberto Ummarino}
\vspace{0.1em}
    }
\affil{%
{Politecnico di Torino, Dipartimento di Scienza Applicata e Tecnologia, corso Duca degli Abruzzi 24, 10129 Torino, Italy}
    }
\affil{National Research Nuclear University MEPhI, Kashirskoe hwy 31, 115409
       Moscow, Russian Federation%
\vspace{-0.025em}
    }%
\affil{\href{mailto:giovanni.ummarino@polito.it}{\texttt{giovanni.ummarino@polito.it}}
    }

\date{}

\makeatletter
\patchcmd{\@maketitle}{\begin{center}}{\begin{adjustwidth}{-0.25in}{-0.25in}\begin{center}}{}{}
\patchcmd{\@maketitle}{\end{center}}{\end{center}\end{adjustwidth}}{}{}
\makeatother

\begin{document}

\maketitle
\smallskip

\begin{abstract}
\noindent
We review and discuss some recent developments on the unconventional interaction between superconducting systems and the local gravitational field.
While it is known that gravitational perturbations (such as gravitational waves) can affect supercondensates and supercurrents dynamics, here we want to focus on the more subtle superfluid back-reaction acting on the surrounding gravitational field, analysing some specific favourable situations.
To this end, we will consider suitable quantum macrosystems in a coherent state, immersed in the static weak Earth's gravitational field, investigating possible slight local alterations of the latter not explained in terms of classical physics.
\end{abstract}

\bigskip

\tableofcontents
\pagebreak


\section{Introduction} \label{sec:Intro}
The gravitational force has the distinctive feature of universal interaction with all forms of matter and energy. It dominates at large-scales where it is well described by general relativity. In the latter theory, gravity is not interpreted as a standard force acting on different masses, but as a direct affection of the geometry of the spacetime: masses generate curvature, which in turn dictates the motion of the masses. The spacetime then plays a dynamical role and it is not a rigid background structure.\par
While classical general relativity gives a consistent description of the large-scale dynamics dominated by gravity, we know that quantum field theory is the fundamental formulation to describe physics at microscopic scale, where the effects gravity are in general negligible. In the last decades, different quantum gravity formulations have been proposed to consistently describe the physics of the particles when the gravitational field is so intense as to affect the motion of elementary particles (presumably in the vicinity of a black hole or a neutron star, as well as in the early stages of the evolution of our Universe). This is clearly an ambitious target, since it will imply a fundamental knowledge about the functioning of the laws of nature. However, experimental verification of this kind of theories is really hard to realize, since this would in general imply very high ranges of energy. Then, direct observation of quantum gravity effects, involving gravitons dynamical interactions with other quantum fields at the microscopic level, is a very difficult task.\par
A different approach could originate from the study of unconventional, macroscopic states of matter. In this regard, one should consider quantum macrosystems existing in nature, like superconductors and superfluids. The latter can be thought as large systems featuring a macroscopical coherent phase, suitably described by order parameters. It could be then possible to formulate (and observe) a possible interplay between the extended, coherent system and the surrounding gravitational field \cite{DeWitt:1966yi,papini1966london,papini1967detection,hirakawa1975superconductors,
Ciubotariu:1991zw,anandan1994relgra,podkletnov1992possibility,
Modanese:1995tx,Modanese:1996zm,agop1996gravitational,li1991effects,
ahmedov1999general,agop2000some,Modanese:2001wv,Wu:2003aq,hathaway2003gravity,
Kiefer:2004hv,Quach:2015qwa,Ummarino:2017bvz,
atanasov2017geometric,atanasov2018gravitation}.
In this regard, the coupling with the current flow without resistance in superconductors was exploited to use the latter as a sensitive detection systems, in particular for gravitational waves \cite{Papini:1970cw,adler1976long,anandan1984relativistic,anandan1985detection,
carelli1985coupling,chan1987superconducting,mashhoon1989detection,
Preparata:1989gd,Peng:1990ca,Peng:1990cc,Peng:1991ua,peng1991electrodynamics,
li2007detection,Minter:2009fx,inan2017interaction,inan2017new,Hammad:2020xta}.
Another remarkable phenomenon, showing quantum effects originating from the interaction of quantum particles with a weak-field gravitational background, is the gravity-induced quantum interference \cite{overhauser1974experimental,Colella:1975dq,Anandan:1977ra,Anandan:1978na,
Anandan:1983fe,Cai:1988sd,Ahluwalia:1996ev,Bhattacharya:1999na,Muntinga:2013pta,
Asenbaum:2016djh}. This effect takes place in the presence of a gravitational potential, to be considered in the Schrodinger equation \cite{Kiefer:1990pt,Sakurai:2011zz} and giving rise to a phase shift for elementary particles.%
\footnote{%
The experimental effect can be measured splitting a nearly monoenergetic beam of thermal particles and considering the produced interference paths: a gravity-induced quantum mechanical phase shift is observed, due to the presence of the Earth’s gravitational field \cite{overhauser1974experimental,Colella:1975dq}.}\par
Inspired by the above results, we then also want to consider the back-reaction of superfluids and supercurrents on the local gravitational field in some specific, favourable situation. The first step to achieve the goal will be to formulate an appropriate theoretical model justifying this anomalous coupling. In the following subsections, we will briefly discuss the most convincing theoretical basis and experimental evidences in favour of the existence of this unconventional interaction.

\subsection{Theoretical foundations}
We now want to characterize a possible interplay between superfluids and the local gravitational field in the framework of a quantum gravity theory or, at least, in a suitable approximation of the latter for weak fields.\par
Let us first consider the classical picture. Clearly, the absence of (gravitational) charges of opposite sign excludes the possibility of counteracting the field inside the medium by a local redistribution, ruling out dielectric-type effects. If we then take the medium to be a standard quantum mechanical system, the smallness of the gravitational coupling strongly suppresses the possibility of a (graviton) excitation for a medium particle and any subsequent affection of the local field. We are then led to consider the interaction of the gravitational field with an anomalous external source, that is, an unconventional state of matter exhibiting quantization on a macroscopic scale, like a Bose condensate or a more generic superfluid.\par
Let us then consider a quantum gravity framework and write the Lagrangian for this coherent macrosystem, immersed in the Earth's gravitational field%
\footnote{we work in the ``mostly plus'' convention, where the Minkowski metric is $\eta_{\mu\nu}=\mathrm{diag}(-1,+1,+1,+1)$, and set $c=\hbar=1$
} \cite{Modanese:1995tx,Modanese:1996zm}:
\begin{equation}
\mathscr{L}\=\mathscr{L}_\textsc{eh}+\mathscr{L}_{\phi}
    \=\frac{1}{8\pi\GN}\,\left(R-2\,\Lambda\right)
    -\frac12\,\invgmetr\,{\dm{\phi}}^{\,\ast}\,\dm[\nu]\phi+\frac12\,m^2\,\phi^{\ast}\phi\:.
\end{equation}
The first term is the Einstein-Hilbert contribution, $R$ being the Ricci scalar and $\Lambda$ the cosmological constant. The other terms describe the dynamics of the medium supercondensate (for example Cooper pairs of mass $m$), that we can characterize as a bosonic field $\phi$ with non-vanishing vacuum expectation value (vev) $\phi_0=\langle0|\phi|0\rangle$. We assume this vev to be forced from the outside to a certain value, as it happens, for instance, in a superconductor subjected to external electromagnetic fields. In the weak gravity limit, the metric $\gmetr$ can be written as
\begin{equation} \label{eq:metricweak}
\gmetr(x)\=\emetr+\hmetr(x)\:,
\end{equation}
sum of the flat Minkowski background $\emetr$ plus perturbations given by the $h_{\mu\nu}(x)$ contribution. Now, we expand the bosonic field as
\begin{equation}
\phi(x)=\phi_0(x)+\bar{\phi}(x)\:,
\end{equation}
where the $\phi_0$ vev depends on the medium characteristics
and can be seen as an external source, while the $\bar{\phi}$ contribution is included in the integration variables. The scalar field $\phi$ then suitably describes a superfluid with ground state density $\phi_0$ fixed by external conditions (for example, Cooper pairs density in a supercondensate in the presence of external EM fields).
In the weak gravity limit, the $\mathscr{L}_{\phi}$ term then reads
\begin{equation}
\mathscr{L}_{\phi}\=\mathscr{L}_{\,\bar{\!\phi}}\+\mathscr{L}_{h}\+\mathscr{L}_0\:.
\end{equation}
In the above expansion, the first term is related to the $\bar\phi$ contributions, involved in the negligible excitation processes related to the graviton emission-absorption mechanism, and several vertices of interaction that turn out to be irrelevant due to the smallness of the gravitational coupling. The second term takes into account the coupling of the condensate with the $h_{\mu\nu}$ metric fluctuations and is written as
\begin{equation}
\mathscr{L}_{h} \:\propto\:
    h^{\mu\nu}\,{\dm{\phi_0}}^{\!\ast}\,\dm[\nu]\phi_0\:,
\label{eq:Lh}
\end{equation}
which determines corrections to the gravitational propagator, which is again a negligible contribution. Finally, the last term determines a local supercondensate contribution to the total effective cosmological term of the form
\begin{equation}
\mathscr{L}_0\=
    -\frac12\,\dm{\phi_0}^{\!\ast}\,\dmup[\mu]\phi_0+\frac12\,m^2\left|\phi_0\right|^2\:,
\label{eq:L0}
\end{equation}
connected to the coherent vacuum energy density and depending on the fixed external source $\phi_0$.
The above coupling has the correct structure to produce possible, localized instabilities in superfluid regions featuring larger condensate density \cite{Modanese:1995tx,Modanese:1996zm}: this could determine detectable effects in spite of the smallness of the gravitational coupling. We should also note that, in the latter instable regions, some physical cutoff or regularizing process should come into play, preventing local contribution of arbitrary intensity.
\footnote{%
This can be considered a gravitational analog of the Casimir effect, where observable evidences originate from inhomogeneities in the vacuum fluctuations. In the latter case, the metallic conductors impose a cutoff on the electromagnetic vacuum fluctuations, while the same role is played here by the coherent superfluid.
}
The field then tends to be pinned, assuming fixed extremal values which are independent from those in the neighbouring regions. One could expect, as a physical effect, some kind of slight partial shielding (``absorption'') locally affecting fields propagation and potentials. As we already pointed out, the introduced superfluid density $\phi_0(x)$ is related to the microscopic structure of the involved sample, as well as to the presence of currents, vortex lattices and electromagnetic fields in the supercondensate.%
\footnote{%
It has also been conjectured that high-frequency electromagnetic fields could provide the required energy to enhance the described gravitational field affection \cite{Modanese:1995tx,Wu:2003aq}.}

We have then described a theoretical quantum gravity model with an unconventional coupling between the local gravitational field and the superfluid. The existence of strong variations of the supercondensate components density (for example, Cooper pairs) produce small regions with higher density, where a criticality condition 
could take place, giving rise to localized instabilities.
This gives us a possible way to elude the weakness of the standard coupling and produce a related affection of the local gravitational field.
The key ingredient is the macroscopic quantum coherence of the condensate, that is taken into account when computing the anomalous interplay, at a fundamental level, between the superfluid and the external gravitational field.

\subsection{Experimental evidences}
\label{subsec:experimental}
The discussed formulation laid the foundations of a theoretical approach to an unconventional coupling between superfluids and gravity, in the framework of a quantum model. This however involves a formalism that makes almost impossible to extract quantitative predictions, to be tested in a laboratory experiment. For this reason, one is then led to also consider many phenomenological researches and evidences, to better understand the proposed interplay and obtain an effective theory leading to more explicit experimental predictions.\par
One the first attempts to formulate an effective quantum model describing the interaction between conductors and the local gravitational field was given in \cite{schiff1966gravitation}, where a quantum-mechanical formalism is developed to calculate an (additional) electric field component, generated in the vicinity of a
conductor by the presence of the Earth's gravity. The main consequence of this formulation is the definition of generalized electric-type fields and potentials, existing near the surface of a conductor and featuring a gravitationally-induced component. We can schematically express this generalized field as
\begin{equation}
\E=\Ee+\E_{\text{ind}}\,, \qquad\qquad  V=\Ve+V_{\text{ind}}\,,\quad
\end{equation}
where $\Ee$ is the standard electric field, while $\E_{\text{ind}}$ is the gravitationally-induced component.
In \cite{witteborn1967experimental,witteborn1968experiments} the induced $\E_{\text{ind}}$ and $V_{\text{ind}}$ were experimentally detected as a direct affection to the free fall of electrons in the presence of conductors. These evidences of existence of generalized fields and potentials where then theoretically analysed and experimentally verified in subsequent works \cite{beams1968potentials,herring1968gravitationally,peshkin1968gravityAP,
peshkin1969gravityPLA,craig1969direct,rieger1970gravitationally,
leung1972electric,lockhart1977evidence}.\par
Analogous concepts were subsequently extended to superconductors, obtaining similar results \cite{anandan1984new,peng1983calculation,Jain:1987zz,peng1991electrodynamics,
li1991effects,li1992gravitational,harris1991analogy,torr1993gravitoelectric,
agop2000local,Tajmar:2002gm,Tajmar:2004ww,ahmedov2005electromagnetic,
dematos2008gravitational,tajmar2008electrodynamics}: generalized gravitoelectric and gravitomagnetic fields can be induced by the presence of a local gravitational field coupled to the supercondensate.\par
In the following section we discuss a formal derivation of a consistent form for this generalized fields and potentials, exploiting a weak field expansion for the local gravitational field. This approach will lead us to the definition of a generalized form for Maxwell equations.\par
\bigskip

\section{Linearized gravity. Gravito--Maxwell fields}%
\label{sec:linearizedgravity}
It is well known that gravity is in general mediated by a symmetric $\gmetr$ tensor field, featuring 10 independent components (potentials).  However, under certain approximations and suitable gauge choice, the gravitational field behaviour can be described in an electromagnetic-like fashion, by means of vector-like field equations instead of the corresponding tensorial expressions. In particular, linearized gravity can be considered as a consistent weak-field limit of the complete tensorial theory, in the regime where non-linear effects can be ignored.%
\footnote{In the linear-order assumption, the gravitational field does not transfer energy to the gravitational sources; this also cancels matter-gravity coupling from the domain of linear approximations.
}\par\smallskip
Let us the consider a nearly--flat spacetime, characterized by the presence of weak and static gravitational field. This means we can consider small perturbation of the Minkowski metric $\eta_{\mu\nu}$ and express the spacetime metric $\gmetr$ as
\begin{equation} \label{eq:gmetr}
\gmetr~\simeq~\emetr+\hmetr\;,
\end{equation}
where the symmetric tensor $\hmetr$ is a small perturbation of the constant flat $\emetr$ in the mostly plus convention, $\emetr=\mathrm{diag}(-1,+1,+1,+1)$. The inverse metric, in linear approximation, is given by
\begin{equation}
\invgmetr~\simeq~\invemetr-\invhmetr\;.
\end{equation}
while the metric determinant can be expanded as
\begin{align}
g\,=\,\det\left[\gmetr\right]
    \,=\,\varepsilon^{\mu\nu\rho\sigma}\gmetr[1][\mu]\,\gmetr[2][\nu]\,\gmetr[3][\rho]\,\gmetr[4][\sigma]
    \,\simeq\,-1-h
\;\quad\Rightarrow\quad\;
\sqrt{-g}\:\simeq\,1+\frac12\,h\;,\qquad\qquad
\end{align}
where $h=\hud[\sigma][\sigma]$.\par\medskip
We are now going to exploit the above weak-limit expansion of the metric to obtain a linearized form for the Einstein and London equations. We will then take advantage of the obtained results to define suitable backgrounds to test the discussed gravity/superfluid interplay.

\subsection{Generalizing Maxwell equations}
Let us put ourselves in an inertial coordinate system. To first order in $\hmetr$, the connection is expanded as
\begin{equation}\label{eq:Gam}
\Gam[\lambda][\mu][\nu]~\simeq~\frac12\,\invemetr[\lambda][\rho]\,
     \left(\dm[\mu]\hmetr[\nu][\rho]+\dm[\nu]\hmetr[\rho][\mu]
     -\dm[\rho]\hmetr[\mu][\nu]\right)\:.
\end{equation}
The Riemann tensor is defined as:
\begin{equation}
\begin{split}
\Ruddd&\=\dm[\lambda]\Gam[\sigma][\mu][\nu]-\dm[\nu]\Gam[\sigma][\mu][\lambda]
        +\Gam[\sigma][\rho][\lambda]\,\Gam[\rho][\nu][\mu]
        -\Gam[\sigma][\rho][\nu]\,\Gam[\rho][\lambda][\mu]\:,
\end{split}
\end{equation}
while the Ricci tensor is obtained from the contraction
\begin{equation}
\Ricci\=\Ruddd[\sigma][\mu][\sigma][\nu]\:,
\end{equation}
and, to linear order in $\hmetr$, it is expressed as
\begin{equation} \label{eq:Ricci}
\begin{split}
\Ricci&\:\simeq\:
    \dm[\sigma]\Gam[\sigma][\mu][\nu]+\dm[\mu]\Gam[\sigma][\sigma][\nu]+\cancel{\Gamma\,\Gamma}-\cancel{\Gamma\,\Gamma}\=
\\[\jot]
    &\=\frac12\,\left(\dm\dmup[\rho]\hmetr[\nu][\rho]+\dm[\nu]\dmup[\rho]\hmetr[\mu][\rho]\right)
      -\frac12\,\dm[\rho]\dmup[\rho]\hmetr-\frac12\,\dm\dm[\nu]h\=
\\[\jot]
   &\=\dmup[\rho]\dm[{(}\mu]\hmetr[\nu{)}][\rho]-\frac12\,\dd^2\hmetr-\frac12\,\dm\dm[\nu]h\:,
\end{split}
\end{equation}
having used eq.\ \eqref{eq:Gam}.\par\smallskip
The Einstein equations are written as:
\begin{equation}
\Ricci-\dfrac12\,\gmetr\,R\=8\pi\GN\;T_{\mu\nu}\:,
\label{eq:Einstein}
\end{equation}
where $R=\invgmetr\Ricci$ is the Ricci scalar. In linear-order approximation, we have
\begin{equation}
\frac12\,\gmetr\,R~\simeq~
    \frac12\,\emetr\,\invemetr[\rho][\sigma]\Ricci[\rho][\sigma]
    \=\frac12\,\emetr\,\left(\dmup[\rho]\dmup[\sigma]\hmetr[\rho][\sigma]
       -\dd^2h\right)\:,
\end{equation}
having used eq.\ \eqref{eq:Ricci}. The l.h.s.\ of \eqref{eq:Einstein} then reads
\begin{equation}
\begin{split}
\Ricci-\dfrac12\,\gmetr\,R~\simeq~
    \dmup[\rho]\dm[{(}\mu]\hmetr[\nu{)}][\rho]-\frac12\,\dd^2\hmetr-\frac12\,\dm\dm[\nu]h
    -\frac12\,\emetr\left(\dmup[\rho]\dmup[\sigma]\hmetr[\rho][\sigma]-\dd^2h\right)\:.
\label{eq:lhsEinstein}
\end{split}
\end{equation}
Let us now introduce the symmetric traceless tensor
\begin{equation}
\bhmetr\=\hmetr-\frac12\,\emetr\,h\:,
\end{equation}
so that \eqref{eq:lhsEinstein} is rewritten as
\begin{equation}
\begin{split}
\Ricci-\dfrac12\,\gmetr\,R\:\simeq\:
    &\frac12\left(\dmup[\rho]\dm\bhmetr[\nu][\rho]+\dmup[\rho]\dm[\nu]\bhmetr[\mu][\rho]
     -\dmup[\rho]\dm[\rho]\bhmetr[\mu][\nu]
     -\emetr\,\dmup[\rho]\dmup[\sigma]\bhmetr[\rho][\sigma]\right)\=
\\[1.5\jot]
    \=&\dmup[\rho]\dm[{[}\nu]\bhmetr[\rho{]}][\mu]+\dmup[\rho]\dmup[\sigma]\emetr[\mu][{[}\sigma]\,\bhmetr[\nu{]}][\rho]\=\qquad
\\[2.5\jot]
    \=&\dmup[\rho]\left(
         \dm[{[}\nu]\bhmetr[\rho{]}][\mu]+\dmup[\sigma]\emetr[\mu][{[}\rho]\,\bhmetr[\nu{]}][\sigma]
         \right)\:.
\end{split}
\end{equation}
We also define the tensor
\begin{equation} \label{eq:Gscr}
\Gscr~\equiv~
\dm[{[}\nu]\bhmetr[\rho{]}][\mu]+\dmup[\sigma]\emetr[\mu][{[}\rho]\,\bhmetr[\nu{]}][\sigma]\:,
\end{equation}
in terms of which the Einstein equations take the compact form:
\begin{equation}
\dmup[\rho]\Gscr\=8\pi\GN\;T_{\mu\nu}\:.
\label{eq:Einstein_dG}
\end{equation}
\par\smallskip

\paragraph{Gauge fixing.}
We now consider the \emph{harmonic coordinate condition}, expressed by the relation \cite{Misner:1974qy,Wald:1984rg}:
\begin{equation}
\dm\left(\sqrt{-g}\,\invgmetr\right)=0
\;\quad\Leftrightarrow\quad\;
\Box x^\mu=0\:,
\label{eq:harmcond}
\end{equation}
that in turn can be rewritten in the form
\begin{equation}
\invgmetr\,\Gam\,=\,0\:,
\label{eq:DeDonder}
\end{equation}
also known as \emph{De Donder gauge}. The requirement of the above coordinate condition \eqref{eq:harmcond} plays then the role of a gauge fixing. %
In particular, in harmonic coordinates the metric satisfies a manifestly Lorenz-covariant condition, so that the De Donder gauge becomes a natural choice. Moreover, if one considers the weak-field expansion of the Einstein-Hilbert action in De Donder gauge, the action itself (as well as the graviton propagator) takes a particularly simple form.\par
Using eqs.\ \eqref{eq:gmetr} and
\eqref{eq:Gam} together with the above gauge fixing \eqref{eq:DeDonder}, in first-order approximation we find:
\begin{equation}
0\:\simeq\: \frac12\,\invemetr\,\invemetr[\lambda][\rho]\left(\dm[\mu]\hmetr[\nu][\rho]+\dm[\nu]\hmetr[\rho][\mu]-\dm[\rho]\hmetr[\mu][\nu]\right)\=
\dm\invhmetr[\mu][\lambda]-\frac12\,\dmup[\lambda]h\:,
\end{equation}
that in turn implies the condition
\begin{equation} \label{eq:gaugecond0}
\dm\invhmetr\simeq\frac12\,\dmup[\nu]h
\;\quad\Leftrightarrow\;\quad
\dmup\hmetr\simeq\frac12\,\dm[\nu]h\:.
\end{equation}
We can also write
\begin{equation}
\dmup\hmetr\=\dmup\left(\bhmetr+\frac12\,\emetr h\right)
\=\dmup\bhmetr+\frac12\,\dm[\nu]h\:,
\end{equation}
so that, using eq.\ \eqref{eq:gaugecond0}, we obtain the \emph{Lorentz gauge condition}:
\begin{equation}
\dmup\bhmetr\:\simeq\:0\:.
\label{eq:Lorentzgauge}
\end{equation}
This condition further simplifies eq. \eqref{eq:Gscr} for $\Gscr$, which takes the simple form
\begin{equation}
\Gscr~\simeq~\dm[{[}\nu]\bhmetr[\rho{]}][\mu]\:,
\end{equation}
and satisfies the relation
\begin{equation}
\dm[{[}\lambda{|}]\Gscr[0][{|}\mu][\nu{]}]\=0
\;\qRq\;
\Gscr[0][\mu][\nu] \propto \dm\mathcal{A}_\nu-\dm[\nu]\mathcal{A}_\mu\:,
\end{equation}
The above expression then suggests the existence of a potential. In the following paragraphs we are going to analyse in detail suitable expressions for fields and potentials defining the desired formalism.
\par\medskip

\paragraph{Gravito--Maxwell equations.}
Let us now define the following fields
\cite{peng1983calculation,agop2000local,
Ummarino:2017bvz,Ummarino:2019cvw,Ummarino:2020loo}%
\footnote{for the sake of simplicity, we initially set the physical charge $e=m=1$}
\begin{subeqs}\label{eq:fields0}
\begin{align}
\Eg&~\equiv~E_i~=\,-\,\frac12\,\Gscr[0][0][i]~=\,-\,\frac12\,\dm[{[}0]\bhmetr[i{]}][0]\:,\\[\jot]
\Ag&~\equiv~A_i\=\frac14\,\bhmetr[0][i]\:,\\[\jot]
\Bg&~\equiv~B_i
                 \=\frac14\,{\varepsilon_i}^{jk}\,\Gscr[0][j][k]\:,
\end{align}
\end{subeqs}
with $i=1,2,3$ and
\begin{equation}
\Gscr[0][i][j]\=\dm[{[}i]\bhmetr[j{]}][0]
\=\frac12\left(\dm[i]\bhmetr[j][0]-\dm[j]\bhmetr[i][0]\right)\=4\,\dm[{[}i]A_{j{]}}\:.
\end{equation}
From the above definitions, it follows that
\begin{equation}
\Bg\=\frac14\,{\varepsilon_i}^{jk}\,4\,\dm[{[}j]A_{k{]}}
   \={\varepsilon_i}^{jk}\,\dm[j]A_k=\nabla\times\Ag\:,
\end{equation}
that also implies
\begin{equation}
\nabla\cdot\Bg\=0\:.
\end{equation}
We then also find
\begin{equation}
\nabla\cdot\Eg\=\dmup[i]E_i\=-\dmup[i]\frac{\Gscr[0][0][i]}{2}
              \=-8\pi\GN\;\frac{T_{00}}{2}
              \=4\pi\GN\:\rhog\;,
\end{equation}
having used eq.\ \eqref{eq:Einstein_dG} and defined the mass density as $\rhog\equiv-T_{00}$\,.\par\smallskip
We then consider the curl of $\Eg$:
\begin{equation}
\begin{split}
\nabla\times\Eg&\={\varepsilon_i}^{jk}\,\dm[j]E_k
               \=-{\varepsilon_i}^{jk}\,\dm[j]\frac{\Gscr[0][0][k]}{2}
               \=-\frac12\,{\varepsilon_i}^{jk}\,\dm[j]\dm[{[}0]\bhmetr[k{]}][0]\=
\\[2\jot]
              &\=-\frac14\,4\;\dm[0]\,{\varepsilon_i}^{jk}\,\dm[j]A_k
               \=-\dm[0]B_i\=-\frac{\dd\Bg}{\dd t}\:.
\end{split}
\end{equation}
Finally, for the curl of $\Bg$ we find
\begin{equation}\label{eq:gravMaxwell4}
\begin{split}
\nabla\times\Bg&\={\varepsilon_i}^{jk}\,\dm[j]B_k
      \=\frac14\,{\varepsilon_i}^{jk}\,
                {\varepsilon_k}^{\ell m}\,\dm[j]\Gscr[0][\ell][m]
      \=\frac14\left({\delta_i}^\ell\delta^{jm}-{\delta_i}^m\delta^{j\ell}\right)\dm[j]\Gscr[0][\ell][m]\=
\\[3\jot]
    &\=\frac12\,\dmup[j]\Gscr[0][i][j]
     \=\frac12\left(\dmup\Gscr[0][i][\mu]+\dm[0]\Gscr[0][i][0]\right)
     \=\frac12\left(\dmup\Gscr[0][i][\mu]-\dm[0]\Gscr[0][0][i]\right)\=
\\[3\jot]
    &\=\frac12\left(8\pi\GN\;T_{0i}-\dm[0]\Gscr[0][0][i]\right)
     \=4\pi\GN\;j_i+\frac{\dd E_i}{\dd t}
     \=4\pi\GN\;\jg\+\frac{\dd\Eg}{\dd t}\:,
\end{split}
\end{equation}
having used again eq.\ \eqref{eq:Einstein_dG} and defined the mass gravito-current density vector as \,$\jg \equiv j_i \equiv T_{0i}$\,.\par\medskip
Summarizing, once defined the fields \eqref{eq:fields0}, one can write the field equations \cite{Braginsky:1976rb,ross1983london,peng1983calculation,thorne1988gravitomagnetism,
mashhoon1989detection,peng1990new,li1991effects,Peng:1991ua,torr1993gravitoelectric,
agop2000local,Ruggiero:2002hz,Tartaglia:2003wx,Ummarino:2017bvz,Vieira:2016csi,
Ummarino:2019cvw,Behera:2017voq,Giardino:2018ffd,Sbitnev:2019iyz,Gallerati:2020tyq,
Williams:2020fgi,Gallerati:2021ops,Toth:2021dut}:
\begin{equation} \label{eq:gravMaxwell}
\begin{split}
&\nabla\cdot\Eg\=4\pi\GN\;\rhog\:,\\[2\jot]
&\nabla\cdot\Bg\=0 \:,\\[2\jot]
&\nabla\times\Eg~=-\dfrac{\dd\Bg}{\dd t} \:,\\[2\jot]
&\nabla\times\Bg\=\frac{4\pi\GN}{c^2}\;\jg
                  \+\frac{1}{c^2}\,\frac{\dd\Eg}{\dd t}\:,
\end{split}
\end{equation}
having restored physical units. The above expressions are formally equivalent to Maxwell equations, $\Eg$ and $\Bg$ being the gravitoelectric and gravitomagnetic field, respectively.%
\footnote{%
For instance, on the Earth's surface, $\Eg$ corresponds to the Newtonian gravitational acceleration while $\Bg$ is related to angular momentum interactions \cite{Braginsky:1976rb,peng1983calculation,peng1990new,agop2000local}.
}
The mass current density vector $\jg$ can also be written as:
\begin{equation}
\jg \= \rhog\,\mathbf{v}\:,
\end{equation}
in terms of the mass density and velocity $\mathbf{v}$.
\par\medskip


\paragraph{Generalized Maxwell equations.}
Inspired by the discussion of Sect.\ \ref{subsec:experimental}, it is now straightforward to extend the above results and introduce generalized electric/magnetic fields, scalar and vector potentials. The latter feature both electromagnetic and gravitational contributions and can be written as:
\begin{equation}
\E=\Ee+\frac{m}{e}\,\Eg\,,\quad\quad
\B=\Be+\frac{m}{e}\,\Bg\,,\quad\quad
V=V_\textrm{e}+\frac{m}{e}\,\Vg\,,\quad\quad
\A=\Ae+\frac{m}{e}\,\Ag\,,
\label{eq:genEMfields}
\end{equation}
where $m$ and $e$ are the electron mass and charge, respectively \cite{schiff1966gravitation}.\par\smallskip
The generalized Maxwell equations then become:
\begin{equation} \label{eq:genMaxwell}
\begin{split}
&\nabla\cdot\E\=\left(\frac1\epsg+\frac{1}{\epsz}\right)\,\rho \:,\\[2\jot]
&\nabla\cdot\B\=0 \:,\\[2\jot]
&\nabla\times\E\:=-\dfrac{\dd\B}{\dd t} \:,\\[2\jot]
&\nabla\times\B\=\left(\mug+\muz\right)\,\jj
                  \+\frac{1}{c^2}\,\dfrac{\dd\E}{\dd t} \:,
\end{split}
\end{equation}
where $\epsz$ and $\muz$ are the standard electric permittivity and magnetic permeability in the vacuum, and where we have set
\begin{equation}
\rhog\=\frac{m}{e}\,\rho\:,\qquad\quad
\jg\=\frac{m}{e}\,\jj\:,
\end{equation}
$\rho$ and $\jj$ being the electric charge density and electric current density, respectively. The introduced vacuum gravitational permittivity $\epsg$ and vacuum gravitational permeability $\mug$ are defined as
\begin{equation}
\epsg=\frac{1}{4\pi\GN}\,\frac{e^2}{m^2}\:,\qquad\quad
\mug=\frac{4\pi\GN}{c^2}\,\frac{m^2}{e^2}\:.
\end{equation}
The obtained generalized Maxwell equations turn out to be a consistent approximation to the complete tensorial theory, valid in the limit of weak gravitational field (like the static, weak Earth's gravity). It is then possible to take advantage of the obtained results and consider suitable situations and parameters regime where the gravitoelectric field plays a fundamental role and/or where gravitomagnetic effects are not negligible.

\subsection{Generalizing London equations}
The London equations for a superconductor in stationary state characterize an analogous Ohm's law (zero resistivity) and Meissner effect (expulsion of the magnetic field from the interior sample) for the superfluid.
They can be explicitly written as \cite{DeGennes2018superconductivity,tinkham2004introduction,
ketterson1999superconductivity}:
\begin{subeqs} \label{eq:London}%
\begin{align}
\Ee&\=\frac{m}{\ns\,e^2}\;\dfrac{\dd\jj}{\dd t}\;;\label{eq:London1}
\\[2\jot]
\Be&\=-\frac{m}{\ns\,e^2}\;\nabla\times\jj \;.\label{eq:London2}
\end{align}
\end{subeqs}
where \;$\jj=\ns\,e\,\vs$\; is the supercurrent and \,$\ns$\, is the superelectron density.\par
The Ampère's law for a superconductor in stationary state (no displacement current) has the form
\begin{equation} \label{eq:Ampere}
\nabla\times\Be\=\muz\,\jj\;,
\end{equation}
so that taking the curl gives
\begin{equation}
\nabla\times\nabla\times\Be
    \=\nabla\left(\xcancel{\nabla\cdot\Be}\right)-\nabla^2\Be
    \=\muz\,\nabla\times\jj\=-\frac{\muz\,\ns\,e^2}{m}\;\Be\:,
\end{equation}
that is,
\begin{equation}
\nabla^2\Be\=\frac{1}{\lambdae^2}\;\Be\:,
\label{eq:Meissner}
\end{equation}
having introduced the penetration depth
\begin{equation}
\lambdae\=\sqrt{\frac{m}{\muz\,\ns\,e^2}}\;.
\label{eq:lambdae}
\end{equation}
The above quantity gives an estimate of the mean distance the magnetic field $\Be$ can penetrate the sample. Since the values of the $\lambdae$ parameter vary from $2\,\nm$ (low $\Tc$ superconductors) to $200\,\nm$ (high $\Tc$ superconductors), the above \cref{eq:Meissner,eq:lambdae} quantitatively characterize the Meissner effect.\par
The two London equations \eqref{eq:London} can be now rewritten in terms of the vector potential $\Ae$ in the (not gauge-invariant) form:
\begin{equation} \label{eq:Londonsumm}
\jj=-\frac{1}{\muz\,\lambdae^2}\;\Ae \qquad\qquad\qquad
\end{equation}
with \,$\Be=\nabla\times\Ae$\, and expressing the electric field as \,$\Ee=-\dfrac{\dd\Ae}{\dd t}$.
\par\medskip

\paragraph{Generalized London equations.}
Let us now take into account gravitational contributions, and consider for the fields and potentials the generalized form \eqref{eq:genEMfields}.
%
%
In particular, we consider the generalized potential $\A$ minimally coupled to the wave function
\begin{equation} \label{eq:psi}
\psi\=\psi_0\,\exp(i\,\varphi)\,, \qquad\qquad \psi_0^2\equiv\abs{\psi}^2=n_s\,.
\end{equation}
The second London equation can be derived from the quantum mechanical current density
\begin{equation}
\jj\=-\frac{i}{2m}\left(\psi^\ast\tilde{\nabla}\psi-\psi\tilde{\nabla}\psi^\ast\right)\:,
\end{equation}
where $\tilde{\nabla}$ is the covariant derivative for the minimal coupling:
\begin{equation}
\tilde{\nabla}\=\nabla-i\,\tilde{g}\,\A\:,
\end{equation}
with unknown coupling constant $\tilde{g}$. We then find for the current
\begin{equation}
\jj\=-\frac{i}{2m}\left(\psi^\ast\nabla\psi-\psi\nabla\psi^\ast\right)
            -\frac{\tilde{g}}{m}\,\A\,\abs{\psi}^2
    \=\frac{1}{m}\,\abs{\psi}^2\left(\nabla\varphi-\tilde{g}\,\A\right)\,.
\end{equation}
Now, taking the curl of the previous expression gives
\begin{equation}
\B\=-\frac{m}{\tilde{g}\,\abs{\psi}^2}\;\nabla\times\jj
   \=-\frac1\zeta\;\nabla\times\jj\;,
\end{equation}
which is the generalized form of the second London equation \eqref{eq:London2} \cite{ross1983london}.\par\smallskip
We now want to fix the values of the $\zeta$ parameter and coupling constant $\tilde{g}$. To this end, let us restrict to the case $\Bg=0$:
\begin{equation}
\B\=\Be+\frac{m}{e}\:\xcancel{\Bg}\=-\frac1\zeta\;\nabla\times\jj\:,
\end{equation}
so that, using \eqref{eq:London2}, \eqref{eq:lambdae} and \eqref{eq:psi}, we find
\begin{equation}
\tilde{g}\,=\,e^2\,, \qquad\quad \frac1\zeta\=\muz\,\lambdae^2\,.
\end{equation}
In order to define an analogue gravitational penetration depth, we now consider the case $\Be=0$:
\begin{equation}
\B\=\xcancel{\Be}+\frac{m}{e}\,\Bg
    \=-\muz\,\lambdae^2\;\nabla\times\jj
    \=-\muz\,\lambdae^2\;\frac{m}{e}\,\nabla\times\jg\:,
\end{equation}
the gravito-Ampère's law \eqref{eq:gravMaxwell} in stationary state reading
\begin{equation}
\nabla\times\Bg\=\mug\;\jg\:.
\end{equation}
Taking the curl of the above equation, we have
\begin{equation}
\nabla\times\nabla\times\Bg\=-\nabla^2\Bg
     \=\mug\,\nabla\times\jg\=-\mug\,\frac{1}{\muz\,\lambdae^2}\;\Bg
     \=-\frac{1}{\lambdag^2}\;\Bg\:,
\end{equation}
having introduced the gravitational penetration depth
\begin{equation} \label{eq:lambdag}
\lambdag\=\sqrt{\frac{\muz\,\lambdae^2}{\mug}}
        \=\sqrt{\frac{c^2}{4\pi\GN\,m\,n_s}}\:.
\end{equation}
Writing now the stationary generalized Ampère's law \eqref{eq:genMaxwell} and using eq.\ \eqref{eq:lambdag} we obtain
\begin{equation} \label{eq:genAmpere}
\nabla\times\B\=\left(\muz+\mug\right)\,\jj
    \=\muz\left(1+\frac{\lambdae^2}{\lambdag^2}\right)\jj\:,
\end{equation}
and taking the curl we find the general form
\begin{equation}
\begin{split}
\nabla^2\B&\=-\muz\left(1+\frac{\lambdae^2}{\lambdag^2}\right)\,\nabla\times\jj
        \=\muz\,\frac{1}{\muz\,\lambdae^2}\,\left(1+\frac{\lambdae^2}{\lambdag^2}\right)\,\B\=
\\[\jot]
        &\=\left(\frac{1}{\lambdae^2}+\frac{1}{\lambdag^2}\right)\,\B
        \=\frac{1}{\lambda^2}\;\B\:,
\end{split}
\end{equation}
where we have introduced the \emph{generalized penetration depth} $\lambda$\,:
\begin{equation}
\lambda\=\frac{\lambdag\,\lambdae}{\sqrt{\lambdag^2+\lambdae^2}}
       \:\simeq\: \lambdae\:, \qquad\qquad \text{with} \quad \frac{\lambdag}{\lambdae}\simeq{10}^{21}\;.
\end{equation}
Finally, we can recast eq.\ \eqref{eq:Londonsumm} in the form
\begin{equation}
\jj\=-\,\zeta\,\A\:,
\end{equation}
with $\B=\nabla\times\A$. Moreover, since charge-conservation requires the condition ${\nabla\cdot\jj=0}$, we obtain for the vector potential
\begin{equation*}
\nabla\cdot\A\=0\:,
\end{equation*}
that is, the so-called \emph{Coulomb gauge} (or \emph{London gauge}).\par\bigskip
In the following sections we are going to consider suitable frameworks where the proposed gravity/superfluid interplay can in principle be detected, precisely characterizing the physical system and optimizing the range of parameters in order to maximize the effect. We will also exploit the described formalism and introduced generalized fields.
\par\bigskip

\section{A simple application: Josephson effect}
The Josephson effect consists in the transmission of supercurrents through thin insulating barriers by means of quantum-mechanical tunnelling \cite{Josephson:1962zz}. The phenomenon can be seen as a general property of coupled superconducting systems and could take place in suitable tunnel junctions, where quantum interference appears.
In particular, when the states of two superconductors are assumed to be coherent (that is, coherent superpositions of states with different numbers of particle pairs) there exists a phase-dependent coupling energy between the two. The latter then implies the possibility of a supercurrent flowing across the junction \cite{anderson1967josephson}.

\subsection{Josephson junction}
If two superconductors are put in contact and the critical current in the contact region is well below those of the individual constituents, the configuration is defined as \emph{weak link}. Once the weak link is formed, coherence is established across the barrier, with a phase difference $\Delta\varphi$ causing interference between the previously independent wavefunctions, so that the system can be described with a single wavefunction as a whole.\par
When two  superconducting samples are connected through a weak Josephson link, the response of the supercondensate (through the corresponding coupling energy) keeps the macroscopic internal coherence of the system, allowing direct observation of coherence-interference phenomena.
%
In particular, a simple manifestation of the Josephson effect can be observed in a circuit closed on a {superconductor-insulator-superconductor} (SIS) junction, to which a constant potential difference $\Delta V$ is applied. The voltage, in turn, produces a sinusoidal superconductive current across the junction with pulsation \cite{Josephson:1962zz,anderson1967josephson,Barone1982physics}
\begin{equation}
\omega=\frac{2\,e\,\Delta V}{\hbar}\:.
\end{equation}
Let us briefly discuss the phenomenon.\par\medskip
\begin{figure}[H]
\captionsetup{skip=5pt,belowskip=15pt,font=small,labelfont=small,format=hang}
\centering
\includegraphics[width=0.9\textwidth,keepaspectratio]{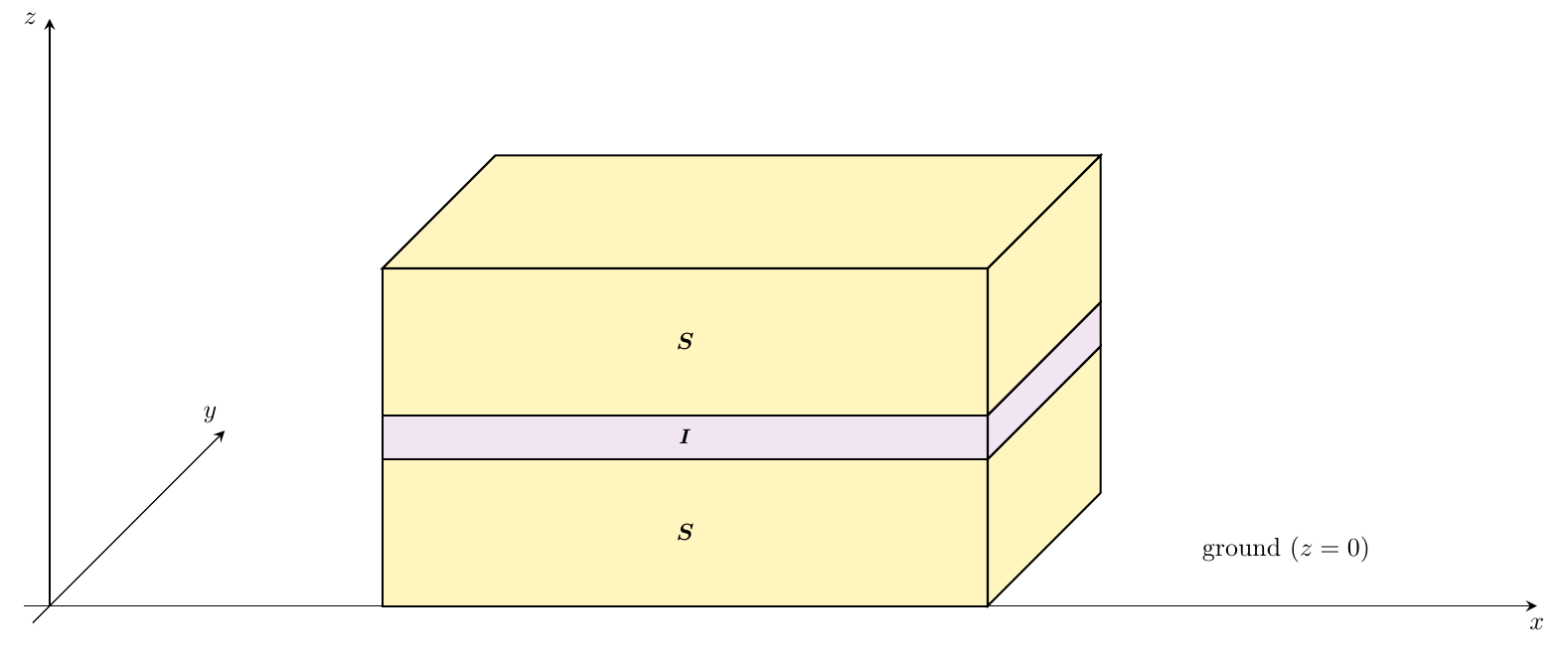}
\caption{SIS junction with axis directed along the $z$ direction, parallel to the local Earth's gravitational field.}
\label{fig:SIS}
\end{figure}

\paragraph{Josephson AC current.}
Let us consider a weak link between two superconductors. The latter, when taken separately, are described by wavefunctions of phases $\varphi_{1}$, $\varphi_{2}$ and amplitudes $|\psi_{1}|$, $|\psi_{2}|$. We can explicitly write
\begin{equation} \label{eq:psi12Jos}
\begin{split}
\psi_{1}\=|\psi_{1}|\,\exp\left(i\,\varphi_{1}\right)
    \=\sqrt{\rho_{1}}\;\exp\left(i\,\varphi_{1}\right)\:,
\\[\jot]
\psi_{2}\=|\psi_{2}|\,\exp\left(i\,\varphi_{2}\right)
    \=\sqrt{\rho_{2}}\;\exp\left(i\,\varphi_{2}\right)\:,
\end{split}
\end{equation}
where $\rho_{1}$, $\rho_{2}$ are the probability amplitudes of Cooper pair densities.\par
Once the weak link is formed, coherence is established across the barrier, and the  phase difference
\begin{equation}
\Delta\varphi\=\varphi_{2}-\varphi_{1}\=\gamma
\end{equation}
determines interference between the (previously independent wavefunctions), so that the system can be described by means of a single wavefunction as a whole.\par\smallskip
A SIS tunnel Josephson junction is a typical weak link consisting of two superconductors (that we take equal for simplicity) of thickness $L$ and surface $A$, separated by a thin oxide layer of thickness ${\ell\ll L}$, see Fig.\ \ref{fig:SIS}.
The time dependent Schrodinger equation can be used to characterize the system evolution as
\begin{equation}
i\hbar\,\frac{\partial\psi}{\partial t}\=\mathcal{E}\,\psi\;.
\end{equation}
As we already discussed, in the weak link takes place a coherent overlap between the two wavefunctions, and an additional term must be added to take into account the interaction. In particular, the rate of change of $\psi_{1}$ is proportional to its coupling to $\psi_{2}$, the same happening for $\psi_{2}$ on the other side. It is then possible to write the relations \cite{Feynman1965Feynman}
\begin{subeqs}\label{eq:changerate}
\begin{align}
i\hbar\,\frac{\partial\psi_{1}}{\partial t} \=\mathcal{E}_{1}\,\psi_{1}+K\,\psi_{2}\;,
\label{eq:changerate1}
\\[2ex]
i\hbar\,\frac{\partial\psi_{2}}{\partial t}\= \mathcal{E}_{2}\,\psi_{2}+K\,\psi_{1}\;.
\label{eq:changerate2}
\end{align}
\end{subeqs}
For the sake of simplicity, we will consider superconductor of the same kind, so that the probability amplitudes of Cooper pair densities are equal, $\rho_{1}=\rho_{2}=\rho$.
The quantities $\mathcal{E}_{1}$ and $\mathcal{E}_{2}$ are the ground state energies of the unperturbed system (i.e.\ when $K=0$) and we choose the zero of energy to be halfway between $\mathcal{E}_{1}$ and $\mathcal{E}_{2}$, the evolution of the system then depending on the difference $\Delta\mathcal{E}=\mathcal{E}_{2}-\mathcal{E}_{1}$ .\par
Now, we put expressions \eqref{eq:psi12Jos} in the evolution relations \eqref{eq:changerate}, assuming each wavefunction having a well-defined Cooper pair density and space-independent macroscopic phase. Separating the real and imaginary part, we find:
\begin{subeqs} \label{eq:changerateReIm}
\begin{align}
\frac{\partial\gamma}{\partial t}&\=\frac{\Delta\mathcal{E}}{\hbar}\=0\;,
\label{eq:changerateRe}
\\[\jot]
\frac{\partial\rho}{\partial t}
        &\=\frac{2\,K}{\hbar}\,\rho\,\sin(\gamma)\;,
\label{eq:changerateIm}
\end{align}
\end{subeqs}
with $\gamma=\varphi_{2}-\varphi_{1}$, the expression being valid in the absence of applied voltage of any kind (electric or gravitational-like).%
\footnote{%
Equation \eqref{eq:changerateIm} is written in the standard Josephson formalism \cite{fossheim2005superconductivity} with a little abuse of notation: the $\rho$-density involved in the time-derivative on the l.h.s.\ refers to the superconducting current density across the interface, while the density on the r.h.s.\ refers to the global density of Cooper pairs in the system; the latter is a conserved (constant) quantity, being the system in the superconductive state.
}.
The supercurrent across the contact then is written
\begin{equation}
J_\text{s}\:=\,-2\,e\,\frac{\partial\rho}{\partial t}
     \:=\,-\frac{4\,e\,K}{\hbar}\,\rho\,\sin(\gamma)\=J_{0}\,\sin(\gamma)\;,
\end{equation}
flowing through the thin layer separating the superconductors and depending on the phase difference across the barrier.\par\smallskip
%
%
If we apply a constant voltage $\Delta V$ across the junction, an oscillatory variation of phase difference takes place. A corresponding AC supercurrent then appears in the weak link, due to the existing finite potential difference in the junction. The phenomenon is a manifestations of the Josephson-Gor'kov principle \cite{gorkov1959microscopic,josephson1965supercurrents,anderson1967josephson}, which simply states that the oscillation frequency of the coherent matter field is driven by the existing chemical potential (corresponding to pairs of electrons in the case of superconductivity). Since the supercurrent is a periodic function of $\Delta\varphi$, AC supercurrents must be associated with any applied voltage difference (general principle of gauge invariance dictates that all physical properties must be periodic functions of the phase with period $2\pi$). Equation \eqref{eq:changerateRe} is then modified in order to take into account the applied voltage, resulting in a time-dependent relation of the form \cite{Josephson:1962zz,josephson1964coupled,josephson1965supercurrents,
anderson1967josephson,Barone1982physics}:
\begin{equation}
\frac{\partial\gamma}{\partial t}\=\frac{2\,e\,\Delta V}{\hbar}\;,
\label{eq:dgamdt}
\end{equation}
relating the phase difference variation on opposite sides to the existing potential difference across the junction. After integration, the above \eqref{eq:dgamdt} gives
\begin{equation}
\gamma(t)\=\gamma_{0}+\frac{2\,e\,\Delta V}{\hbar}t\;,
\end{equation}
$\gamma_{0}$ being an integration constant. The supercurrent density turns out to be
\begin{equation}
J_\text{s}\=J_{0}\,\sin\left(\gamma_{0}+\frac{2\,e\,\Delta V}{\hbar}\,t\right)\;.
\end{equation}
and, below the critical temperature, the amplitude of the corresponding tunnelling supercurrent $I_{0}=J_{0}\,A$ is temperature-dependent and is expressed by the Ambegaokar–Baratoff formula \cite{Ambegaokar:1963zz1,Ambegaokar:1963zz2}:
\begin{equation}
I_{0}\=\frac{\pi\,\Delta_{\textsc{s}}(T)}{2\,e\,R_\textsc{n}}\,\tanh\left(\frac{\Delta_\textsc{s}(T)}{2\,k_\textsc{b}\,T}\right)\;,
\end{equation}
where $R_\textsc{n}$ is the junction resistance in the normal state and $\Delta_\textsc{s}(T)$ is the superconductive gap.
The described AC signal, coming from the applied DC voltage, may be understood as the result of the energy conversion of electron pairs into photons \cite{Saxena2009proximity}.\par\smallskip
In the following we will examine the possibility of a Josephson-like effect induced by the weak-static Earth's gravitational field, also analysing suitable experimental settings and parameters optimization.

\subsection{Josephson effect induced by gravity}
We have discussed and motivated in the previous sections the introduction of generalized electric field and potential of the form
\begin{equation}
\E\=\Ee+\frac{m}{e}\:\Eg\,,\qquad\quad
V\=\Ve+\frac{m}{e}\,V_\text{g}\,.
\end{equation}
If we now restrict to a simple situation in which it is present only the Earth's static gravitational field ($\Ee=0$), we have that
\begin{equation}
\E\=\frac{m}{e}\:\Eg\=\frac{m}{e}\:\textbf{g}\:,
\end{equation}
while the corresponding potential difference reads
\begin{equation}
\Delta V\=\frac{m}{e}\,\Delta V_\text{g}
    \=\int_{0}^{\ell}\!dz\;\frac{m}{e}\,g\=\frac{m}{e}\,g\,\ell\;,
\end{equation}
having chosen the $z$-axis along the direction of the gravitational field, see Fig.\ \ref{fig:SIS}. The resulting induced Josephson current \cite{Ummarino:2020loo} has then the form
\begin{equation}
I_\text{s}(t)
    \=I_0\,\sin\left(\frac{2\,e\,\Delta V}{\hbar}\,t+\varphi\right)
    \=I_0\,\sin(\omega\,t+\varphi)\;.
\end{equation}
We also expect the induced effect to disappear when the junction is rotated in a position where the normal vector to the surface is perpendicular to the gravitational field direction.

\begin{figure}[H]
\centering
\vspace{2em}
\includegraphics[width=0.85\textwidth,keepaspectratio]{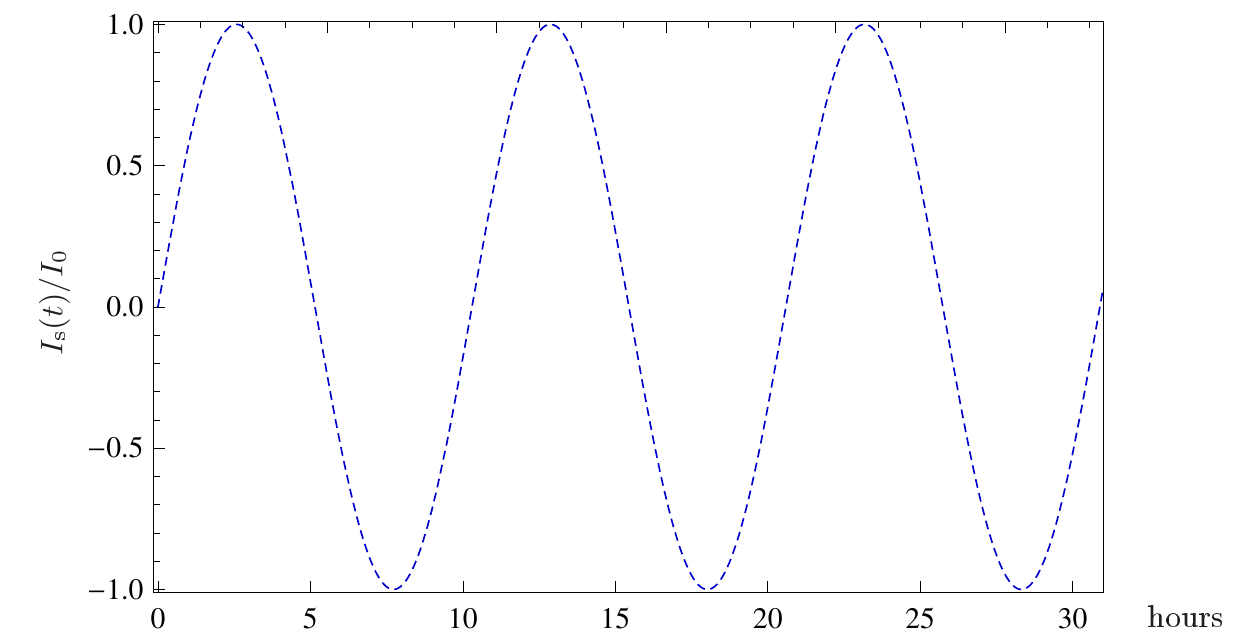}
\caption{\sloppy%
    Time dependence of the Josephson current for an insulating layer of thickness $\ell=1\,\text{nm}$.}
\label{fig:SISL1}
\end{figure}

\paragraph{Experimental settings.}
\sloppy
Let us first consider a junction involving high-$\Tc$ superconductors (HTSC). The latter have a coherence length $\xi$ of the order of $10^{-9}\,\mathrm{m}$, that fixes the thickness $\ell$ of the insulating layer to be $\ell\lesssim\xi$. Then, if we consider the pulsation
\begin{equation}
\omega\=\frac{2\,e\,\Delta V}{\hbar}\=\frac{2\,m\,g\,\ell}{\hbar}\:,
\end{equation}
\sloppy
a junction with an insulating layer of thickness $\ell\simeq1\,\nm$ would result in ${\omega\simeq 1.7\times10^{-4}\,\mathrm{s}^{-1}}$, determining a corresponding period for the Josephson current ${T=2\pi/\omega\simeq3.7\times10^4\,\mathrm{s}}$.
This implies that the distinctive oscillatory behaviour can be observed only in very stable junctions, since a reasonable duration of the experiment turns out to be longer than one day, see Fig.\ \ref{fig:SISL1}.\par
\begin{figure}[H]
\centering
\includegraphics[width=0.85\textwidth,keepaspectratio]{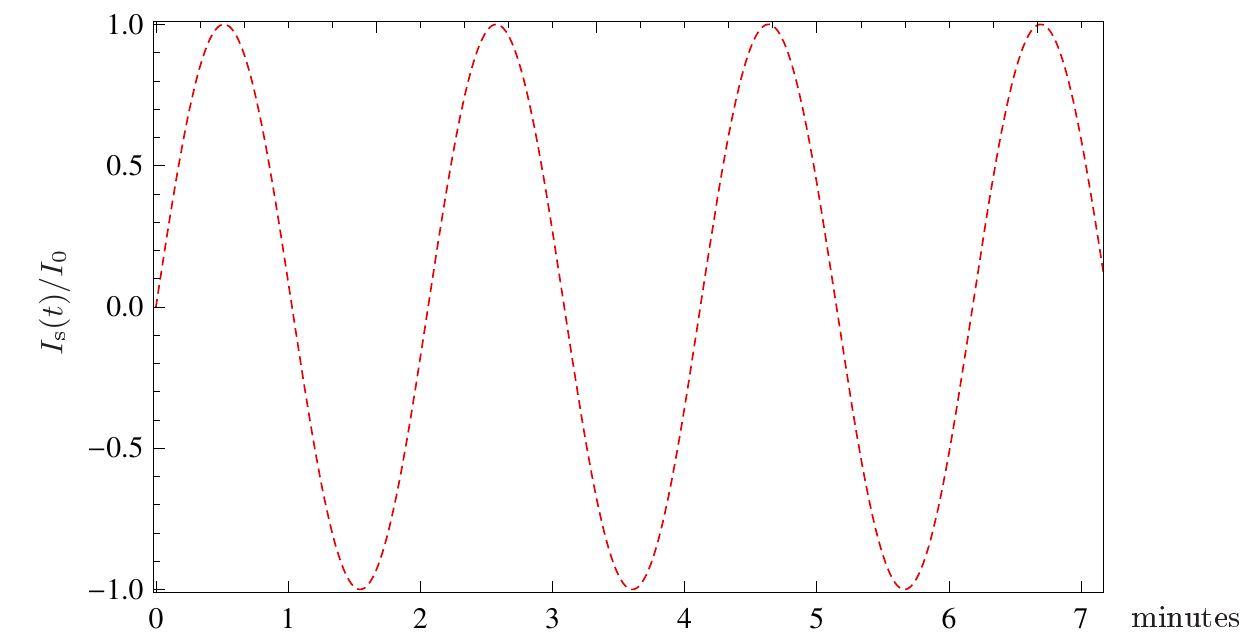}
\caption{\sloppy%
    Time dependence of the Josephson current for an insulating layer of thickness $\ell=300\,\text{nm}$.}
\label{fig:SISL2}
\end{figure}

To reduce the time duration of the experiment, it is necessary to increase the voltage $V_\text{g}$. Clearly, it is impossible to vary the intensity of the local gravitational field, so that the only strategy left is to have larger $\ell$. This means we need tunnelling junctions working in the presence of a thicker insulating layer: this is possible using low-$\Tc$ superconductors (LTSC), the latter can feature a larger coherent length, of the order of $10^{3}\,\nm$. In this case, we can take an insulating layer of thickness $\ell\simeq300\,\nm$ and obtain for the voltage $V_\text{g}\simeq1.67\times10^{-17}\,\mathrm{Volt}$. The pulsation turns out to be $\omega\simeq0.05\,\mathrm{s}^{-1}$ and the corresponding period $T\simeq123\,\mathrm{s}$, strongly reducing the experiment duration, see Fig.\ \ref{fig:SISL2}.\par
From a practical point of view, it would be preferable to work with experimental setups stable enough to allow accurate oscillations measurements, but that, at the same time, give rise to a Josephson current of detectable intensity. If we increase the junction thickness using low-$\Tc$ superconductors, the time duration for the experiment decreases and a stable setting is then possible, but the associated Josephson current becomes very weak and difficult to measure.
Currently, the best choice to observe experimental evidences is to realize the most stable setup possible with HTCS, then making long-time measurements of stronger Josephson currents.\par
%
\bigskip
%

We have seen how the proposed theoretical model gives the possibility to investigate the discussed interplay between gravitation and a superconductive condensate. For the simple case of the Josephson junction, the difficulties lie in the experimental setup, that has to be stable in time to allow for careful observations of the oscillatory behaviour, and very sensitive to the induced voltage. In the following sections, we are going to analyse a more detailed microscopic description of the superfluid, exploiting a mean-field theory formulation for the system thermodynamics, including the effects of thermal fluctuations. In particular, we will analyse how the local gravitational field can be affected by the presence of a supercondensate exploiting the time-dependent Ginzburg--Landau equations in the regime of fluctuations.
\par\bigskip

\section{Affecting the field just outside the sample. Ginzburg--Landau formulation}
\label{sec:outside}
We now want to better characterize the interaction between the superfluid and the local gravitational field. To this end, we need a microscopic quantum model describing the supercondensate behaviour. However, the formalism that characterizes the material superconductive state is in general very complicated, so that extracting quantitative predictions (or even just qualitative descriptions) for the interplay turns out to be an almost impossible task.\par
A simpler framework for analysing the interaction mechanism is given by a superconducting sample in the vicinity of its critical temperature $\Tc$. In particular, for $T$ near $\Tc$, the system can be described by the Ginzburg--Landau equations, for which analytic solutions could be found.

\subsection{Thermodynamic fluctuations vs.\ mean field theory}
The physics of low-temperature condensed matter systems is based on two fundamental notions: the low-energy long-living excitations (quasiparticles) and the mean field approximation. For instance, the BCS theory of superconductivity \cite{Bardeen:1957mv} is a paradigmatic example of the exploitation of both approaches mentioned.\par
Physical situations which cannot be consistently described in terms of the quasiparticle method or the mean field approximation are called \emph{fluctuations}. The regime in which the fluctuations come into play is, in general, a very narrow temperature range around the critical temperature.%
\footnote{On the contrary, for high temperature cuprate superconductors, organic superconductors, iron pnictides, low dimensional and amorphous superconducting systems, the situation changes radically due to the very small value of the coherence length, so that the temperature range of fluctuation is considerably larger.
}
In particular, many effects on the superconducting phase occur while the system is still in the normal phase (just above the critical temperature) and originate from the appearance of the superconducting fluctuations themselves. In this regard, diamagnetic susceptibility, conductivity, heat capacity and other physical quantities may increase considerably near the transition temperature.\par
If we consider a range of temperature sufficiently far from the critical $\Tc$, the fluctuations regime ceases and the physics of the system is described in terms of a mean field formulation. The latter approach approximates the physics by averaging over the degrees of freedom of the system, that is, by approximating all the interactions acting on a single component with a single averaged effect. The technique allows to map a multi-body problem onto a one-body problem.
In particular, the thermodynamic properties of the system are obtained by treating the order parameter as spatially constant, the spatial fluctuations being negligible. Many predictions can therefore be obtained by exploiting a much simpler mathematical formulation: this is a great advantage when dealing with new and unconventional systems, for which a complete description is not known.

\paragraph{GL equations.}
As we have briefly discussed, the analysis of condensed matter systems in general involves the study of complicated, many electron states. A certain number of phenomenological approaches, based on classical field theory, were then developed to address the problem. A possibility is to consider a slowly-varying density of fields, carrying sufficient quantum information to write down an energy function for the system to be minimized: this corresponds to the celebrated Ginzburg--Landau (GL) formulation \cite{Ginzburg:1950sr,GL2009Springer}, based on a mean field approach. Its most notable use is in the theory of superconductors, where a complex scalar field $\psi$ is used to characterize the density of the superconducting paired electrons. Even if the GL approach is, in general, superseded by the more fundamental BCS theory, it is a powerful tool in the vicinity of the critical temperature, where a more fundamental theory is lacking or the formulation is too complicated.

\subsection{Ginzburg--Landau formulation}
Let us consider a superconductive sample near its critical temperature. At the microscopical level, thermodynamic fluctuations of the order parameter $\psi(\x,t)$ describing superconducting electrons occur, giving rise to localized regions of accelerated charge carriers \cite{ginzburg1961some,thouless1960perturbation,shier1966superconducting,
glover1967ideal,strongin1968curie}. From a physical point of view, $\psi$ can be thought as the pseudowavefunction characterizing the motion of the center of mass of the Cooper pairs.
The average size of these regions is much greater than the mean free path
for a certain range of temperature above $\Tc$, while it decreases for larger temperature \cite{ferrell1967predicted}. Moreover, we are going to consider sufficiently dirty materials, so that the effects of the fluctuations can be observed over a sizable range of temperature \cite{shier1966superconducting}.%
\footnote{%
In order to have a sufficiently large temperature interval, the electronic mean free path characterizing the material in the normal state should be less than $10\,\Ang$.}\par\smallskip
We now want to characterize more in detail the behaviour of the superconductive sample, also analysing its possible interaction with the surrounding gravitational field. If the sample is put at a temperature $T$ slightly greater than $\Tc$ but sufficiently far from the transition point (mean field regime), the system can be described in terms of linearized time-dependent Ginzburg--Landau equations%
\footnote{Being the order parameter very small in the thermodynamic fluctuations regime, a linear order formulation can be exploited.
}.
The latter can be expressed in the gauge-invariant form as \cite{schmid1966time,hurault1969nonlinear,schmid1969diamagnetic}:
\begin{equation}
\Gamma\left(\hbar\,\partial_t-2\,i\,e\,\phi\right)\psi\:=\,
    \frac{1}{2m}\left(\hbar\,\nabla\phi-2\,i\,e\,\mathbf{A}\right)^2\psi
    +\alpha\,\psi\qquad\quad
(T>\Tc)\;\:.\quad
\label{eq:TDGL_linearized}
\end{equation}
where $\psi(\x,t)$ is the order parameter, $\phi(\x,t)$ the electric potential and $A(\x,t)$ is the vector potential. We also introduce the quantities:
\begin{equation}
\Gamma=\frac{\alpha}{\epsilon(T)}\,\frac{\pi}{8\,\text{k}_\textsc{b\,}\Tc}\,,\qquad\:
\epsilon(T)=\sqrt{\frac{T-\Tc}{\Tc}}\,,\qquad\:
\xi(T)=\frac{\xi_0}{\sqrt{\epsilon(T)}}\,,\qquad\:
\alpha=\frac{\hbar^{2}}{2\,m\,\xi(T)^{\,2}}\,,\quad
\label{eq:GLparam}
\end{equation}
$\xi_0$ being the BCS intrinsic coherence length, roughly characterizing the smallest size of a wave packet formed by superconducting charge carriers.
It plays a role analogous to the mean free path in the nonlocal electrodynamics of normal metals and is in general larger in metal superconductors.%
\footnote{%
In spite of the fact that the two electrons in a Cooper pair can be far apart from each other, other electrons belonging to different Cooper pairs are usually closer.
}
The temperature-dependent Ginzburg--Landau coherence length $\xi(T)$ provides a measure of the distance over which the order parameter can vary without undue energy increase, for a given temperature $T$. Alternatively, it can be thought of as a characterization of the distance from the surface over which the order parameter is close to its bulk value.\par
We now consider the following ansatz for the solution:
\begin{equation}
\psi(\x,t)\=f(\mathbf{x},t)\:\exp\big(i\,g(\x,t)\big)\:,
\end{equation}
and obtain from \eqref{eq:TDGL_linearized} the relations
\begin{subeqs}\label{eq:dtfdtg}
\begin{align}
\Gamma\,\hbar\,\dt[f]&\=
    \alpha\,f-\frac{1}{2}m\,\vs^{2}\,f+\frac{\hbar^{2}}{2m}\,\Delta f\;,
\label{eq:dtf}
\\[2.5\jot]
\Gamma\,\hbar\,f\,\dt[g]&\=
    2\,e\,\Gamma\,\phi\,f-\frac{\hbar^{2}}{2m}\,f\,
    \Delta g-2\,\hbar\;\vvs\cdot\nabla f\:.
\label{eq:dtg}
\end{align}
\end{subeqs}
The superfluid speed $\vvs$ has the form
\begin{equation}
\vvs\=\frac{1}{m}\left(\hbar\,\nabla g
                    +2\,\frac{e}{c}\,\mathbf{A}\right)\:,
\label {eq:vs}
\end{equation}
and the associated supercurrent density $\js$ reads
\begin{equation}
\js\:=-2\,\frac{e}{m}\,|\psi|^{2}
    \left(h\,\nabla g+2\,\frac{e}{c}\,\mathbf{A}\right)
    \:=-2\,e\,f^{2}\,\mathbf{v}_\text{s}\:.
\end{equation}

\paragraph{Thermodynamic fluctuations.}
The presence of a thermal energy of the order of \,$\sim \kB\,T$ implies that the system could fluctuate in different low-lying states with a non-zero probability. Let us then use $f_k$ to define the value of $f$ for a fluctuation of the wave vector $\kk$. The above \eqref{eq:dtfdtg} can be recast in the form
\begin{subeqs} \label{eq:dtfk_dtvs}
\begin{align}
\Gamma\,\hbar\,\dt[f_{k}]&\=
    \alpha\,f_{k}-\frac{\hbar^{2}}{2m}\,k^{2}\,f_{k}-\frac{1}{2}\,m\,\vs^2\,f_{k}\;,
\label{eq:dtfk}
\\[1.5\jot]
\dt[\vvs]&\:=\,-2\,\frac{e}{m}\,\E\;
\label{eq:dtvs}
\end{align}
\end{subeqs}
having used eqs.\ \eqref{eq:vs}, \eqref{eq:dtg} and
\begin{equation}
\nabla\phi=-\E-\dt[\A]\:.
\end{equation}
Eq.\ \eqref{eq:dtvs} can be easily integrated and the resulting expression for the superfluid speed can be used in \eqref{eq:dtfk} giving
\begin{equation}
\Gamma\,\hbar\,\dt[f_{k}]\=
    \left(\alpha-\frac{\hbar^{2}}{2m}\,k^{2}
        -2\,\frac{e^2}{m}\,E^{2}\,t^{2}\right)f_{k}\:.
\end{equation}
We then find for $f_{k}$ 
\begin{equation}
f_{k}(t)\=f_{k}(0)\,\exp\left(\frac{\left(\alpha-\frac{\hbar^2}{2m}k^{2}\right)t
    -\frac{2}{3}\,\frac{e^2}{m}\,E^{2}\,t^3}{\Gamma\,\hbar}\right)\:,
\end{equation}
with
\begin{equation}
f_{k}^{2}(0)=\frac{\kB\,T}{2\left(|\alpha|+\frac{\hbar^2}{2m}k^{2}\right)}\:,
\end{equation}
and the associated current density $\jsk(t)$ can be written as
\begin{equation}
\jsk(t)\=\frac{4\,e^{2}}{m}\,\E\,\;t\;f_{k}^{2}(0)\,
    \exp\left(2\,\frac{\left(\alpha-\frac{\hbar^2}{2m}k^{2}\right)t-\frac{2}{3}\,\frac{e^2}{m}\,E^{2}\,t^3}{\Gamma\,\hbar}\right)\:.
\end{equation}
Finally, the explicit expression for the physical supercurrent density $\js$ \cite{Ummarino:2019cvw} can be found integrating over $\kk$:
\begin{equation}
\js(t)\=
    \frac{1}{8\pi^{3}}\,\int_{0}^{+\infty}\!\!dk\:4\pi\,k^{2}\,\jsk(k,t)\:,
\label{eq:js}
\end{equation}
where we have considered a three-dimensional sample of dirty material, whose dimensions are larger than the correlation length.

\paragraph{Generalized EM fields.}
The above expression for the supercurrent density allows to extract the explicit form of the generalized electromagnetic fields and potentials characterizing the physical evolution of the system. First of all, the vector potential $\A(x,y,z,t)$ is obtained from
\begin{equation}
\A(x,y,z,t)\=\frac{\mu_0}{4\pi}\int
    \frac{\js(t')\;\,dx'\,dy'\,dz'}{\sqrt{(x-x')^{2}+(y-y')^{2}+(z-z')^{2}}}\;,
\label{eq:A_from_js}
\end{equation}
where $t'$ is the retarded time
\begin{equation}
t'\=t-\frac{\sqrt{(x-x')^{2}+(y-y')^{2}+(z-z')^{2}}}{c}\:.
\end{equation}
The generalized electric field $\E(x,y,z,t)$ \eqref{eq:genEMfields} is obtained from:
\begin{equation}
\E(x,y,z,t)\:=\,-\dt[\A(x,y,z,t)]+\frac{m}{e}\,\g\:.
\label{eq:E_from_A}
\end{equation}
As we can appreciate, the generalized gravito--Maxwell $\E(x,y,z,t)$ features two contributions. In particular, the second term is the standard, constant weak Earth's gravity contribution. On the other hand, the unconventional first term originates from the presence of the (non-constant) supercurrent density and can determine a local, additional contribution to the constant gravitational field $\g$. The final result clearly depends on the superconducting sample shape and dimensions, as well as on the space point (outside the sample) where the gravitational fluctuation is measured.

\subsection{Expected effects}\label{subsec:outside_exp_eff}
Let us now study in detail a suitable experimental setting to evaluate the proposed interplay. Here we consider a superconductive disk at a temperature higher but very close to $T_\text{c}$. The sample is kept in the normal state by a weak magnetic field, that is then turned off at the time $t=0$, where the superconductive transition occurs. The axis of the disk is aligned with the direction of the gravitational field, the bases being parallel to the ground.\par
The chosen temperature regime ($T\gtrsim\Tc$) corresponds to the thermodynamic fluctuations regime we discussed in the previous section, so that we can exploit the corresponding results for the supercurrent and generalized EM fields expressions, see \cref{eq:js,eq:A_from_js,eq:E_from_A}. We are interested in the gravitational correction along the axis of the disk, just above the upper base of cylindrical sample.\par
\begin{figure}[H]
\centering
\vspace{2em}
\includegraphics[width=0.85\textwidth,keepaspectratio]{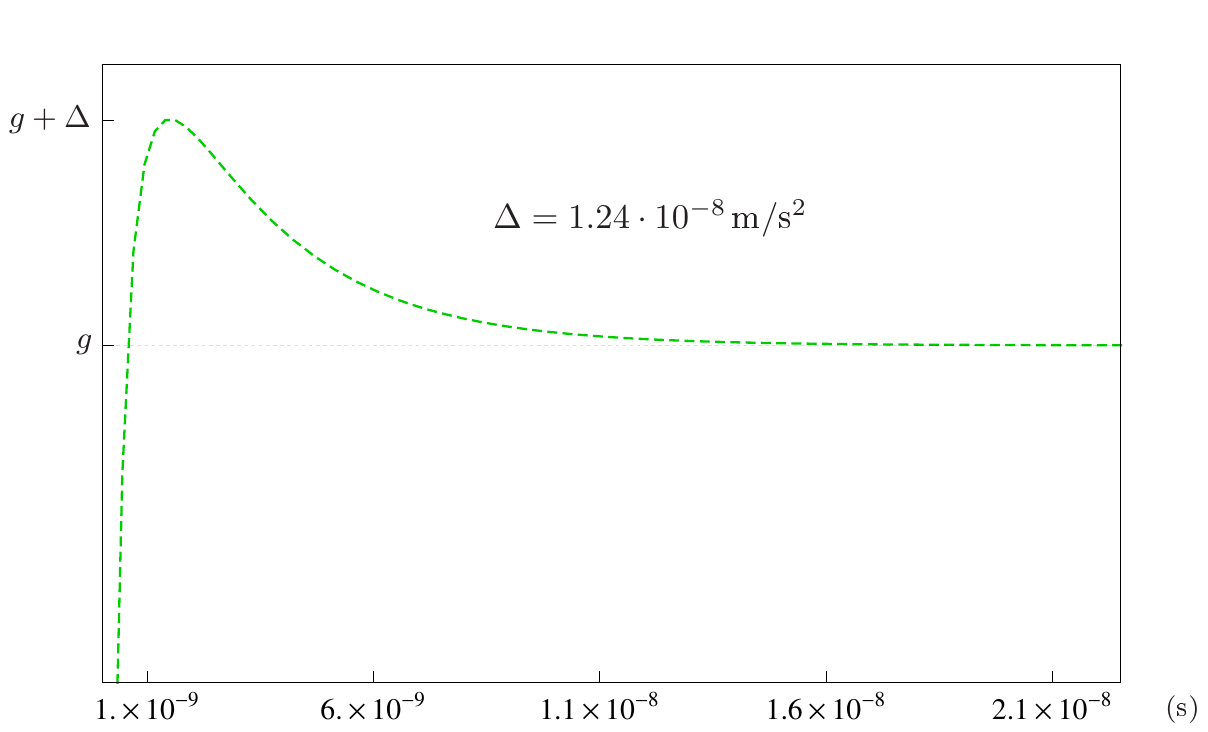}
\caption{\sloppy%
    Local gravitational field variation as a function of time for a In sample \big(\,${\xi_0=360\,\mathrm{nm}}$, \,${T_\text{c}=3.410\,\mathrm{K}}$, \,${\Delta T=10^{-3}\,\mathrm{K}}$ \,\cite{poole1999handbook}\,\big) measured along the axis of a superconductive disk at fixed distance ${d=0.1\,\mathrm{cm}}$ above the base surface. The disk radius is ${R=15\,\mathrm{cm}}$ and the disk thickness is ${h=3\,\mathrm{cm}}$.}
\label{fig:Intime}
\end{figure}

\begin{figure}[H]
\centering
\includegraphics[width=0.85\textwidth,keepaspectratio]{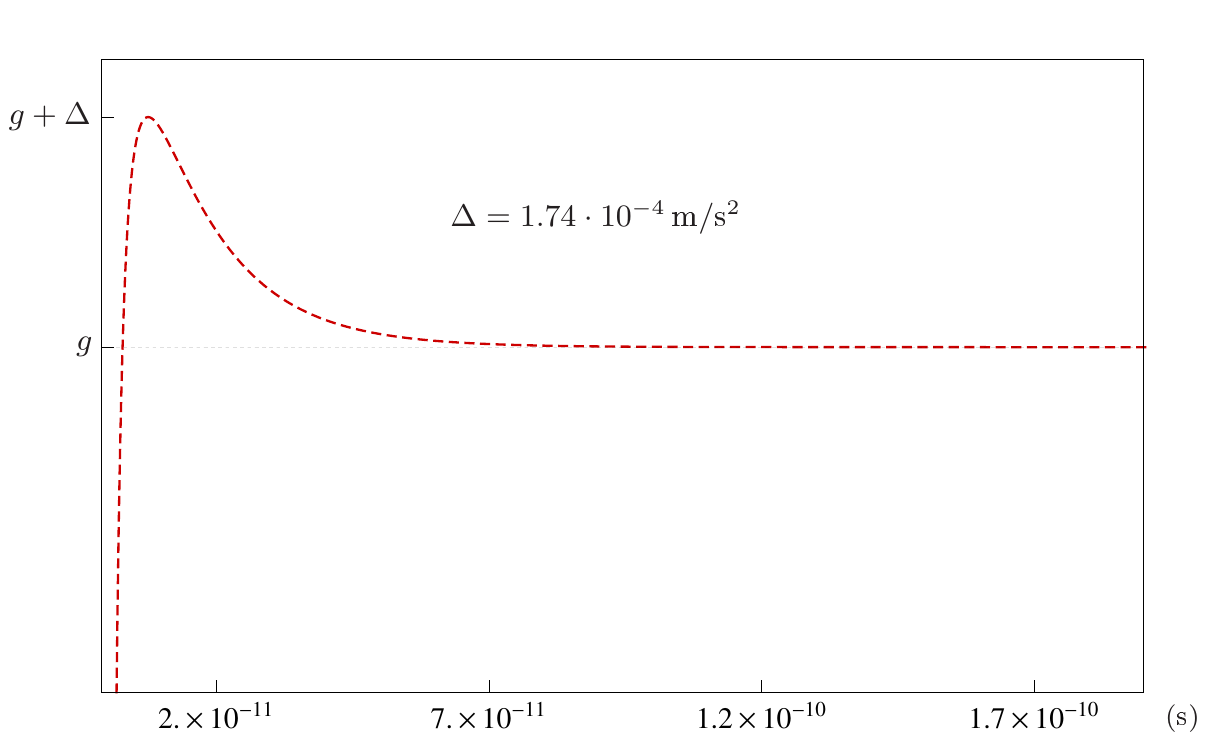}
\caption{\sloppy%
    Local gravitational field variation as a function of time for a Ba${}_{0.4}$K${}_{0.6}$Fe${}_{2}$As${}_{2}$ sample \big(\,${\xi_0=1.20\,\mathrm{nm}}$, \,${T_\text{c}=37.0\,\mathrm{K}}$, \,${\Delta T=0.1\,\mathrm{K}}$ \,\cite{welp2009anisotropic}\,\big) measured along the axis of a superconductive disk at fixed distance ${d=0.1\,\mathrm{cm}}$ above the base surface. The disk radius is ${R=15\,\mathrm{cm}}$ and the disk thickness is ${h=3\,\mathrm{cm}}$.}
\label{fig:BaKFeAstime}
\end{figure}

\smallskip
First, we consider the local alteration of the gravitational as a function of time. In Fig.\ \ref{fig:Intime} we show the computed effect for a In sample. The latter is a low-$\Tc$ metallic superconductor, then featuring a large intrinsic coherence length $\xi_0$. The same analysis is then performed in Fig.\ \ref{fig:BaKFeAstime} for a Ba${}_{0.4}$K${}_{0.6}$Fe${}_{2}$As${}_{2}$ sample, an high-$\Tc$ superconductor with small $\xi_0$. In both cases, the variation is measured along the disk axis, at a fixed distance $d$ above the base surface. We can note that the local gravitational field is initially reduced with respect to the unperturbed value; then it increases up to a maximum $g+\Delta$\, for $t=\tau_{0}$\, and it finally relaxes to the standard unperturbed value $g$.%
\footnote{%
We can also note that, for a very short time interval, the local field seems to change sign: this can be prevented by means of appropriate physical cutoffs, excluding the arbitrary growth of instabilities which would give rise to negative values \cite{Modanese:1995tx}.
}\par
\begin{figure}[H]
\centering
\includegraphics[width=0.85\textwidth,keepaspectratio]{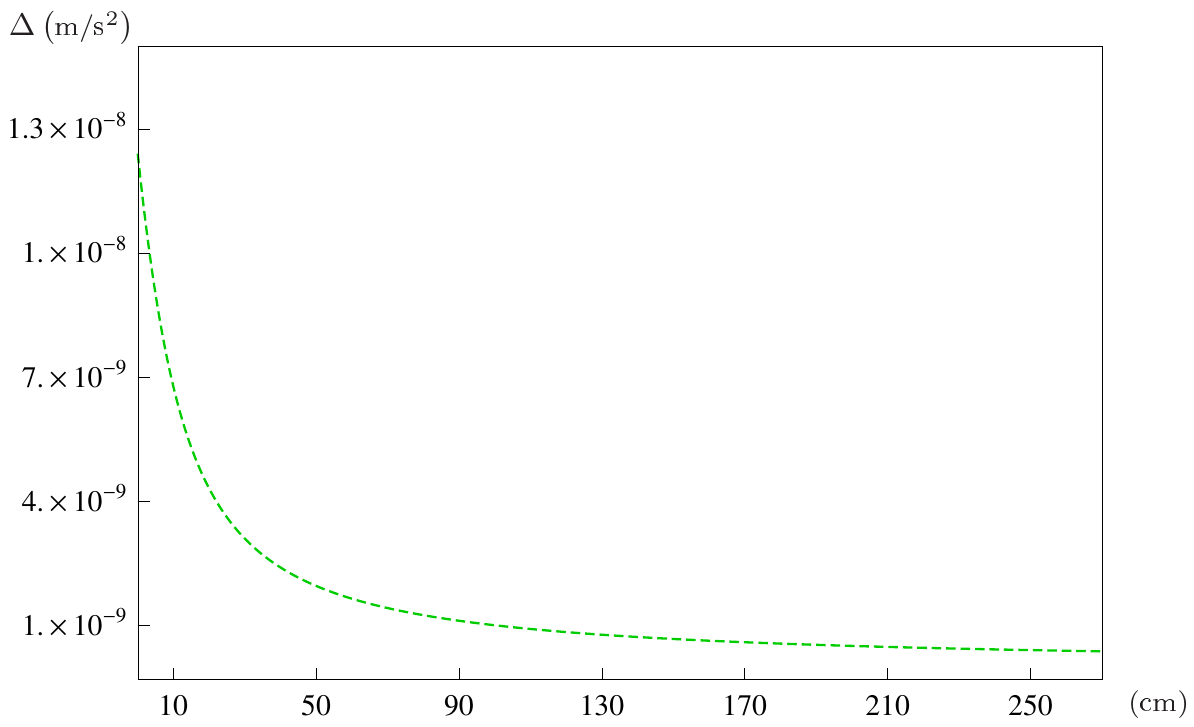}
\caption{\sloppy%
    Local gravitational field variation as a function of distance for the same In sample, measured along the disk axis above the base surface, at fixed time ${t=\tau_0=1.64\,\mathrm{ns}}$. The disk radius is ${R=15\,\mathrm{cm}}$ and the disk thickness is ${h=3\,\mathrm{cm}}$.}
\label{fig:Indistance}
\end{figure}

\begin{figure}[H]
\centering
\includegraphics[width=0.85\textwidth,keepaspectratio]{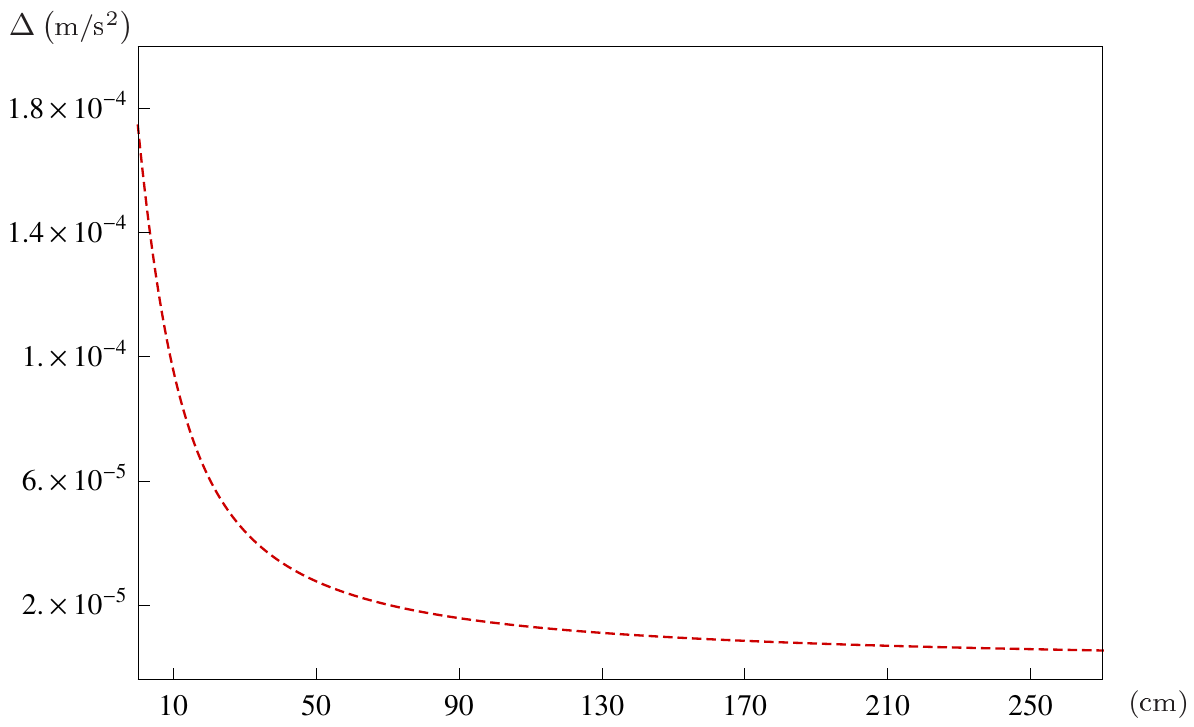}
\caption{\sloppy%
    Local gravitational field variation as a function of distance for the same Ba${}_{0.4}$K${}_{0.6}$Fe${}_{2}$As${}_{2}$ sample, measured along the disk axis above the base surface at fixed time ${t=\tau_0=7.50\times10^{-3}\,\mathrm{ns}}$. The disk radius is ${R=15\,\mathrm{cm}}$ and the disk thickness is ${h=3\,\mathrm{cm}}$.}
\label{fig:BaKFeAsdistance}
\end{figure}

\smallskip
We then focus on the local alteration as a function of the distance from the sample for fixed time. In particular, we choose to maximize the effect putting ourselves at $t=\tau_0$\,. In Figs.\ \ref{fig:Indistance} and \ref{fig:BaKFeAsdistance} it is shown the variation, measured along the axis of the disk above the base surface, for the same In and Ba${}_{0.4}$K${}_{0.6}$Fe${}_{2}$As${}_{2}$ samples. In both cases, the effect is stronger in the vicinity of the sample, as it seems reasonable.\par\smallskip
%
From a preliminary qualitative analysis, it is possible to show that the maximum perturbation value $\Delta$ of the local field is proportional to inverse of the coherence length,
\begin{equation}
\Delta\:\propto\:\xi(T)^{-1}\:.
\end{equation}
This suggests that a stronger affection can be obtained by using high--$\Tc$ superconductors (the latter featuring smaller coherence length) and can be appreciated comparing the strength of the perturbation for low and high--$\Tc$ superconducting samples in the presented Figures.\par
On the other hand, it is easily demonstrated that the maximal effect occurs after a time interval
\begin{equation}
\tau_{0}\,\propto\,(T-T_\text{c})^{-1}\:.
\end{equation}
This means that the time range in which the perturbation takes place can be extended keeping the sample at a temperature close to the transition temperature. From this point of view, if we want to be very close to the effective critical $\Tc$, it could be easier to consider a low--$\Tc$ sample, being the temperature transition range very narrow for the latter. However, this in turns results in a reduced alteration of the local field, since, close to $\Tc$, the Ginzburg--Landau coherence length $\xi(T)$ diverges, see eq.\ \eqref{eq:GLparam}.\par\smallskip
In light of the above discussion, an optimized experimental settings should involve a large high--$T_\text{c}$ superconducting sample at a temperature very close to $T_\text{c}$. The latter condition could help in extending the time range in which the effect takes place, while choosing an high--$T_\text{c}$ superconductor would determine an enhanced local alteration due to the short intrinsic coherence length. Finally, large dimensions for the sample give a larger integration range and a resulting stronger contribution.\par
The above considerations show how a careful arrangement of the experimental setup is very important, since the material parameters and the sample geometry, dimensions and temperature directly affect the magnitude of the interaction and the related time scales. In this regard, the very short time intervals in which the effect occurs complicate direct measurements.\par\medskip

\bigskip
\bigskip
\bigskip

In the following chapter, we will consider the possible affection of the local gravitational field in the sample interior exploiting again the effective framework of the gravito--Maxwell formulation combined with the Ginzburg--Landau formalism. The analysis will suggest that, in the superfluid region, a slight affection of the local field could take place, as we have discussed in Sect.\ \ref{sec:Intro} considering the formal quantum gravity point of view.
A possibility to enhance the effect comes from the presence of suitable electric and magnetic fields, determining the formation of moving vortices and giving rise to a further interaction with the local gravitational field.
\par\bigskip

\section{Affecting the field inside the sample. Vortex lattice}
Now we want to consider the possible alteration of the local static gravitational field in the region inside the superfluid. To this end, we will exploit the time-dependent Ginzburg--Landau equations for the supercondensate order
parameter, looking for analytic solutions in the weak field condition.
First, we will restrict to the simpler case of an isolated isotropic superconductor immersed in the Earth's gravity in the absence of external EM fields.
Then, we will analyse a more complicated setup, switching on suitable electric and magnetic fields: this will give rise to the formation of a vortex lattice inside the superfluid, possibly determining stronger effects for the proposed interplay.

\subsection{Time-dependent Ginzburg--Landau formulation}
Let us consider the case of a superconducting sample on the Earth surface. We already pointed out that the situation leads to the appearance of effective, generalized Maxwell fields. In particular, the local static weak gravitational field is treated as the gravitational component of the generalized gravitoelectric field, exploiting the formal analogy discussed in the previous Sect.\ \ref{sec:linearizedgravity}.\par
The chosen physical system can be characterized in terms of time-dependent Ginzburg--Landau equations (TDGL). The latter are derived minimizing the total Gibbs free energy of the system \cite{tinkham2004introduction,
ketterson1999superconductivity,DeGennes2018superconductivity}, and can be written in a general explicit form as \cite{ullah1991effect,tang1995time,du1996high,lin1997ginzburg,fleckinger1998dynamics,
kopnin1999time,ghinovker1999explosive}:
\begin{subeqs} \label{eq:TDGL}
\begin{align}
&\frac{\hbar^2}{2\,m}\left(i\,\nabla
        +\frac{2\,e}{\hbar}\,\A\right)^{\!2}\psi
      \,-\,a\,\psi\,+\,b\,\abs{\psi}^2\psi
    \:=\,-\,\frac{\hbar^2}{2\,m\,\mathcal{D}}\left(\frac{\dd}{\dd t}
      \,+\,\frac{2\,i\,e}{\hbar}\,\phi\right)\,\psi\;,
\label{subeq:TDGL1}
\\[2\jot]
&\nabla\times\nabla\times\A\,-\,\nabla\times\B
    \=\mu_0\,\big(\jj_\text{n}+\jj_\text{s}\big)\,,
\end{align}
\end{subeqs}
where $\jj_\text{n}$ and $\jj_\text{s}$ are expressed as
\begin{equation}
\begin{split}
\jj_\text{n}&\:=-\,\sigma\left(\frac{\dd\A}{\dd t}+\nabla\phi\right)\:,
\\[3\jot]
\jj_\text{s}&\:=-\,i\,\hbar\,\frac{e}{m}\left(\psi^*\,\nabla\psi-\psi\,\nabla\psi^*\right)
    -\frac{4\,e^2}{m}\,\abs{\psi}^2\A\:.
\end{split}
\end{equation}
and correspond to the contributions of the normal current and supercurrent densities, respectively. In the above expressions, $\sigma$ is the conductivity in the normal phase, $\mathcal{D}$ is the diffusion coefficient, $\B$ is the applied field and the vector potential $\A$ is minimally coupled to $\psi$.
The coefficients $a$ and $b$ in \eqref{subeq:TDGL1} can be written as:
\begin{equation}
a\=a(T)\=a_{0}\,(T-\Tc)\,,\qquad\qquad
b\=b(\Tc)\,,\qquad\qquad
\end{equation}
where is $\Tc$ the critical temperature of the superconductor, while $a_0$ and $b$ are positive constant quantities. We can write consistent boundary and initial conditions for the system as
%
\begin{align}
  \left.
  \begin{aligned}
  \left(i\,\nabla\psi+\frac{2\,e}{\hbar}\,\A\,\psi\right)\cdot\mathbf{n}=0&  \cr
  \hfill\nabla\times\A\cdot\mathbf{n}=\B\cdot\mathbf{n}&  \\[1.5\jot]
  \hfill\A\cdot\mathbf{n}=0&
  \end{aligned}
  \;\;\right\} \; \text{on }\dd\Omega\times(0,t)\;,
\qquad\quad
  \left.
  \begin{aligned}
  \psi(x,0)&\=\psi_0(x) \cr
  \A(x,0)&\=\A_0(x)
  \end{aligned}
  \!\!\!\!\right\} \; \text{on }\Omega\;,\qquad
\label{eq:boundary}
\end{align}
%
where $\dd\Omega$ is the boundary of a smooth and simply connected domain in $\mathbb{R}^\ms{\textrm{N}}$.

\paragraph{Dimensionless TDGL.}
The above eqs.\ \eqref{eq:TDGL} can be recast in a useful dimensionless form. To this end, we define the following quantities:
\begin{equation} \label{eq:param}
\begin{split}
&\Psi^2(T)=\frac{\abs{a(T)}}{b}\,,\qquad\;
\xi(T)=\frac{\hbar}{\sqrt{2\,m\,|a(T)|}}\,,\qquad\;
\lambda(T)=\sqrt{\frac{b\,m}{4\,\mu_0\,|a(T)|\,e^2}}\,,\qquad\;
\kappa=\frac{\lambda(T)}{\xi(T)}\,,\quad
\\[2\jot]
&\tau(T)=\frac{\lambda^2(T)}{\mathcal{D}}\,,\qquad\;
\eta=\mu_0\,\sigma\,\mathcal{D}\,,\qquad\;\;
\Bc(T)=\sqrt{\frac{\mu_0\,\abs{a(T)}^2}{b}}=
    \frac{\hbar}{2\sqrt{2}\,e\,\lambda(T)\,\xi(T)}\,,
\end{split}
\end{equation}
where $\lambda(T)$, $\xi(T)$ and $\Bc(T)$ are the penetration depth, coherence length and thermodynamic critical field, respectively. We also introduce the dimensionless quantities
\begin{equation}
t'=\:\frac{t}{\tau}\,,\qquad\;
x'=\:\frac{x}{\lambda}\,,\qquad\;
y'=\:\frac{y}{\lambda}\,,\qquad\;
\psi'=\:\frac{\psi}{\Psi}\,,\qquad
\label{eq:dimlesscoord}
\end{equation}
and the new dimensionless fields and currents
\begin{equation}
\A'=\frac{\A\,\kappa}{\sqrt{2}\,\Bc\,\lambda}\,,\qquad
\phi'=\frac{\phi\,\kappa}{\sqrt{2}\,\Bc\,\mathcal{D}}\,,\qquad
\E'=\frac{\E\,\lambda\,\kappa}{\sqrt{2}\,\Bc\,\mathcal{D}}\,,\qquad
\B'=\frac{\B\,\kappa}{\sqrt{2}\,\Bc}\,,\qquad
\jj'=\frac{\jj\,\mu_0\,\lambda\,\kappa}{\sqrt{2}\,\Bc}\,.
\label{eq:dimlessfields}
\end{equation}
We then insert the above \cref{eq:dimlesscoord,eq:dimlessfields} in eqs.\ \eqref{eq:TDGL} (we also drop the primes for the sake of notational simplicity) and get the dimensionless TDGL equations in a bounded, smooth and simply connected domain in $\mathbb{R}^\ms{\textrm{N}}$ \cite{tang1995time,lin1997ginzburg}:
\begin{subeqs}\label{eq:dimlessTDGL}
\begin{align}
&\frac{\dd\psi}{\dd t} \,+\, i\,\phi\,\psi
    \,+\,\kappa^2\left(\abs{\psi}^2-1\right)\,\psi
    \,+\,\left(i\,\nabla+\A\right)^2 \psi\=0 \:,
\label{subeq:dimlessTDGLn1}
\\[2\jot]
&\nabla\times\nabla\times\A\,-\,\nabla\times\B
    \=\jj_\text{n}+\jj_\text{s}
    \:=\,-\,\eta\,\left(\frac{\dd\A}{\dd t}+\nabla\phi\right)
    -\frac{i}{2}\left(\psi^*\nabla\psi-\psi\nabla\psi^*\right)
    -\abs{\psi}^2\A\:,
\label{subeq:dimlessTDGLn2}
\end{align}
\end{subeqs}
while the boundary and initial conditions \eqref{eq:boundary} in the dimensionless form read
\begin{align}
  \left.
  \begin{aligned}
  \left(i\,\nabla\psi+\A\,\psi\right)\cdot\mathbf{n}=0&\cr
  \nabla\times\A\cdot\mathbf{n}=\B\cdot\mathbf{n}&\cr
  \A\cdot\mathbf{n}=0&
  \end{aligned}
  \!\!\!\right\} \; \text{on }\dd\Omega\times(0,t)\;;
\qquad\quad
  \left.
  \begin{aligned}
  \psi(x,0)&\=\psi_0(x)\cr
  \A(x,0)&\=\A_0(x)
  \end{aligned}
  \!\!\!\!\right\} \; \text{on }\Omega\;.\qquad
\label{eq:dimlessboundary}
\end{align}
\smallskip

\subsection{Isolated superconductor in weak gravitational field}
Let us now now try to solve the above equations for a superconductor immersed in the Earth's static gravity in the absence of external electromagnetic fields \eqref{eq:genEMfields}:
\begin{equation}
\Ee=0\,,\quad\; \Be=0
\qquad\Longrightarrow\qquad
\E=\frac{m}{e}\,\Eg\,,\quad\; \B=0\,,\qquad
\end{equation}
having also set to zero the $\Bg$ contribution, that is negligible in the Solar system \cite{mashhoon1989detection,ljubivcic1992proposed}.\par\smallskip

\paragraph{Solving TDGL equations.}
A convenient gauge choice for subsequent calculations turns out to be $\phi=0$, i.e.\  the vanishing of the scalar potential.%
\footnote{%
Clearly, any alternative gauge shall not influence any physical results, being the equations gauge-invariant.
}
From a physical point of view, this choice also reflects the absence of localized charges inside the superfluid, while contributions to the total gravitational field originating from the sample mass are clearly totally irrelevant. The dimensionless TDGL then explicitly read \cite{Ummarino:2021vwc}:
%
\begin{subeqs}\label{eq:TDGLphi0}
\begin{align}
\frac{\partial\psi}{\partial t}\:&=\,
    -\left(i\,\nabla+\A\right)^{2}\psi\-\kappa^2\left(\abs{\psi}^2-1\right)\psi\:,
\label{subeq:TDGLphi01}
\\[2\jot]
\eta\,\frac{\partial\A}{\partial t}\:&=
    \,-\,\nabla\times\nabla\times\A\,+\,\nabla\times\B
    \,-\,\abs{\psi}^2\left(\A-\nabla\theta\right)\,,
\label{subeq:TDGLphi02}
\end{align}
\end{subeqs}
where $\psi\equiv\psi(\x,t)$ is a complex function that we can write as
\begin{equation}
\psi\=\abs{\psi}\,\exp(i\,\theta)\=\Real\,\psi+i\,\Img\,\psi\=\psi_1+i\,\psi_2\:,
\end{equation}
so that \eqref{subeq:TDGLphi01} splits into two distinct equations for the real and imaginary parts $\psi_1$ and $\psi_2$.\par
\begin{figure}[H]
\captionsetup{skip=5pt,belowskip=15pt,font=small,labelfont=small,format=hang}
\centering
\includegraphics[width=0.9\textwidth,keepaspectratio]{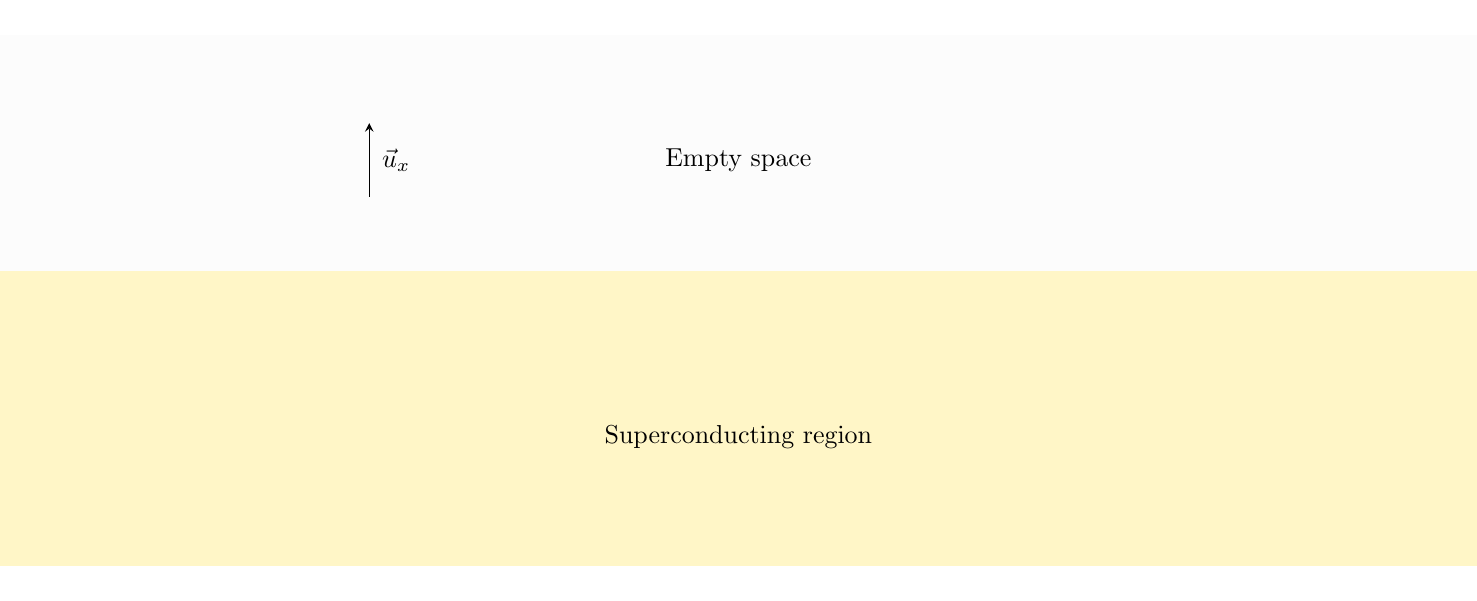}
\caption{Half-infinite superconductor approximation. The Earth's gravitational field is parallel to the $\ux$ direction.}
\label{fig:half_inf_plane}
\end{figure}

\medskip
Let us now restrict to a one-dimensional field configuration, so that one has
\begin{equation}
\nabla\rightarrow\,\partial/\partial x\,, \qquad\quad
\A\rightarrow\, A_x\equiv A\,.
\end{equation}
In this simplified framework, the \eqref{eq:TDGLphi0} read:
\begin{equation}
\begin{split}
\dt[\psi_1]&\=
    \ddx[\psi_1]+A\,\dx[\psi_2]+\psi_2\,\dx[A]-\psi_1\,A^2
    -\kappa^2\left(\abs{\psi_1}^2+\abs{\psi_2}^2-1\right)\psi_1\:,
\\[2\jot]
\dt[\psi_2]&\=
    \ddx[\psi_2]-A\,\dx[\psi_1]-\psi_1\,\dx[A]-\psi_2\,A^2
    -\kappa^2\left(\abs{\psi_1}^2+\abs{\psi_2}^2-1\right)\psi_2\:,
\\[2\jot]
\eta\,\dt[A]&\=-\left(\psi_2\,\dx[\psi_1]-\psi_1\,\dx[\psi_2]\right)
    -\left(\psi_1^2+\psi_2^2\right)A\:,
\end{split}
\end{equation}
since, in one dimension, \,$\nabla^2 A\,=\,\tfrac{\partial}{\partial x}\,\left(\nabla\cdot\A\right)$\, and then
\begin{equation}
\nabla\times\nabla\times\A\;=\;\nabla\,\left(\nabla\cdot\A\right)-\nabla^2 A
    \overset{\tts{1d}}{\;=\;}0\:.
\end{equation}
Then, let us consider an ideal, half-infinite superconductive region, see Fig.\ \ref{fig:half_inf_plane}. The $\ux$ direction is orthogonal to the superconducting separation surface, corresponding to the $yz$ plane and parallel to the ground, so that for $x>0$ we find an empty space, while the superfluid region is located at $x\leq0$. The whole setting is immersed in the Earth's uniform and static gravitational field, that is captured by the gravitoelectric component
\begin{equation}
\Eg^{\textsc{ext}}=-g\,\ux\:,
\label{eq:Egdim}
\end{equation}
$g$ being the standard gravity acceleration.\par\medskip
The \emph{dimensional} form of the gravitoelectric field \emph{inside} the superfluid region
\begin{equation}
\Eg=-\dt[\A_\text{g}(t)]\:,
\label{eq:EgdtA}
\end{equation}
while \eqref{eq:Egdim} suggests for the external (\emph{outside}) gravitational vector potential the form
\begin{equation}
\A_\text{g}^{\!\textsc{ext}}(t)=g\left(C+t\right)\,\ux\:,
\end{equation}
$C$ being a constant.\par
In the 1-D setup, the generalized external potential in the \emph{dimensionless} form reads
\begin{equation}
A^{\textsc{ext}}\=\frac{m}{e}\,A_\text{g}^{\textsc{ext}}\,\frac{\kappa}{\sqrt{2}\,\Bc\,\lambda}
    \=\gstar\left(c_1+t\right)\:,
\end{equation}
where we have dropped the primes for notational simplicity. Using \eqref{eq:param}, we can also explicitly write
\begin{equation}
c_1\,=\,\cfrac{C}{\tau}\;,\qquad\quad\;\;
\gstar\=\frac{m\,\kappa\,\lambda(T)\,g}{\sqrt{2}\,e\,\mathcal{D}\,\Bc(T)}\:\ll\:1\:.
\end{equation}
Next, we express $\psi_1$, $\psi_2$ and $A$ as:
\begin{subeqs}
\begin{align}
\psi_1(x,t)&\=\psi_{10}(x)+\gstar\,\gamma_1(x,t)\:,
\\[1.5\jot]
\psi_2(x,t)&\=\psi_{20}(x)+\gstar\,\gamma_2(x,t)\:,
\\[1.5\jot]
A(x,t)&\=\gstar\,\beta(x,t)\:,\label{subeq:A_gstBeta}
\end{align}
\end{subeqs}
where $\psi_{10}$ and $\psi_{20}$ characterize the unperturbed system and satisfy
\begin{subeqs}
\begin{align}
0&\=\frac{1}{\kappa^2}\,\ddx[\psi_{10}]+\psi_{10}
    -\psi_{10}\left(\psi_{10}^2+\psi_{20}^2\right)\:,
\label{subeq:psi10}
\\[1.5\jot]
0&\=\frac{1}{\kappa^2}\,\ddx[\psi_{20}]+\psi_{20}
    -\psi_{20}\left(\psi_{10}^2+\psi_{20}^2\right)\:.
\end{align}
\end{subeqs}
the $\psi_{10}$ and $\psi_{20}$ behaviour therefore being described by equations of the same type.\par
We now choose to set
\begin{equation}
\psi_{20}=0 \:\quad\Rightarrow\quad\:
\psi_{0}=\psi_{10}+i\,\psi_{20}=\psi_{10}\,\in\mathbb{R}\:,
\end{equation}
so that \eqref{subeq:psi10} reads
\begin{equation}
0\=\frac{1}{\kappa^2}\,\ddx[\psi_{10}]+\psi_{10}-\psi_{10}^3\:,
\end{equation}
and is solved by \cite{ketterson1999superconductivity}
\begin{equation}
\psi_{10}=\tanh\left(\frac{\kappa x}{\sqrt{2}}\right)\:.
\end{equation}
We are therefore left with the following set of equations:
\begin{subeqs}\label{eq:TDGL0gam1gam2beta}
\begin{align}
\dt[\gamma_1]&\=
    \ddx[\gamma_1]+\kappa^2\left(1-3\,\psi_{10}^2\right)\gamma_1\:,
\\[2\jot]
\dt[\gamma_2]&\=
    \ddx[\gamma_2]+\kappa^2\left(1-3\,\psi_{10}^2\right)\gamma_2
    -\beta\,\dx[\psi_{10}]-\psi_{10}\,\dx[\beta]\:,
\\[2\jot]
\eta\,\dt[\beta]&\=
    -\left(\gamma_2\,\dx[\psi_{10}]-\psi_{10}\,\dx[\gamma_2]\right)
    -\psi_{10}^2\,\beta\:,
\label{subeq:TDGL0beta}
\end{align}
\end{subeqs}
the last \eqref{subeq:TDGL0beta} implying that $\beta(x,t)$ does not depend on $\gamma_1(x,t)$.\par\smallskip
If we now decide to put ourselves away from borders, we can set
$\psi_{10}\simeq1$ in \eqref{eq:TDGL0gam1gam2beta}, resulting in
\begin{subeqs}
\begin{align}
\dt[\gamma_1]&\:\simeq\:
    \ddx[\gamma_1]-2\,\kappa^2 \gamma_1\:,
\\[2\jot]
\dt[\gamma_2]&\:\simeq\:
    \ddx[\gamma_2]-2\,\kappa^2 \gamma_2-\dx[\beta]\:,
\\[2\jot]
\eta\,\dt[\beta]&\:\simeq\:
    \dx[\gamma_2]-\beta\:.
\end{align}
\end{subeqs}
We then find for $\beta$ the solution
\begin{equation}
\beta(x,t)\=e^{-\tfrac{t}{\eta}}\,\left(b_1(x)
    +\frac{1}{\eta}\;\int^t_0\!\!dt\;e^{\tfrac{t}{\eta}}\;
    \dx[\gamma_2(x,t)]\right)\:.
\end{equation}
where $b_1(x)=c_1$, as it is implied by eq.\ \eqref{subeq:A_gstBeta} for $t\rightarrow0$
.\par\smallskip
%
Let us imagine that the sample transition to the superconducting state occurs at $t=0$. We also make the natural assumption that, before the transition, no alteration of the gravitational field takes place (material in the normal state), the gravitational field assuming the same value inside and outside the sample region for $t<0$.
This results in the following boundary and initial conditions:
\begin{equation}\label{eq:boundarycond}
\begin{split}
\begin{alignedat}{3}
&\psi(0,t)=0\,,\qquad\quad
&&\psi(x,0)=\psi_{10}(x)\,,\qquad\quad
&&\dx[\psi_{1}](x,0)=0\,,
\\[\jot]
&\gamma_{1}(0,t)=0\,,\qquad\quad
&&\gamma_{1}(x,0)=0\,,\qquad\quad
&&\dx[\gamma_{1}](x,0)=0\,,
\\[\jot]
&\gamma_{2}(0,t)=0\,,\qquad\quad
&&\gamma_{2}(x,0)=0\,,\qquad\quad
&&\dx[\gamma_{2}](x,0)=0\,,
\end{alignedat}
\end{split}
\end{equation}
together with the condition for $\beta$
\begin{equation}
\lim_{\;t\to 0}\,\gstar\,\dt[\beta](x,t)\=\gstar\:.
\end{equation}
implying that the interplay occurs only in the presence of a superconducting phase.\par
In order to fix the dimensionless constant $c_1$, we use \cref{subeq:A_gstBeta,eq:EgdtA,subeq:TDGL0beta} to write the relation between $E_\textrm{g}$ and $\beta$ as
\begin{equation}
\frac{E_\textrm{g}}{\gstar}\=-\dt[\beta]\=
    \frac{1}{\eta}\left(\gamma_2\,\dx[\psi_{10}]-\psi_{10}\,\dx[\gamma_2]\right)
    +\frac{\psi_{10}^2}{\eta}\,\beta\:.
\end{equation}
%
%
To satisfy the hypothesis that any affection of the gravitational field occurs only after the appearance of a superconducting phase ($t>0$), we assume
\begin{equation}
\lim_{\;t\to 0^-}\,\frac{E_\textrm{g}}{\gstar}\=1\:,
\end{equation}
while from the initial conditions in \eqref{eq:boundarycond} we also have
\begin{equation}
\lim_{\;t\to 0}\,\gamma_2(x,t)\=0\,,\qquad\quad
\lim_{\;t\to 0}\,\dx[\gamma_2](x,t)\=0\,.
\end{equation}
We then obtain 
\begin{equation}
t\rightarrow0: \quad\;
1\=\frac{\psi_{10}^2}{\eta}\:\beta(x,0)\=
    \frac{\psi_{10}^2}{\eta}\:\frac{A^{\textsc{ext}}(0)}{\gstar}\=
    \frac{\psi_{10}^2}{\eta}\:c_1
\;\quad\Longrightarrow\quad\;
c_1=\frac{\eta}{\psi_{10}^2}\:.\qquad
\end{equation}
This $c_1$ constant is ineffective in empty space, while it is responsible for the desired, unconventional effects in the presence of the superconductor.\par
Finally, we can write the final form for $\beta(x,t)$ away from borders $(\psi_{10}\simeq1,\, c_1\simeq\eta)$:
\begin{equation}
\beta(x,t)\=e^{-\tfrac{t}{\eta}}\,\left(\eta
    +\frac{1}{\eta}\;\int^t_0\!\!dt\;e^{\tfrac{t}{\eta}}\;
    \dx[\gamma_2(x,t)]\right)\:,
\end{equation}
from which we obtain the ratio
\begingroup%
\setlength{\belowdisplayskip}{15pt plus 5pt minus 1pt}%
\begin{equation}
\frac{\Eg}{\gstar}\=-\frac{\partial \beta(x,t)}{\partial t}\=
    \frac{1}{\eta}\,e^{-\tfrac{t}{\eta}}\,\left(\eta
    +\frac{1}{\eta}\;\int^t_0\!\!dt\;e^{\tfrac{t}{\eta}}\;
    \dx[\gamma_2(x,t)]\right)-\frac{1}{\eta}\frac{\partial \gamma_2(x,t)}{\partial x}\:.
\label{eq:ratio_Eg_g}
\end{equation}
\endgroup
The discussed formulation characterize more explicitly the proposed interplay between gravity and supercondensates in the presented, simplified setup. First, we see that the external gravitational vector potential seems to play a role in the superconducting transition: in particular, the external constant $c_1$ tends to assume fixed values depending on the specific properties of the sample undergoing the superconducting transition. On the other hand, we expect the back-reaction on the local gravitational to take place only after the transition itself, when the vector potential begins to ``perceive'' the presence of a superfluid phase.

\paragraph{Expected effects.}
The above \eqref{eq:ratio_Eg_g} for the ratio $\Eg/\gstar$ can be used to estimate the value of gravitational field inside the superconductor just after the superconducting phase transition:
\begin{equation}
t \simeq 0^{+}\::\qquad
    \frac{\Eg}{\gstar}\:\simeq\:1-\frac{t}{\eta}-\frac{1}{\eta}\frac{\partial \gamma_2(x,0^+)}{\partial x}\:.
\label{eq:Eg_g_t0}
\end{equation}
In the superconducting state, the alteration of the local field depends on physical characteristic of the involved sample. In particular, \eqref{eq:Eg_g_t0} shows that the relevant quantities are $\eta$, and the spatial derivative of $\gamma_2$.\par
In order to enhance the interaction, we should maximize the variation $\tfrac{\partial\gamma_2}{\partial x}$, an effect than can be achieved by introducing suitable disorder in the material sample.%
\footnote{%
This can be obtained, for instance, by means of chemical doping or proton
irradiation.
}
A maximized effect would also require small values for $\eta$.
The latter is proportional to the product of the diffusion coefficient $\mathcal{D}$ times the conductivity just above $\Tc$, see \eqref{eq:param}. This would suggest to consider materials that are bad conductors in the normal state and have low Fermi energies (for example, cuprates).\par
%
Finally, we have to take into account the (usually very small) time scales in which the effect occurs, expressed by the $\tau$ coefficient
\begin{equation}
\tau(T)=\frac{\lambda^2(T)}{\mathcal{D}}\,,\qquad\quad
\text{with}\quad\lambda(T)\simeq\frac{\lambda_0}{\sqrt{\frac{\Tc-T}{\Tc}}}\:.
\label{eq:tau_lambda}
\end{equation}
The latter can be maximized with a reduced diffusion coefficient and large penetration length, as occurs in superconducting cuprates with internal disorder.\par\smallskip
\sloppy
Performing measurements at a temperature close to $\Tc$ would give rise to enhanced effects: for example, in the case of \,Bi${}_2$Sr${}_2$CaCu${}_2$O${}_8$\, \big(\,${\Tc\simeq109\,\Kelv}$,\: ${\lambda_0\simeq500\,\nm}$,\: $\sigma^{-1}\simeq3.6\times10^{-6}\,\Omega\,\mt$,\: ${\mathcal{D}\simeq10^{-3}\,\mt^2/\s}$,\: ${\xi_0\simeq1.4\,\nm}$\, \cite{poole1999handbook}\,\big)\, for ${T\simeq105\,\Kelv}$ we find
\begin{equation}
T\simeq105\,\Kelv \::\qquad
    \tau\simeq6.8\times10^{-9}\,,\quad\:
    \eta\simeq3.5\times10^{-4}\,.\qquad\qquad
\end{equation}
This would determine a reduction of the local gravitational field of the order of $2\times10^{-5}$, see eq.\ \eqref{eq:Eg_g_t0} neglecting the last term.%
\footnote{%
Non irradiated high--$\Tc$ superconductors (like BSCCO) usually feature low disorder, resulting in reduced values for the spatial derivative of $\gamma_2$. 
}\par\smallskip
%
%
The above analysis shows how a perceptible affection of the local field inside the sample is possible even in a simplified setup (zero EM fields). Experimental difficulties may still arise from the short time intervals in which the effect manifests itself (see prevoius Subsect.\ \ref{subsec:outside_exp_eff}). Also in this case, an appropriate choice of the material parameters is essential, in order to enhance the interaction and extend the time ranges to workable scales.\par\smallskip
In the following section, we will analyse a more complicated setup involving external electric and magnetic fields, which in turn determine the presence of moving vortices. The new configuration will not only result in an additional affection of the local gravitational field, but also in the appearance of a new
component of the generalized electric field inside the sample, parallel to the superconductor surface.

\medskip

\subsection{Switching on EM fields. Vortex lattice}
We now consider a superconducting sample with finite thickness $L$ and very large dimensions along $\uz$ and $\uy$ directions. The sample is immersed in an external magnetic field $\B_0$ and has a square lattice of vortices, whose axes are directed along $\B_0$. We choose the latter as
\begin{equation}
\B_0=B_0\,\uz\:,
\end{equation}
together with a vector potential $\A$ of the form
\begin{equation}
\A=B_0\,x\,\uy\:.
\end{equation}
Here we decide to work in the Coulomb gauge $\nabla\cdot\A=0$, where
\begin{equation}
\nabla^2\A\=-\mu_0\left(\jj_\text{n}+\jj_\text{s}\right)\:.
\end{equation}
We also allow for the presence of a constant external (standard) electric field $\E_0^{{}^\mlt{(e)}}$ along the $\ux$ direction. Given the simultaneous presence of the Earth's static gravity, the situation gives rise to a \emph{generalized} static field $\E_0$ of the form
\begin{equation}
\E_0\=\E_0^{{}^\mlt{(e)}}+\E_0^{{}^\mlt{(g)}}
    \=\left(E_0^{{}^\mlt{(e)}}-E_0^{{}^\mlt{(g)}}\right)\,\ux
    \=\left(E_0^{{}^\mlt{(e)}}-\frac{m}{e}\,g\right)\,\ux\=E_0\,\ux\:,\quad
\end{equation}
and a related scalar potential
\begin{equation}
\phi_0\:=\,-\,E_0\,x\:.
\label{eq:phivort}
\end{equation}
As in the previous case, the transition takes place at $t=0$. In particular, for $t<0$ we also have  $T<\Tc$ and $B>\Bctwo$, while at $t=0$ we still have $T<\Tc$ but $B\simeq\Bctwo$. The external vector potential (outside the superfluid) is denoted by $\A_0$, and coincides with the inside value for $t<0$ (sample in the normal state and very weakly diamagnetic material).\par

\paragraph{Linearized TDGL.}
In the new setup with non-zero external EM fields, it is possible to write an analytic approximate solution of the TDGL \eqref{eq:TDGL} for the order parameter as
\begin{equation}
\psi(x,y,t)\=
    \sum_{n=-\infty}^{\infty}\! c_n\,\exp\left(i\,q\,n\left(y+\frac{E_0}{B_0}\:t\right)\!\right)\;\:
    \exp\left(-\frac{1}{2\,\xi(T)}\left(x-\frac{\hbar\,q\,n}{2\,e\,B_0}\right)^{\!2}
        +\,i\;\frac{e\,E_0\,\xi^2(T)}{\hbar\,\mathcal{D}}\left(x-\frac{\hbar\,q\,n}{2\,e\,B_0}\right)\right)\:.
\end{equation}
The expression is valid for an external magnetic field $B_0\lesssim\Bctwo$ and is then a solution of linearized TDGL equations \cite{kopnin1993flux,kopnin2001theory} describing the behaviour of an ordered vortex lattice, moving under the influence of the external $E_0$.\par
The above solution does not necessarily hold for different values of the magnetic field (for instance, $B_0\sim\Bcone$), where the order parameter values are bigger and the linearized approximation does not hold. Moreover, close to $\Bctwo$, the vortices are densely packed and the distance between them can be estimate to be of the order the coherence length $\xi(T)$. This can be then use to precisely characterize the vortex lattice, while this is not possible for generic values of $B_0$ \cite{ketterson1999superconductivity}.%
\footnote{%
The presence of the external electric fields causes vortices motion and determines dissipative phenomena even in the superconducting state; it is possible to prevent it and anchor the vortices (vortex pinning) by introducing defects in the sample, thus reducing or eliminating energy dissipation \cite{kopnin1993flux}.
}\par
From an experimental point of view, in high-$\Tc$ superconductors the formation of a square lattice seems to be energetically favourable, and in the following we will restrict to this possibility.%
\footnote{%
This is not the case for low-$\Tc$ superconductors, where a triangular lattice formation usually occurs.
}
We denote by $q$ the distance between adjacent vortices, that, for a square lattice, reads \cite{hoffmann2012ginzburg}
\begin{equation}
q\:\simeq\:\frac{2\pi}{\xi(T)}\:,
\end{equation}
and the general $c_n$ coefficients could be replaced by the correspondent $\cbox$ expression for the square lattice:
\begin{equation}
c_n\;\rightarrow\;\:\cbox=\frac{2\sqrt{2\pi}}{\xi^2(T)}\:,
\end{equation}
the $\cbox$ coefficients being then independent of $n$.
\par\bigskip

\paragraph{Dimensionless framework.}
Let us now consider the useful introduced dimensionless formulation. Working in the dimensionless version of the chosen Coulomb gauge \,$\nabla'\cdot\A'=0$\,, it is possible to write a first-order expression for the dimensionless order parameter satisfying a linearized form for adimensional TDGL equations \eqref{eq:dimlessTDGL} as \cite{Ummarino:2021tpz}:
\begin{equation}
\psi(x,y,t)\=
    \sum_{n=-\infty}^{\infty}\! \abs{c_n}\,\exp\left(i\,q\,n\left(y+\frac{E_0}{B_0}\:t\right)\!\right)\;\:
    \exp\left(-\frac{\kappa^2}{2}\left(x-n\,x_0\right)^2
        +i\,\frac{E_0}{\kappa}\left(x-n\,x_0\right)\right)\:,
\label{eq:psisum}
\end{equation}
with
\begin{equation}
\abs{\psi}^2\=\sum_{n=-\infty}^{\infty}\!\abs{c_n}^2\,\exp\left(-\kappa^2\left(x-n\,x_0\right)^2\right)\:.
\end{equation}
The equations for the vector potential components read
\begin{equation} \label{eq:vectpot0}
\begin{split}
\frac{\dd^2 A_x(x,t)}{\dd x^2}&\=
    \eta\,\left(\frac{\dd A_x(x,t)}{\dd t}-E_0\right) +\left(A_x(x,t)-\frac{E_0}{\kappa}\right)
        \sum_{n=-\infty}^{\infty}\!\abs{c_n}^2\,\exp\Big(-\kappa^2(x-n\,x_0)^2\Big)\:,
\\[1ex]
\frac{\dd^2 A_y(x,t)}{\dd x^2}&\=
    \eta\,\frac{\dd A_y(x,t)}{\dd t} +\sum_{n=-\infty}^{\infty}\!\big(A_y(x,t)-2\,\pi\,\kappa\,n\,\big)\,\abs{c_n}^2\,\exp\Big(-\kappa^2(x-n\,x_0)^2\Big)\:,
\\[1ex]
\frac{\dd^2 A_z(x,t)}{\dd x^2}&\=
    \eta\,\frac{\dd A_z(x,t)}{\dd t} +\sum_{n=-\infty}^{\infty}\!\abs{c_n}^2\,\exp\Big(-\kappa^2(x-n\,x_0)^2\Big)\:.
\end{split}
\end{equation}
Let us now consider an expansion to linear order in $E_0$. In order to obtain a more explicit solution for the order parameter \eqref{eq:psisum}, we have to estimate the summations
\begin{equation}
\begin{split}
&\sum_{n=-\infty}^{\infty}\!\abs{c_n}^2\,\exp\left(-\kappa^2\left(x-n\,x_0\right)^2\right)\,,\qqquad
\\[1.25ex]
&\sum_{n=-\infty}^{\infty}\!n\,\abs{c_n}^2\,\exp\left(-\kappa^2\left(x-n\,x_0\right)^2\right)\,.
\end{split}
\label{eq:summations}
\end{equation}
Since we are interested in high-$\Tc$ superconductors featuring a square vortex lattice, we replace the general coefficients $c_n$ with the correspondent $\cbox$ that, in the considered framework, reads \cite{hoffmann2012ginzburg}
\begin{equation}
\cbox^2\=2\,\sqrt{2\pi}\,\kappa^2\:.
\end{equation}
being then a constant function of $\kappa=\lambda/\xi$\,.\par
For high-$\Tc$ superconductors, the $\kappa$ parameter is usually large, $\kappa^2\,\ms{\gtrsim}\,10^4$: this in turn implies for the above \eqref{eq:summations}
\begin{equation}
\begin{split}
&\sum_{n=-\infty}^{\infty}\!\abs{c_n}^2\,\exp\left(-\kappa^2\left(x-n\,x_0\right)^2\right)
    \=\cbox^2\,e^{-\kappa^2x^2}\sum_{n=-\infty}^{\infty}\!e^{-\kappa^2n^2x_0^2}\;e^{2\,x\,x_0\,n\,\kappa^2}
    \:\simeq\:\cbox^2\,e^{-\kappa^2x^2}\:,
\\[1.25ex]
&\sum_{n=-\infty}^{\infty}\!n\,\abs{c_n}^2\,\exp\left(-\kappa^2\left(x-n\,x_0\right)^2\right)
    \:\simeq\:0\:,
\end{split}
\end{equation}
where the summation on the first line receives a non-negligible contribution only from the $n=0$ term.\par
The equation \eqref{eq:vectpot0} for the vector potential can be now recast as
\begin{equation} \label{eq:vectpot1}
\begin{split}
\frac{\dd^2 A_x(x,t)}{\dd x^2}&\=
    \eta\,\left(\frac{\dd A_x(x,t)}{\dd t}-E_0\right) +\left(A_x(x,t)-\frac{E_0}{\kappa}\right)\,\cbox^2\,e^{-\kappa^2x^2}\:,
\\[1.25ex]
\frac{\dd^2 A_y(x,t)}{\dd x^2}&\=
    \eta\,\frac{\dd A_y(x,t)}{\dd t}+A_y(x,t)\,\cbox^2\,e^{-\kappa^2x^2}\:,
\\[1.25ex]
\frac{\dd^2 A_z(x,t)}{\dd x^2}&\=
    \eta\,\frac{\dd A_z(x,t)}{\dd t}+\cbox^2\,e^{-\kappa^2x^2}\:.
\end{split}
\end{equation}
Since we are considering high-$\Tc$ superconductors ($\kappa^2\,\ms{\gtrsim}\,10^4$) it is also possible to approximate
\begin{equation}
e^{-\kappa^2x^2}\:\simeq\:\frac{\sqrt{\pi}}{\kappa}\,\delta(x)\:,
\end{equation}
so that the above expressions read
\begin{subeqs}\label{eq:vectpot2}
\begin{align}
\frac{\dd A_x(x,t)}{\dd t}\:&\simeq\:
    \frac{1}{\eta}\,\frac{\dd^2 A_x(x,t)}{\dd x^2}
    -\left(A_x(x,t)-\frac{E_0}{\kappa}\right)\,\cbox^2\,\frac{\sqrt{\pi}}{\eta\,\kappa}\,\delta(x)+E_0\:,
\label{subeq:Ax}
\\[1.25ex]
\frac{\dd A_y(x,t)}{\dd t}\:&\simeq\:
    \frac{1}{\eta}\,\frac{\dd^2 A_y(x,t)}{\dd x^2}
    -A_y(x,t)\:\cbox^2\,\frac{\sqrt{\pi}}{\eta\,\kappa}\,\delta(x)\:,
\label{subeq:Ay}
\\[1.25ex]
\frac{\dd A_z(x,t)}{\dd t}\:&\simeq\:
    \frac{1}{\eta}\,\frac{\dd^2 A_z(x,t)}{\dd x^2}
    -\cbox^2\,\frac{\sqrt{\pi}}{\eta\,\kappa}\,\delta(x)\:.
\label{subeq:Az}
\end{align}
\end{subeqs}
The initial conditions for the vector potential components are:
%
%
\begin{equation}
A_x\left(x,\,0\right)=0\,,\qquad\quad
A_y\left(x,\,0\right)=B_0\,x\,,\qquad\quad
A_z\left(x,\,0\right)=0\,,\quad
\label{eq:initcond}
\end{equation}
and the generalized electric field $\E$ inside the superfluid is given by
\begin{equation}
\E\=-\frac{\dd\A}{\dd t}-\nabla\phi\:.
\label{eq:EdtAdxphi}
\end{equation}

\par

\paragraph{Averaged solutions.}
We now consider the spatial averaged effects, determined by the presence of generalize field and poptentials, inside the supercondensate region. This can be obtained integrating the vector potential components \eqref{eq:vectpot2} over the $x$-variable \cite{sanders2007averaging}.\par
First, we integrate eq.\ \eqref{subeq:Az} over $x$ in the interval $x\in[-L/2,\,L/2]$, obtaining
\begin{equation}
\frac{\dd\bar{A}_z(t)}{\dd t}
    \=-\cbox^2\,\frac{\sqrt{\pi}}{\eta\,\kappa\,L}\:,
\end{equation}
having introduced the averaged component
\begin{equation}
\bar{A}_z(t)\=\frac{1}{L}\int\limits_{{-L/2}}^{\;{L/2}}\!\!dx\:A_z(x,t)\:,
\label{eq:barAzint}
\end{equation}
and taking advantage of symmetric conditions for the first derivatives with respect to $x$. Let us also keep in mind that we are dealing with the \emph{dimensionless} quantities, having dropped the primes for the sake of notational simplicity.%
\footnote{%
In particular, the $x$ coordinate correspond to the dimensionless $x'$ of \eqref{eq:dimlesscoord}, while one would explicitly have for the dimensionless thickness $L'=L/\lambda$,\, $L$ being the physical thickness and $\lambda$ the penetration depth.
}
The above \eqref{eq:barAzint} is solved by
\begin{equation}
\bar{A}_z(t)\=-\cbox^2\,\frac{\sqrt{\pi}}{\eta\,\kappa\,L}\,t+\bar{A}_z(0)
    \=-\cbox^2\,\frac{\sqrt{\pi}}{\eta\,\kappa\,L}\,t\:,
\label{eq:barAz}
\end{equation}
where initial conditions \eqref{eq:initcond} implies $\bar{A}_z(0)=0$. The averaged, generalized electric field $\bar{E}_z$ component is then given by
\begin{equation}
\bar{E}_z\=\cbox^2\,\frac{\sqrt{\pi}}{\eta\,\kappa\, L}
    \=\frac{2\sqrt{2}\,\pi\,\kappa}{\eta\,L}\:.
\label{eq:avEz}
\end{equation}
having used \eqref{eq:EdtAdxphi} and \eqref{eq:phivort}.\par
The averaged differential equation for the $\bar{A}_y(t)$ component, defined in the same way as \eqref{eq:barAzint}, is obtained from \eqref{subeq:Ay} and reads
\begin{equation}
\frac{\dd\bar{A}_y(t)}{\dd t}
    \=-\bar{A}_y(t)\:\cbox^2\,\frac{\sqrt{\pi}}{\eta\,\kappa\,L}\:,
\end{equation}
having used the approximation $A_y(0,t)\simeq \bar{A}_y(t)$. The resulting averaged component reads
\begin{equation}
\bar{A}_y(t)\=\bar{A}_y(0)\:\exp\left(-\cbox^2\,\frac{\sqrt{\pi}}{\eta\,\kappa\,L}\:t\right)
    \=0\:,
\end{equation}
having again used initial condition \eqref{eq:initcond}. This also implies that the electric field $\bar{E}_y(t)$ component is vanishing,
\begin{equation}
\bar{E}_y(t)\=0\:.
\end{equation}
The equation for the vertical component comes from the \eqref{subeq:Ax} expression and reads
\begin{equation}
\frac{\dd \bar{A}_x(t)}{\dd t}
    \=-\left(\frac{\bar{A}_x(t)}{L}-\frac{E_0}{\kappa}\right)\,\cbox^2\,\frac{\sqrt{\pi}}{\eta\,\kappa}+E_0\:,
\end{equation}
Using again the approximation $A_x(0,t)\simeq \bar{A}_x(t)$ and the initial conditions \eqref{eq:initcond}, we find for the $\bar{A}_x(t)$ solution
\begin{equation}
\begin{split}
\bar{A}_x(t)&\=\bar{A}_x(0)\,\exp\left(-\cbox^2\,\frac{\sqrt{\pi}}{\eta\,\kappa\,L}\,t\right)
    +E_0\left(\frac{L}{\kappa}+\frac{\eta\,\kappa\,L}{\cbox^2\sqrt{\pi}}\right)\,
    \left(1-\exp\left(-\cbox^2\,\frac{\sqrt{\pi}}{\eta\,\kappa\,L}\,t\right)\,\right)\=
\\[1ex]
    &\=E_0\left(\frac{L}{\kappa}+\frac{\eta\,\kappa\,L}{\cbox^2\sqrt{\pi}}\right)\,
    \left(1-\exp\left(-\cbox^2\,\frac{\sqrt{\pi}}{\eta\,\kappa\,L}\,t\right)\,\right)
\end{split}
\end{equation}
Finally, the averaged $E_x(t)$ component along the vertical direction for the generalized electric field comes from formulas \eqref{eq:EdtAdxphi} and \eqref{eq:phivort} and reads
\begin{equation}\label{eq:Excorr}
\begin{split}
\bar{E}_x(t)&\=E_0
    \,-\,E_0\left(\frac{L}{\kappa}+\frac{\eta\,\kappa\,L}{\cbox^2\sqrt{\pi}}\right)\,
    \cbox^2\,\frac{\sqrt{\pi}}{\eta\,\kappa\,L}\;\exp\left(-\cbox^2\,\frac{\sqrt{\pi}}{\eta\,\kappa\,L}\,t\right)\=
\\[1ex]
&\=E_0
   \,-\,E_0\left(\frac{2\sqrt{2}\,\pi}{\eta}+1\right)\;\exp\left(-\frac{2\sqrt{2}\,\pi\,\kappa}{\eta\,L}\,t\right)\:.
\end{split}
\end{equation}

\par\smallskip

\paragraph{Expected effects.}
The analysis of the averaged effect inside the supercondensate region shows some interesting predictions.\par\smallskip
The first effect is the emergence of a new component of the (generalized) electric field, parallel to the superconductor surface and directed along the external applied magnetic field. The value of this new contribution is found using the (dimensionless) result \eqref{eq:avEz} together with formulas \eqref{eq:dimlessfields}, and in dimensional units reads
\begin{equation}
E_{z}\=\frac{4\pi\,\Bc(T)\,\mathcal{D}}{\eta\,L}\:.
\end{equation}
\sloppy
If we consider a \,Bi${}_2$Sr${}_2$Ca${}_3$Cu${}_3$O${}_{10}$\, sample \big(\,${\Tc\simeq107\,\Kelv}$,\: ${\lambda_0\simeq2.4\times10^{-7}\,\mt}$,\: ${\xi_0\simeq1\,\nm}$,\: ${\sigma^{-1}\simeq3.6\times10^{-6}\,\Omega\,\mt}$,\: ${\mathcal{D}\simeq10^{-3}\,\mt^2/\s}$\: \cite{weigand2010mixed,piriou2008effect}\,\big) of thickness $L=15\,\cm$ at a temperature $T=102\,\text{K}$, this would correspond to a resulting field
\begin{equation}
T\simeq102\,\Kelv \::\qquad\;
    E_{z}\=\frac{4\pi\,\Bc(T)\,\mathcal{D}}{\eta\,L}
         \=\frac{4\pi\,\Bc(T)}{\mu_0\,\sigma\,L}
         \:\simeq\:77\,\frac{\textrm{V}}{\mt}\;,\qquad\qquad
\end{equation}
with $\Bc(T)\simeq0.32\,\textrm{Tesla}$.\par\smallskip
The second expected effect is affection of the local gravitational field along the $x$ direction in the supercondensate region. The averaged effect is expressed by eq.\ \eqref{eq:Excorr}, from which it is possible to appreciate the predicted, temporary alteration of the local field.\par
\begin{figure}[H]
\centering
\vspace{2em}
\includegraphics[width=0.85\textwidth,keepaspectratio]{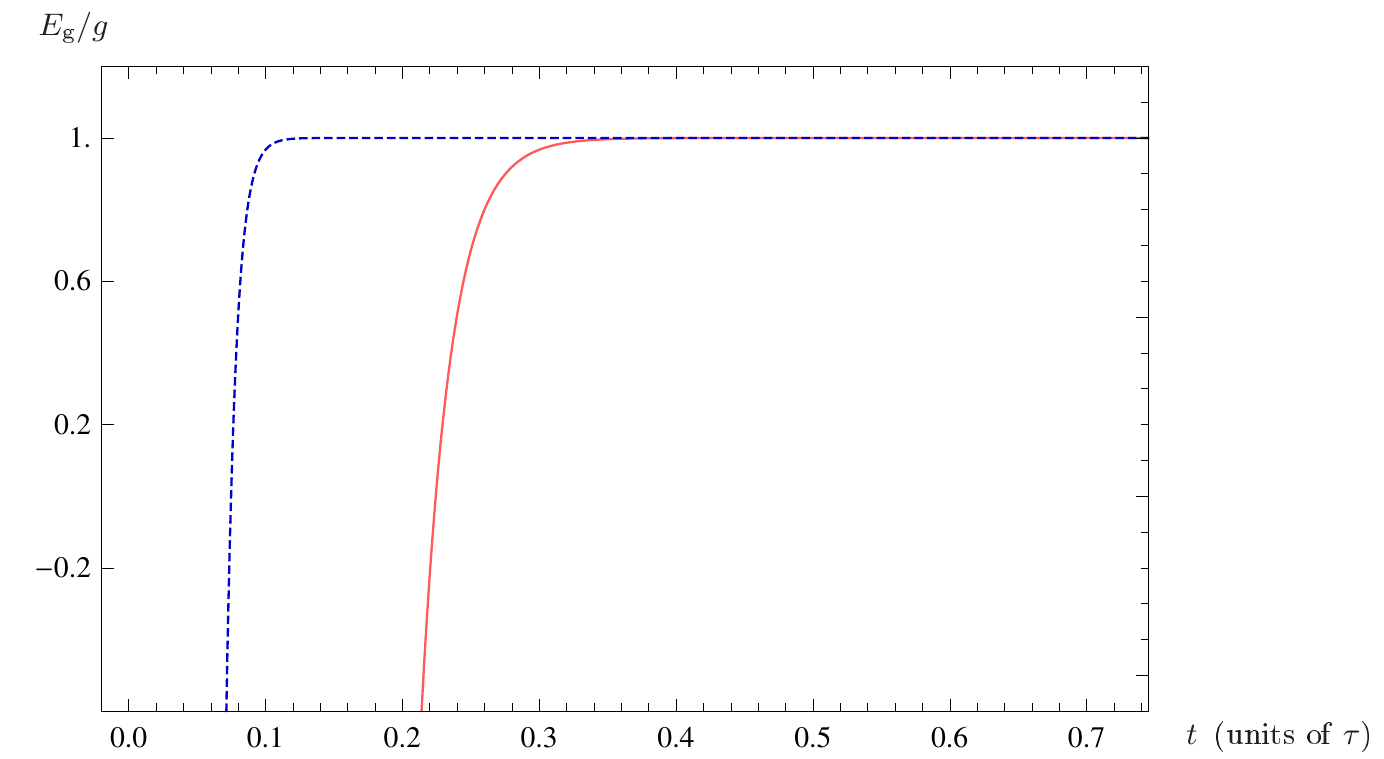}
\caption{\sloppy%
    Local field variation as a function of time for a \,Bi${}_2$Sr${}_2$Ca${}_3$Cu${}_3$O${}_{10}$\, sample \big(\,${\Tc\simeq107\,\Kelv}$,\; ${\lambda_0\simeq2.4\times10^{-7}\,\mt}$,\; ${\sigma^{-1}\simeq3.6\times10^{-6}\,\Omega\,\mt}$,\; ${\mathcal{D}\simeq10^{-3}\,\mt^2/\s}$,\; ${\xi_0\simeq1\,\nm}$\: \cite{weigand2010mixed,piriou2008effect}\,\big) at a temperature $T=102\,\text{K}$. The red solid line refers to a sample of thickness $L=15\,\cm$, while the blue dashed line shows the result for $L=5\,\cm$.}
\label{fig:BSCCO_outside}
\end{figure}

In Fig.\ \ref{fig:BSCCO_outside} we plot the field variation \emph{inside} the superfluid region for two samples of different dimensions. In analogy with the results of Section \ref{sec:outside} about the local alteration \emph{outside} the material, we can see that, for very short time scales, the gravitational field has a non-negligible reduction. Clearly, sample dimensions and chemical composition play a key role in maximizing the effect.

First, we can appreciate that larger samples (i.e.\ larger values of $L$) would determine an increase of the time scales in which the effect manifests itself. In the same way, \eqref{eq:Excorr} suggests that large values of the $\eta$ parameter, a sample characteristics, determine an analogous increase of time ranges. The analysis then shows that $L$ and $\eta$ parameters determine similar effects: choosing a sample of reduced dimension (small $L$) of disordered material (small $\eta$, bad conductors in the normal state) would results in very short time scales, with a slight enhancement of the effect. Again, appropriate physical cutoffs should come into play, preventing non-physical growth of instabilities within the supercondensate, which would in turn lead to local field alterations of arbitrary intensity.\par
%
Since experimental issues would reside in the very short observation times, it is useful to take advantage of effects determined by the internal disorder. The effects of the latter can be easily understand, since material disorder causes an increase of the $\lambda$ penetration depth. This, in turn, dictates an extension of the typical time scale $\tau$ of duration being $\tau\,\propto\,\lambda^2$,
see definitions \eqref{eq:param}, \eqref{eq:tau_lambda}.\par
Finally, if the system is put at temperatures very close to $\Tc$, there is again an increase of the $\lambda$ parameter and related larger time scales. In the latter case, however, the effects of thermal fluctuations should also be taken into account \cite{larkin2005theory}. The described effects occur in analogy with what we found in Sect.\ \ref{sec:outside} for the affection of the local field just outside the sample.\par

\bigskip

\section{Conclusions}
A deeper intertwining of different scientific areas has always proved to be a powerful tool for improving our understanding of many fascinating physical aspects of our world, see e.g.\ \cite{Zurek:1996sj,volovik1990superfluid,Volovik:2000ua, Baeuerle:1996zz,Ruutu:1995qz,Garay:1999sk,Jacobson:1998ms,
Barcelo:2000tg,novello2002artificial,Barcelo:2005fc,Carusotto:2008ep,
Mannarelli:2008jq,Boada:2010sh,Gallerati:2018dgm,
Capozziello:2018mqy,Andrianopoli:2019sip,Gallerati:2021htm,zaanen2015holographic,
Franz:2018cqi,Kolobov:2021ynv,Sbitnev:2022khm,Gallerati:2021rtp,lambiase2021interaction}. The intriguing existence of an interplay between gravity and superconductivity has been investigated by many researchers in the last decades, due to the enormous conceptual implications and many possible applications. In particular, the interaction has been theoretically predicted by numerous authors, with very different approaches and techniques. The phenomenon was then successfully tested in relation to the effects of gravitational perturbation on supercurrents and supercondensates, having used the latter as ``gravitational antennas'' for the detection of gravitational waves.\par
In this review, we mainly focused on the possible back-reaction exerted by the superfluid on the surrounding gravitational field, trying to provide qualitative and quantitative predictions about the extent of the proposed effect. Inspired by theoretical and experimental studies on gravity-induced generalized fields in superconductors, we studied the possible alterations exploiting a gravito--Maxwell formalism, integrated with the Ginzburg--Landau theory of phase transitions for superconducting systems. The latter formalism is a phenomenological theory, the superconducting materials being characterizing by parameters which, in principle, can be optimized to enhance specific effects.\par
Clearly, there is still a lot of work to be done in order to better define the ranges and magnitude of the effect, as well as to determine optimal situations from an experimental point of view. In this regard, a crucial role would be played by suitable samples geometry, external electromagnetic fields of adequate frequency and appropriate characteristics of the material. In the future, 2D materials with variable number of layers should also be taken into consideration, in order to exploit their peculiar properties \cite{clark2004unconventional,uchihashi2016two}.


%


\pagebreak
\hypersetup{linkcolor=blue}
\phantomsection 
\addtocontents{toc}{\protect\addvspace{3.5pt}}
\addcontentsline{toc}{section}{References} 
\bibliographystyle{mybibstyle}
\bibliography{bibliografia} 

\providecommand{\href}[2]{#2}\begingroup\begin{thebibliography}{100}

\bibitem{DeWitt:1966yi}
Bryce~S. DeWitt, \textit{``{Superconductors and gravitational drag}''}, Phys.\
  Rev.\ Lett.\ \textbf{16} (1966) 1092--1093.

\bibitem{papini1966london}
G.~Papini, \textit{``{London moment of rotating superconductors and
  Lense-Thirring fields of general relativity}''}, Il Nuovo Cimento B
  \textbf{45} (1966), n.~1, 66--68.

\bibitem{papini1967detection}
G.~Papini, \textit{``{Detection of inertial effects with superconducting
  interferometers}''}, Phys. Lett. A \textbf{24} (1967), n.~1, 32--33.

\bibitem{hirakawa1975superconductors}
H.~Hirakawa, \textit{``Superconductors in gravitational field''}, Phys. Lett. A
  \textbf{53} (1975), n.~5, 395--396.

\bibitem{Ciubotariu:1991zw}
C.D. Ciubotariu, \textit{``{Absorption of gravitational waves}''}, Phys. Lett.
  A \textbf{158} (1991) 27--30.

\bibitem{anandan1994relgra}
J.~Anandan, \textit{``Relativistic gravitation and superconductors''}, Class.
  Quant. Grav. \textbf{11} (1994), n.~6A, 23.

\bibitem{podkletnov1992possibility}
E.~Podkletnov and R.~Nieminen, \textit{``A possibility of gravitational force
  shielding by bulk \textrm{YBa${}_2$Cu${}_3$O${}_{7-\mathrm{X}}$}
  superconductor''}, Physica C: Superconductivity \textbf{203} (1992), n.~3-4,
  441--444.

\bibitem{Modanese:1995tx}
Giovanni Modanese, \textit{``{Theoretical analysis of a reported weak
  gravitational shielding effect}''}, Europhys.\ Lett. \textbf{35} (1996)
  413--418,
  [\href{http://arxiv.org/abs/hep-th/9505094}{\texttt{hep-th/9505094}}].

\bibitem{Modanese:1996zm}
Giovanni Modanese, \textit{``{Role of a `local' cosmological constant in
  Euclidean quantum gravity}''}, Phys. Rev. D \textbf{54} (1996) 5002--5009,
  [\href{http://arxiv.org/abs/hep-th/9601160}{\texttt{hep-th/9601160}}].

\bibitem{agop1996gravitational}
M.~Agop, C.G. Buzea, V.~Griga, C.D. Ciubotariu, C.~Stan and D.~Jatomir,
  \textit{``Gravitational paramagnetism, diamagnetism and gravitational
  superconductivity''}, Australian journal of physics \textbf{49} (1996), n.~6,
  1063--1074.

\bibitem{li1991effects}
N.~Li and D.G. Torr, \textit{``Effects of a gravitomagnetic field on pure
  superconductors''}, Phys. Rev. D \textbf{43} (1991), n.~2, 457.

\bibitem{ahmedov1999general}
B.J. Ahmedov, \textit{``General relativistic thermoelectric effects in
  superconductors''}, General Relativity and Gravitation \textbf{31} (1999),
  n.~3, 357--369.

\bibitem{agop2000some}
M.~Agop, P.D. Ioannou and F.~Diaconu, \textit{``Some implications of
  gravitational superconductivity''}, Progress of Theoretical Physics
  \textbf{104} (2000), n.~4, 733--742.

\bibitem{Modanese:2001wv}
Giovanni Modanese, \textit{``{Local contribution of a quantum condensate to the
  vacuum energy density}''}, Mod. Phys. Lett. A \textbf{18} (2003) 683--690,
  [\href{http://arxiv.org/abs/gr-qc/0107073}{\texttt{gr-qc/0107073}}].

\bibitem{Wu:2003aq}
Ning Wu, \textit{``{Gravitational shielding effects in gauge theory of
  gravity}''}, Commun. Theor. Phys. \textbf{41} (2004) 567--572,
  [\href{http://arxiv.org/abs/hep-th/0307225}{\texttt{hep-th/0307225}}].

\bibitem{hathaway2003gravity}
G.~Hathaway, B.~Cleveland and Y.~Bao, \textit{``Gravity modification experiment
  using a rotating superconducting disk and radio frequency fields''}, Physica
  C: Superconductivity \textbf{385} (2003), n.~4, 488.

\bibitem{Kiefer:2004hv}
C.~Kiefer and C.~Weber, \textit{``{On the interaction of mesoscopic quantum
  systems with gravity}''}, Annalen Phys. \textbf{14} (2005) 253--278,
  [\href{http://arxiv.org/abs/gr-qc/0408010}{\texttt{gr-qc/0408010}}].

\bibitem{Quach:2015qwa}
James~Q. Quach, \textit{``{Gravitational Casimir effect}''}, Phys. Rev. Lett.
  \textbf{114} (2015), n.~8, 081104,
  [\href{http://arxiv.org/abs/1502.07429}{\texttt{arXiv:1502.07429}}].
  [Erratum: Phys. Rev. Lett. \textbf{118} (2017) 139901].

\bibitem{Ummarino:2017bvz}
Giovanni~Alberto Ummarino and Antonio Gallerati, \textit{``{Superconductor in a
  weak static gravitational field}''}, Eur. Phys. J. C \textbf{77} (2017),
  n.~8, 549,
  [\href{http://arxiv.org/abs/1710.01267}{\texttt{arXiv:1710.01267}}].

\bibitem{atanasov2017geometric}
Victor Atanasov, \textit{``{The geometric field (gravity) as an
  electro-chemical potential in a Ginzburg-Landau theory of
  superconductivity}''}, Physica B: Condensed Matter \textbf{517} (2017)
  53--58.

\bibitem{atanasov2018gravitation}
Victor Atanasov, \textit{``{Gravitation at the Josephson junction}''}, Adv.
  Cond. Matt. Phys. \textbf{2018} (2018) 1--7.

\bibitem{Papini:1970cw}
G.~Papini, \textit{``{Superconducting and normal metals as detectors of
  gravitational waves}''}, Lett. Nuovo Cim. \textbf{4S1} (1970) 1027--1032.

\bibitem{adler1976long}
Ronald~J. Adler, \textit{``Long conductors as antennae for gravitational
  radiation''}, Nature \textbf{259} (1976), n.~5541, 296--297.

\bibitem{anandan1984relativistic}
J.~Anandan, \textit{``Relativistic thermoelectromagnetic gravitational effects
  in normal conductors and superconductors''}, Phys. Lett. A \textbf{105}
  (1984), n.~6, 280--284.

\bibitem{anandan1985detection}
J~Anandan, \textit{``Detection of gravitational radiation using superconducting
  circuits''}, Phys. Lett. A \textbf{110} (1985), n.~9, 446--450.

\bibitem{carelli1985coupling}
P.~Carelli, M.G. Castellano, C.~Cosmelli, V.~Foglietti and I.~Modena,
  \textit{``Coupling of a high-sensitivity superconducting amplifier to a
  gravitational-wave antenna''}, Phys. Rev. A \textbf{32} (1985), n.~6, 3258.

\bibitem{chan1987superconducting}
H.A. Chan and H.J. Paik, \textit{``Superconducting gravity gradiometer for
  sensitive gravity measurements. i. theory''}, Phys. Rev. D \textbf{35}
  (1987), n.~12, 3551.

\bibitem{mashhoon1989detection}
Bahram Mashhoon, Ho~Jung Paik and Clifford~M Will, \textit{``Detection of the
  gravitomagnetic field using an orbiting superconducting gravity gradiometer.
  theoretical principles''}, Phys. Rev. D \textbf{39} (1989), n.~10, 2825.

\bibitem{Preparata:1989gd}
Giuliano Preparata, \textit{``{\,`Superradiance' Effects in a Gravitational
  Antenna}''}, Mod. Phys. Lett. A \textbf{5} (1990) 1.

\bibitem{Peng:1990ca}
H.~Peng, \textit{``{The effects of gravitational waves on a superconducting
  antenna and its sensitivity}''}, Gen. Rel. Grav. \textbf{22} (1990) 33--43.

\bibitem{Peng:1990cc}
H.~Peng and D.G. Torr, \textit{``{The Electric field induced by a gravitational
  wave in a superconductor: A Principle for a new gravitational wave
  antenna}''}, Gen. Rel. Grav. \textbf{22} (1990) 53--59.

\bibitem{Peng:1991ua}
H.~Peng, Y.S. Chin and G.~Lind, \textit{``{Interaction between gravity and
  moving superconductors}''}, Gen. Rel. Grav. \textbf{23} (1991) 1231--1250.

\bibitem{peng1991electrodynamics}
H.~Peng, D.G. Torr, E.K. Hu and B.~Peng, \textit{``Electrodynamics of moving
  superconductors and superconductors under the influence of external
  forces''}, Phys. Rev. B \textbf{43} (1991), n.~4, 2700.

\bibitem{li2007detection}
F.~Li and R.M.L. Baker~Jr., \textit{``Detection of high-frequency gravitational
  waves by superconductors''}, Int. J. Mod. Phys. B \textbf{21} (2007),
  n.~18n19, 3274--3278.

\bibitem{Minter:2009fx}
Stephen~J. Minter, Kirk Wegter-McNelly and Raymond~Y. Chiao, \textit{``{Do
  Mirrors for Gravitational Waves Exist?}''}, Physica E \textbf{42} (2010) 234,
  [\href{http://arxiv.org/abs/0903.0661}{\texttt{arXiv:0903.0661}}].

\bibitem{inan2017interaction}
N.A. Inan, J.J. Thompson and R.Y. Chiao, \textit{``Interaction of gravitational
  waves with superconductors''}, Fortschritte der Physik \textbf{65} (2017),
  n.~6-8, 1600066.

\bibitem{inan2017new}
N.A. Inan, \textit{``A new approach to detecting gravitational waves via the
  coupling of gravity to the zero-point energy of the phonon modes of a
  superconductor''}, Int. J. Mod. Phys. D \textbf{26} (2017), n.~12, 1743031.

\bibitem{Hammad:2020xta}
Fay\c{c}al Hammad and Alexandre Landry, \textit{``{A simple superconductor
  quantum interference device for testing gravity}''}, Mod. Phys. Lett. A
  \textbf{35} (2020), n.~20, 2050171,
  [\href{http://arxiv.org/abs/2005.05798}{\texttt{arXiv:2005.05798}}].

\bibitem{overhauser1974experimental}
A.W. Overhauser and R.~Colella, \textit{``Experimental test of gravitationally
  induced quantum interference''}, Phys. Rev. Lett. \textbf{33} (1974), n.~20,
  1237.

\bibitem{Colella:1975dq}
R.~Colella, A.W. Overhauser and S.A. Werner, \textit{``{Observation of
  gravitationally induced quantum interference}''}, Phys. Rev. Lett.
  \textbf{34} (1975) 1472--1474.

\bibitem{Anandan:1977ra}
J.~Anandan, \textit{``{Gravitational and Rotational Effects in Quantum
  Interference}''}, Phys. Rev. D \textbf{15} (1977) 1448--1457.

\bibitem{Anandan:1978na}
J.~Anandan, \textit{``{Interference, Gravity and Gauge Fields}''}, Nuovo Cim. A
  \textbf{53} (1979) 221.

\bibitem{Anandan:1983fe}
J.~Anandan, \textit{``{Gravitationally Coupled Electromagnetic Systems and
  Quantum Interference}''}, Class. Quant. Grav. \textbf{1} (1984) L51.

\bibitem{Cai:1988sd}
Y.Q. Cai and G.~Papini, \textit{``{Particle Interferometry in Weak
  Gravitational Fields}''}, Class. Quant. Grav. \textbf{6} (1989) 407.

\bibitem{Ahluwalia:1996ev}
D.V. Ahluwalia and C.~Burgard, \textit{``{Gravitationally induced quantum
  mechanical phases and neutrino oscillations in astrophysical
  environments}''}, Gen. Rel. Grav. \textbf{28} (1996) 1161--1170,
  [\href{http://arxiv.org/abs/gr-qc/9603008}{\texttt{gr-qc/9603008}}].

\bibitem{Bhattacharya:1999na}
Tanmoy Bhattacharya, Salman Habib and Emil Mottola, \textit{``{Gravitationally
  induced neutrino oscillation phases in static space-times}''}, Phys. Rev. D
  \textbf{59} (1999) 067301.

\bibitem{Muntinga:2013pta}
H.~M\"untinga et~al., \textit{``{Interferometry with Bose-Einstein Condensates
  in Microgravity}''}, Phys. Rev. Lett. \textbf{110} (2013), n.~9, 093602,
  [\href{http://arxiv.org/abs/1301.5883}{\texttt{arXiv:1301.5883}}].

\bibitem{Asenbaum:2016djh}
Peter Asenbaum, Chris Overstreet, Tim Kovachy, Daniel~D. Brown, Jason~M. Hogan
  and Mark~A. Kasevich, \textit{``{Phase Shift in an Atom Interferometer due to
  Spacetime Curvature across its Wave Function}''}, Phys. Rev. Lett.
  \textbf{118} (2017), n.~18, 183602,
  [\href{http://arxiv.org/abs/1610.03832}{\texttt{arXiv:1610.03832}}].

\bibitem{Kiefer:1990pt}
Claus Kiefer and Tejinder~P. Singh, \textit{``{Quantum gravitational
  corrections to the functional Schrodinger equation}''}, Phys. Rev. D
  \textbf{44} (1991) 1067--1076.

\bibitem{Sakurai:2011zz}
Jun~John Sakurai and Jim Napolitano, \textit{``{Modern Quantum Mechanics}''};
  Cambridge University Press, Cambridge, UK (2017).

\bibitem{schiff1966gravitation}
L.I. Schiff and M.V. Barnhill, \textit{``Gravitation-induced electric field
  near a metal''}, Physical Review \textbf{151} (1966), n.~4, 1067.

\bibitem{witteborn1967experimental}
F.C. Witteborn and W.M. Fairbank, \textit{``Experimental comparison of the
  gravitational force on freely falling electrons and metallic electrons''},
  Phys. Rev. Lett. \textbf{19} (1967), n.~18, 1049.

\bibitem{witteborn1968experiments}
F.C. Witteborn and W.M. Fairbank, \textit{``Experiments to determine the force
  of gravity on single electrons and positrons''}, Nature \textbf{220} (1968),
  n.~5166, 436--440.

\bibitem{beams1968potentials}
J.W. Beams, \textit{``Potentials on rotor surfaces''}, Phys.\ Rev.\ Lett.
  \textbf{21} (1968), n.~15, 1093.

\bibitem{herring1968gravitationally}
C.~Herring, \textit{``Gravitationally induced electric field near a conductor,
  and its relation to the surface-stress concept''}, Phys. Rev. \textbf{171}
  (1968), n.~5, 1361.

\bibitem{peshkin1968gravityAP}
M.~Peshkin, \textit{``Gravity-induced electric field near a conductor''},
  Annals of Physics \textbf{46} (1968), n.~1, 1--11.

\bibitem{peshkin1969gravityPLA}
M.~Peshkin, \textit{``Gravity-induced electric field near a conductor''}, Phys.
  Lett. A \textbf{29} (1969), n.~4, 181--182.

\bibitem{craig1969direct}
Paul~P. Craig, \textit{``Direct observation of stress-induced shifts in contact
  potentials''}, Phys.\ Rev.\ Lett. \textbf{22} (1969), n.~14, 700.

\bibitem{rieger1970gravitationally}
T.J. Rieger, \textit{``Gravitationally induced electric field in metals''},
  Physical Review B \textbf{2} (1970), n.~4, 825.

\bibitem{leung1972electric}
M.C. Leung, \textit{``Electric fields induced by gravitational fields in
  metals''}, Il Nuovo Cimento B \textbf{7} (1972), n.~2, 220--224.

\bibitem{lockhart1977evidence}
J.M. Lockhart, F.C. Witteborn and W.M. Fairbank, \textit{``{Evidence for a
  temperature-dependent surface shielding effect in Cu}''}, Phys. Rev. Lett.
  \textbf{38} (1977), n.~21, 1220.

\bibitem{anandan1984new}
J.~Anandan, \textit{``New relativistic gravitational effects using
  charged-particle interferometry''}, Gen. Rel. Grav. \textbf{16} (1984), n.~1,
  33--41.

\bibitem{peng1983calculation}
H.~Peng, \textit{``On calculation of magnetic-type gravitation and
  experiments''}, General Relativity and Gravitation \textbf{15} (1983), n.~8,
  725--735.

\bibitem{Jain:1987zz}
A.K. Jain, J.E. Lukens and J.S. Tsai, \textit{``{Test for relativistic
  gravitational effects on charged particles}''}, Phys. Rev. Lett. \textbf{58}
  (1987) 1165--1168.

\bibitem{li1992gravitational}
N.~Li and D.G. Torr, \textit{``Gravitational effects on the magnetic
  attenuation of superconductors''}, Phys. Rev. B \textbf{46} (1992), n.~9,
  5489.

\bibitem{harris1991analogy}
E.G. Harris, \textit{``Analogy between general relativity and electromagnetism
  for slowly moving particles in weak gravitational fields''}, American Journal
  of Physics \textbf{59} (1991), n.~5, 421--425.

\bibitem{torr1993gravitoelectric}
D.G. Torr and N.~Li, \textit{``Gravitoelectric-electric coupling via
  superconductivity''}, Found. Phys. Lett. \textbf{6} (1993), n.~4, 371--383.

\bibitem{agop2000local}
M.~Agop, C.G. Buzea and P.~Nica, \textit{``Local gravitoelectromagnetic effects
  on a superconductor''}, Physica C: Superconductivity \textbf{339} (2000),
  n.~2, 120--128.

\bibitem{Tajmar:2002gm}
M.~Tajmar and C.J. De~Matos, \textit{``{Gravitomagnetic field of a rotating
  superconductor and of a rotating superfluid}''}, Physica C \textbf{385}
  (2003) 551--554,
  [\href{http://arxiv.org/abs/gr-qc/0203033}{\texttt{gr-qc/0203033}}].

\bibitem{Tajmar:2004ww}
M.~Tajmar and C.J. de~Matos, \textit{``{Extended analysis of gravitomagnetic
  fields in rotating superconductors and superfluids}''}, Physica C
  \textbf{420} (2005) 56,
  [\href{http://arxiv.org/abs/gr-qc/0406006}{\texttt{gr-qc/0406006}}].

\bibitem{ahmedov2005electromagnetic}
B.J. Ahmedov and V.G. Kagramanova, \textit{``Electromagnetic effects in
  superconductors in stationary gravitational field''}, International Journal
  of Modern Physics D \textbf{14} (2005), n.~05, 837--847.

\bibitem{dematos2008gravitational}
Clovis~Jacinto De~Matos, \textit{``Gravitational force between two electrons in
  superconductors''}, Physica C: Superconductivity \textbf{468} (2008), n.~3,
  229--232.

\bibitem{tajmar2008electrodynamics}
Martin Tajmar, \textit{``Electrodynamics in superconductors explained by proca
  equations''}, Phys. Lett. A \textbf{372} (2008), n.~18, 3289--3291.

\bibitem{Misner:1974qy}
Charles~W. Misner, K.S. Thorne and J.A. Wheeler, \textit{``{Gravitation}''}; W.
  H. Freeman, San Francisco, USA (1973).

\bibitem{Wald:1984rg}
Robert~M. Wald, \textit{``{General Relativity}''}; Chicago Univ. Pr., Chicago,
  USA (1984).

\bibitem{Ummarino:2019cvw}
Giovanni~Alberto Ummarino and Antonio Gallerati, \textit{``{Exploiting weak
  field gravity-Maxwell symmetry in superconductive fluctuations regime}''},
  Symmetry \textbf{11} (2019), n.~11, 1341,
  [\href{http://arxiv.org/abs/1910.13897}{\texttt{arXiv:1910.13897}}].

\bibitem{Ummarino:2020loo}
Giovanni~Alberto Ummarino and Antonio Gallerati, \textit{``{Josephson AC effect
  induced by weak gravitational field}''}, Class. Quant. Grav. \textbf{37}
  (2020), n.~21, 217001,
  [\href{http://arxiv.org/abs/2009.04967}{\texttt{arXiv:2009.04967}}].

\bibitem{Braginsky:1976rb}
Vladimir~B. Braginsky, Carlton~M. Caves and Kip~S. Thorne,
  \textit{``{Laboratory Experiments to Test Relativistic Gravity}''}, Phys.
  Rev. D \textbf{15} (1977) 2047.

\bibitem{ross1983london}
D.K. Ross, \textit{``{The London equations for superconductors in a
  gravitational field}''}, Journal of Physics A \textbf{16} (1983), n.~6, 1331.

\bibitem{thorne1988gravitomagnetism}
K.S. Thorne, \textit{``{Gravitomagnetism, jets in quasars, and the stanford
  gyroscope experiment}''}, in {\em Near Zero: New Frontiers of Physics},
  pp.~573--586, W.H. Freeman {\&} co., New York (1988).

\bibitem{peng1990new}
H.~Peng, \textit{``A new approach to studying local gravitomagnetic effects on
  a superconductor''}, General Relativity and Gravitation \textbf{22} (1990),
  n.~6, 609--617.

\bibitem{Ruggiero:2002hz}
Matteo~Luca Ruggiero and Angelo Tartaglia, \textit{``{Gravitomagnetic
  effects}''}, Nuovo Cim. B \textbf{117} (2002) 743--768,
  [\href{http://arxiv.org/abs/gr-qc/0207065}{\texttt{gr-qc/0207065}}].

\bibitem{Tartaglia:2003wx}
Angelo Tartaglia and Matteo~Luca Ruggiero, \textit{``{Gravitoelectromagnetism
  versus electromagnetism}''}, Eur. J. Phys. \textbf{25} (2004) 203--210,
  [\href{http://arxiv.org/abs/gr-qc/0311024}{\texttt{gr-qc/0311024}}].

\bibitem{Vieira:2016csi}
R.S. Vieira and H.B. Brentan, \textit{``{Covariant theory of gravitation in the
  framework of special relativity}''}, Eur. Phys. J. Plus \textbf{133} (2018)
  165, [\href{http://arxiv.org/abs/1608.00815}{\texttt{arXiv:1608.00815}}].

\bibitem{Behera:2017voq}
Harihar Behera, \textit{``{Comments on gravitoelectromagnetism of Ummarino and
  Gallerati in “Superconductor in a weak static gravitational field” vs
  other versions}''}, Eur. Phys. J. C \textbf{77} (2017), n.~12, 822,
  [\href{http://arxiv.org/abs/1709.04352}{\texttt{arXiv:1709.04352}}].

\bibitem{Giardino:2018ffd}
Sergio Giardino, \textit{``{A novel covariant approach to
  gravito-electromagnetism}''}, Braz. J. Phys. \textbf{50} (2020), n.~3,
  372--378,
  [\href{http://arxiv.org/abs/1812.07371}{\texttt{arXiv:1812.07371}}].

\bibitem{Sbitnev:2019iyz}
Valeriy~I. Sbitnev, \textit{``{Quaternion algebra on 4D superfluid quantum
  space-time. Gravitomagnetism}''}, Found. Phys. \textbf{49} (2019), n.~2,
  107--143,
  [\href{http://arxiv.org/abs/1901.09098}{\texttt{arXiv:1901.09098}}].

\bibitem{Gallerati:2020tyq}
Antonio Gallerati, \textit{``{Interaction between superconductors and weak
  gravitational field}''}, J. Phys. Conf. Ser. \textbf{1690} (2020), n.~1,
  012141, [\href{http://arxiv.org/abs/2101.00418}{\texttt{arXiv:2101.00418}}].

\bibitem{Williams:2020fgi}
L.~L. Williams and N.~Inan, \textit{``{Maxwellian mirages in general
  relativity}''}, New J. Phys. \textbf{23} (2021), n.~5, 053019,
  [\href{http://arxiv.org/abs/2012.08077}{\texttt{arXiv:2012.08077}}].

\bibitem{Gallerati:2021ops}
Antonio Gallerati, \textit{``{Local affection of weak gravitational field from
  supercondensates}''}, Phys. Scripta \textbf{96} (2021), n.~6, 064001.

\bibitem{Toth:2021dut}
Gabor~Zsolt Toth, \textit{``{Energy-momentum tensor and duality symmetry of
  linearized gravity in a Maxwellian formalism}''},
  \href{http://arxiv.org/abs/2108.02124}{\texttt{arXiv:2108.02124}}.

\bibitem{DeGennes2018superconductivity}
Pierre-Gilles De~Gennes, \textit{``Superconductivity of metals and alloys''};
  Taylor \& Francis Ltd, London, UK (2018).

\bibitem{tinkham2004introduction}
Michael Tinkham, \textit{``Introduction to superconductivity''}; Dover
  Publications Inc., New York, USA (2004).

\bibitem{ketterson1999superconductivity}
J.B. Ketterson and S.N. Song, \textit{``Superconductivity''}; Cambridge
  University Press, Cambridge, UK (1999).

\bibitem{Josephson:1962zz}
B.D. Josephson, \textit{``{Possible new effects in superconductive
  tunnelling}''}, Phys. Lett. \textbf{1} (1962) 251--253.

\bibitem{anderson1967josephson}
P.W. Anderson, \textit{``{The Josephson Effect and Quantum Coherence
  Measurements in Superconductors and Superfluids}''}, in {\em Progress in Low
  Temperature Physics}, vol.~5, pp.~1--43, Elsevier, Amsterdam, NL (1967).

\bibitem{Barone1982physics}
Antonio Barone and Gianfranco Patern{\`{o}}, \textit{``{Physics and
  Applications of the Josephson Effect}''}; John Wiley {\&} Sons, {New York,
  USA} (1982).

\bibitem{Feynman1965Feynman}
R.P. Feynman, R.B. Leighton and M.~Sands, \textit{``{The Josephson
  junction}''}, in {\em The Feynman Lectures on Physics}, vol.~III, ch.~21,
  sect.\ 21-9, Addison-Wesley Publ. Comp., New York, USA (1965).

\bibitem{fossheim2005superconductivity}
Kristian Fossheim and Asle Sudb{\o}, \textit{``Superconductivity: physics and
  applications''}; John Wiley {\&} Sons Ltd, New Jersey, USA (2004).

\bibitem{gorkov1959microscopic}
L.P. Gor’kov, \textit{``{Microscopic derivation of the Ginzburg-Landau
  equations in the theory of superconductivity}''}, Sov. Phys. JETP \textbf{9}
  (1959), n.~6, 1364--1367.

\bibitem{josephson1965supercurrents}
B.D. Josephson, \textit{``Supercurrents through barriers''}, Advances in
  Physics \textbf{14} (1965), n.~56, 419--451.

\bibitem{josephson1964coupled}
B.D. Josephson, \textit{``Coupled superconductors''}, Rev. Mod. Phys.
  \textbf{36} (1964), n.~1, 216.

\bibitem{Ambegaokar:1963zz1}
Vinay Ambegaokar and Alexis Baratoff, \textit{``{Tunneling Between
  Superconductors}''}, Phys.\ Rev.\ Lett. \textbf{10} (1963) 486.

\bibitem{Ambegaokar:1963zz2}
Vinay Ambegaokar and Alexis Baratoff, \textit{``{Tunneling Between
  Superconductors (Errata)}''}, Phys.\ Rev.\ Lett. \textbf{11} (1963) 104.

\bibitem{Saxena2009proximity}
Ajay~Kumar Saxena, \textit{``{The Proximity and Josephson Effects}''}, in {\em
  High-Temperature Superconductors}, pp.~147--198, Springer Berlin Heidelberg
  (2009).

\bibitem{Bardeen:1957mv}
John Bardeen, L.N. Cooper and J.R. Schrieffer, \textit{``{Theory of
  superconductivity}''}, Phys. Rev. \textbf{108} (1957) 1175--1204.

\bibitem{Ginzburg:1950sr}
V.L. Ginzburg and L.D. Landau, \textit{``{On the Theory of
  superconductivity}''}, Zh. Eksp. Teor. Fiz. \textbf{20} (1950) 1064--1082.

\bibitem{GL2009Springer}
V.L. Ginzburg and L.D. Landau, \textit{``{On the Theory of
  Superconductivity}''}, in {\em On Superconductivity and Superfluidity},
  pp.~113--137, Springer Berlin Heidelberg (2009).

\bibitem{ginzburg1961some}
V.L. Ginzburg, \textit{``Some remarks on phase transitions of the second kind
  and the microscopic theory of ferroelectric materials''}, Soviet Phys. Solid
  State \textbf{2} (1961) 1824--1834.

\bibitem{thouless1960perturbation}
David~J. Thouless, \textit{``Perturbation theory in statistical mechanics and
  the theory of superconductivity''}, Annals of Physics \textbf{10} (1960),
  n.~4, 553--588.

\bibitem{shier1966superconducting}
J.S. Shier and D.M. Ginsberg, \textit{``Superconducting transitions of
  amorphous bismuth alloys''}, Physical Review \textbf{147} (1966), n.~1, 384.

\bibitem{glover1967ideal}
R.E. Glover, \textit{``Ideal resistive transition of a superconductor''},
  Physics Letters A \textbf{25} (1967), n.~7, 542--544.

\bibitem{strongin1968curie}
M.~Strongin, O.F. Kammerer, J.~Crow, R.S. Thompson and H.L. Fine,
  \textit{``{\,`Curie-Weiss' behavior and fluctuation phenomena in the
  resistive transitions of dirty superconductors}''}, Phys. Rev. Lett.
  \textbf{20} (1968), n.~17, 922.

\bibitem{ferrell1967predicted}
R.A. Ferrell and H.~Schmidt, \textit{``Predicted critical behavior near the
  superconducting phase transition''}, Physics Letters A \textbf{25} (1967),
  n.~7, 544--545.

\bibitem{schmid1966time}
Albert Schmid, \textit{``A time dependent ginzburg-landau equation and its
  application to the problem of resistivity in the mixed state''}, Physik der
  Kondensierten Materie \textbf{5} (1966), n.~4, 302--317.

\bibitem{hurault1969nonlinear}
J.P. Hurault, \textit{``Nonlinear effects on the conductivity of a
  superconductor above its transition temperature''}, Physical Review
  \textbf{179} (1969), n.~2, 494.

\bibitem{schmid1969diamagnetic}
Albert Schmid, \textit{``Diamagnetic susceptibility at the transition to the
  superconducting state''}, Physical Review \textbf{180} (1969), n.~2, 527.

\bibitem{poole1999handbook}
Charles~K. Poole, Horacio~A. Farach and Richard~J. Creswick, \textit{``Handbook
  of superconductivity''}; Academic press, San Diego, USA (1999).

\bibitem{welp2009anisotropic}
U.~Welp, R.~Xie, A.E. Koshelev, W.K. Kwok, H.Q. Luo, Z.S. Wang, G.~Mu and
  Hai-Hu Wen, \textit{``{Anisotropic phase diagram and strong coupling effects
  in Ba${}_{1-x}$K${}_{x}$Fe${}_{2}$As${}_{2}$ from specific-heat
  measurements}''}, Phys. Rev. B \textbf{79} (2009), n.~9, 094505.

\bibitem{ullah1991effect}
S.~Ullah and A.T. Dorsey, \textit{``{Effect of fluctuations on the transport
  properties of type-II superconductors in a magnetic field}''}, Phys. Rev. B
  \textbf{44} (1991), n.~1, 262.

\bibitem{tang1995time}
Q.~Tang and S~Wang, \textit{``{Time dependent Ginzburg-Landau equations of
  superconductivity}''}, Physica D: Nonlinear Phenomena \textbf{88} (1995),
  n.~3-4, 139--166.

\bibitem{du1996high}
Qiang Du and Paul Gray, \textit{``{High-kappa limits of the time-dependent
  Ginzburg-Landau model}''}, SIAM Journal on Applied Mathematics \textbf{56}
  (1996), n.~4, 1060--1093.

\bibitem{lin1997ginzburg}
Fang-Hua Lin and Qiang Du, \textit{``{Ginzburg-Landau vortices: dynamics,
  pinning, and hysteresis}''}, SIAM Journal on Mathematical Analysis
  \textbf{28} (1997), n.~6, 1265--1293.

\bibitem{fleckinger1998dynamics}
Jacqueline Fleckinger-Pell{\'e}, Hans~G Kaper and Peter Tak{\'a}{\v{c}},
  \textit{``{Dynamics of the Ginzburg-Landau equations of
  superconductivity}''}, Nonlinear Analysis: Theory, Methods \& Applications
  \textbf{32} (1998), n.~5, 647--665.

\bibitem{kopnin1999time}
N.B. Kopnin and E.V. Thuneberg, \textit{``{Time-dependent Ginzburg-Landau
  analysis of inhomogeneous normal-superfluid transitions}''}, Phys. Rev. Lett.
  \textbf{83} (1999), n.~1, 116.

\bibitem{ghinovker1999explosive}
M.~Ghinovker, I.~Shapiro and B.~Ya Shapiro, \textit{``Explosive nucleation of
  superconductivity in a magnetic field''}, Phys. Rev. B \textbf{59} (1999),
  n.~14, 9514.

\bibitem{ljubivcic1992proposed}
A~Ljubi{\v{c}}i{\'c} and BA~Logan, \textit{``A proposed test of the general
  validity of mach's principle''}, Physics Letters A \textbf{172} (1992),
  n.~1-2, 3--5.

\bibitem{Ummarino:2021vwc}
G.~A. Ummarino and A.~Gallerati, \textit{``{Possible alterations of local
  gravitational field inside a superconductor}''}, Entropy \textbf{23} (2021),
  n.~2, 193,
  [\href{http://arxiv.org/abs/2102.01489}{\texttt{arXiv:2102.01489}}].

\bibitem{kopnin1993flux}
N.B. Kopnin, B.I. Ivlev and V.A. Kalatsky, \textit{``{The flux-flow Hall effect
  in type II superconductors. An explanation of the sign reversal}''}, J. Low
  Temp. Phys. \textbf{90} (1993), n.~1, 1--13.

\bibitem{kopnin2001theory}
Nikolai Kopnin, \textit{``Theory of nonequilibrium superconductivity''},.

\bibitem{hoffmann2012ginzburg}
K.H. Hoffmann and Qi~Tang, \textit{``{Ginzburg-Landau phase transition theory
  and superconductivity}''}; Springer Basel AG, Basel, Switzerland (2012).

\bibitem{Ummarino:2021tpz}
Giovanni~Alberto Ummarino and Antonio Gallerati, \textit{``{Superconductor in
  static gravitational, electric and magnetic fields with vortex lattice}''},
  Results Phys. \textbf{30} (2021) 104838,
  [\href{http://arxiv.org/abs/2110.07335}{\texttt{arXiv:2110.07335}}].

\bibitem{sanders2007averaging}
Jan~A. Sanders, Ferdinand Verhulst and James Murdock, \textit{``Averaging
  methods in nonlinear dynamical systems''},.

\bibitem{weigand2010mixed}
M.~Weigand, M.~Eisterer, E.~Giannini and H.W. Weber, \textit{``{Mixed state
  properties of Bi${}_2$Sr${}_2$Ca${}_2$Cu${}_3$O${}_{10+\delta}$ single
  crystals before and after neutron irradiation}''}, Phys. Rev. B \textbf{81}
  (2010), n.~1, 014516.

\bibitem{piriou2008effect}
A.~Piriou, Y.~Fasano, E.~Giannini and {\O}.~Fischer, \textit{``{Effect of
  oxygen-doping on Bi${}_2$Sr${}_2$Ca${}_2$Cu${}_3$O${}_{10+\delta}$ vortex
  matter: crossover from electromagnetic to Josephson interlayer coupling}''},
  Phys. Rev. B \textbf{77} (2008), n.~18, 184508.

\bibitem{larkin2005theory}
Anatoli Larkin and Andrei Varlamov, \textit{``{Theory of fluctuations in
  superconductors}''},.

\bibitem{Zurek:1996sj}
W.H. Zurek, \textit{``{Cosmological experiments in condensed matter
  systems}''}, Phys. Rept. \textbf{276} (1996) 177--221,
  [\href{http://arxiv.org/abs/cond-mat/9607135}{\texttt{cond-mat/9607135}}].

\bibitem{volovik1990superfluid}
G.E. Volovik, \textit{``Superfluid 3he-b and gravity''}, Physica B: Condensed
  Matter \textbf{162} (1990), n.~3, 222--230.

\bibitem{Volovik:2000ua}
G.E. Volovik, \textit{``{Superfluid analogies of cosmological phenomena}''},
  Phys. Rept. \textbf{351} (2001) 195--348,
  [\href{http://arxiv.org/abs/gr-qc/0005091}{\texttt{gr-qc/0005091}}].

\bibitem{Baeuerle:1996zz}
C.~Baeuerle, {\relax Yu}.~M. Bunkov, S.~N. Fisher, H.~Godfrin and G.~R.
  Pickett, \textit{``{Laboratory simulation of cosmic string formation in the
  early Universe using superfluid He-3}''}, Nature \textbf{382} (1996)
  332--334.

\bibitem{Ruutu:1995qz}
V.M.H. Ruutu, V.B. Eltsov, A.J. Gill, T.W.B. Kibble, M.~Krusius, Yu.G. Makhlin,
  B.~Placais, G.E. Volovik and Wen Xu, \textit{``{Big bang simulation in
  superfluid He-3-b: Vortex nucleation in neutron irradiated superflow}''},
  Nature \textbf{382} (1996) 334,
  [\href{http://arxiv.org/abs/cond-mat/9512117}{\texttt{cond-mat/9512117}}].

\bibitem{Garay:1999sk}
L.~J. Garay, J.~R. Anglin, J.~I. Cirac and P.~Zoller, \textit{``{Black holes in
  Bose-Einstein condensates}''}, Phys. Rev. Lett. \textbf{85} (2000)
  4643--4647,
  [\href{http://arxiv.org/abs/gr-qc/0002015}{\texttt{gr-qc/0002015}}].

\bibitem{Jacobson:1998ms}
T.~A. Jacobson and G.~E. Volovik, \textit{``{Event horizons and ergoregions in
  He-3}''}, Phys. Rev. D \textbf{58} (1998) 064021,
  [\href{http://arxiv.org/abs/cond-mat/9801308}{\texttt{cond-mat/9801308}}].

\bibitem{Barcelo:2000tg}
Carlos Barcelo, Stefano Liberati and Matt Visser, \textit{``{Analog gravity
  from Bose-Einstein condensates}''}, Class. Quant. Grav. \textbf{18} (2001)
  1137, [\href{http://arxiv.org/abs/gr-qc/0011026}{\texttt{gr-qc/0011026}}].

\bibitem{novello2002artificial}
{Novello, Mario and Visser, Matt and Volovik, Grigory E.},
  \textit{``{Artificial black holes}''}; {World Scientific}, {Singapore}
  ({2002}).

\bibitem{Barcelo:2005fc}
Carlos Barcelo, Stefano Liberati and Matt Visser, \textit{``{Analogue
  gravity}''}, Living Rev. Rel. \textbf{8} (2005) 12.

\bibitem{Carusotto:2008ep}
Iacopo Carusotto, Serena Fagnocchi, Alessio Recati, Roberto Balbinot and
  Alessandro Fabbri, \textit{``{Numerical observation of Hawking radiation from
  acoustic black holes in atomic Bose-Einstein condensates}''}, New J. Phys.
  \textbf{10} (2008) 103001,
  [\href{http://arxiv.org/abs/0803.0507}{\texttt{arXiv:0803.0507}}].

\bibitem{Mannarelli:2008jq}
Massimo Mannarelli and Cristina Manuel, \textit{``{Transport theory for cold
  relativistic superfluids from an analogue model of gravity}''}, Phys. Rev. D
  \textbf{77} (2008) 103014,
  [\href{http://arxiv.org/abs/0802.0321}{\texttt{arXiv:0802.0321}}].

\bibitem{Boada:2010sh}
O.~Boada, A.~Celi, J.~I. Latorre and M.~Lewenstein, \textit{``{Dirac Equation
  For Cold Atoms In Artificial Curved Spacetimes}''}, New J. Phys. \textbf{13}
  (2011) 035002,
  [\href{http://arxiv.org/abs/1010.1716}{\texttt{arXiv:1010.1716}}].

\bibitem{Gallerati:2018dgm}
Antonio Gallerati, \textit{``{Graphene properties from curved space Dirac
  equation}''}, Eur. Phys. J. Plus \textbf{134} (2019), n.~5, 202,
  [\href{http://arxiv.org/abs/1808.01187}{\texttt{arXiv:1808.01187}}].

\bibitem{Capozziello:2018mqy}
Salvatore Capozziello, Richard Pincak and Emmanuel~N. Saridakis,
  \textit{``{Constructing superconductors by graphene Chern-Simons
  wormholes}''}, Annals Phys. \textbf{390} (2018) 303--333.

\bibitem{Andrianopoli:2019sip}
L.~Andrianopoli, B.~L. Cerchiai, R.~D'Auria, A.~Gallerati, R.~Noris,
  M.~Trigiante and J.~Zanelli, \textit{``{$\mathcal{N}$-extended $D = 4$
  supergravity, unconventional SUSY and graphene}''}, JHEP \textbf{01} (2020)
  084, [\href{http://arxiv.org/abs/1910.03508}{\texttt{arXiv:1910.03508}}].

\bibitem{Gallerati:2021htm}
A.~Gallerati, \textit{``{Supersymmetric theories and graphene}''}, PoS
  \textbf{390} (2021) 662,
  [\href{http://arxiv.org/abs/2104.07420}{\texttt{arXiv:2104.07420}}].

\bibitem{zaanen2015holographic}
{Zaanen, Jan and Liu, Yan and Sun, Ya-Wen and Schalm, Koenraad},
  \textit{``{Holographic duality in condensed matter physics}''}; {Cambridge
  University Press}, {Cambridge, UK} ({2015}).

\bibitem{Franz:2018cqi}
M.~Franz and M.~Rozali, \textit{``{Mimicking black hole event horizons in
  atomic and solid-state systems}''}, Nature Rev. Mater. \textbf{3} (2018)
  491--501,
  [\href{http://arxiv.org/abs/1808.00541}{\texttt{arXiv:1808.00541}}].

\bibitem{Kolobov:2021ynv}
Victor~I. Kolobov, Katrine Golubkov, Juan~Ram\'on Mu\~noz~de Nova and Jeff
  Steinhauer, \textit{``{Observation of stationary spontaneous Hawking
  radiation and the time evolution of an analogue black hole}''}, Nature Phys.
  \textbf{17} (2021), n.~3, 362--367.

\bibitem{Sbitnev:2022khm}
Valeriy~I. Sbitnev, \textit{``{Quaternion Algebra on 4D Superfluid Quantum
  Space-Time. Dirac\textquoteright{}s Ghost Fermion Fields.}''}, Found. Phys.
  \textbf{52} (2022), n.~1, 19.

\bibitem{Gallerati:2021rtp}
Antonio Gallerati, \textit{``{Negative-curvature spacetime solutions for
  graphene}''}, J. Phys. Condens. Matter \textbf{33} (2021), n.~13, 135501,
  [\href{http://arxiv.org/abs/2101.03010}{\texttt{arXiv:2101.03010}}].

\bibitem{lambiase2021interaction}
Gaetano Lambiase and Giorgio Papini, \textit{``The interaction of spin with
  gravity in particle physics''}; Springer Nature, Cham, Switzerland AG (2021).

\bibitem{clark2004unconventional}
J.W. Clark, V.A. Khodel, M.V. Zverev and V.M. Yakovenko,
  \textit{``Unconventional superconductivity in two-dimensional electron
  systems with long-range correlations''}, Physics Reports \textbf{391} (2004),
  n.~3-6, 123--156.

\bibitem{uchihashi2016two}
Takashi Uchihashi, \textit{``Two-dimensional superconductors with atomic-scale
  thickness''}, Superconductor Science and Technology \textbf{30} (2016), n.~1,
  013002.

\end{thebibliography}\endgroup


\end{document}